\documentclass[rmp,aps,twocolumn,nofootinbib]{revtex4}
\usepackage{graphics}
\usepackage{epsfig}
 \usepackage[normalem]{ulem}
\usepackage{float}
\usepackage[usenames]{color}
\usepackage{amsmath}
\usepackage{booktabs}
 \usepackage{multirow}
\usepackage{blindtext}
\usepackage{longtable}
\usepackage{array}

\renewcommand{\d}{{\rm d}}

\newcommand{\del}[1]{{  }}
\renewcommand{\P}{p}

\newcommand{\zcf}{{z_{\mbox{\rm\tiny CF}}}}
\newcommand{\kT}{k_{\rm B}T}

\newcommand{\pcf}{{p_{\mbox{\rm\tiny CF}}}}
\newcommand{\pb}{{p_{\mbox{\rm\tiny b}}}}
\newcommand{\fcf}{{f_{\mbox{\rm\tiny CF}}}}
\newcommand{\fb}{{f_{\mbox{\rm\tiny b}}}}
\newcommand{\nucf}{{\nu_{\mbox{\rm\tiny CF}}}}
\newcommand{\nub}{{\nu_{\mbox{\rm\tiny b}}}}
\newcommand{\lambdacf}{{\lambda_{\mbox{\rm\tiny CF}}}}
\newcommand{\lambdab}{{\lambda_{\mbox{\rm\tiny b}}}}
 
\newcommand{\be}{\begin{equation}}
\newcommand{\ee}{\end{equation}}
\newcommand{\bea}{\begin{eqnarray}}
\newcommand{\eea}{\end{eqnarray}}

\def\inbar{\,\vrule height1.5ex width.4pt depth0pt}
\def\IR{\relax{\rm I\kern-.18em R}}
\def\IC{\relax\hbox{$\inbar\kern-.3em{\rm C}$}}

\begin{document}
\title{Modeling semiflexible polymer networks}

\author{C. P. Broedersz}
\affiliation{Lewis-Sigler Institute for Integrative Genomics and Joseph Henry Laboratories of Physics, Princeton University, Princeton, NJ 08544, USA}
\email{c.broedersz@lmu.de}

\author{F. C. MacKintosh}
\affiliation{Department of Physics and Astronomy, Vrije Universiteit, Amsterdam, The Netherlands}
\email{fcmack@gmail.com}

\begin{abstract}  
Here, we provide an overview of theoretical approaches to semiflexible polymers and their networks. Such semiflexible polymers have large bending rigidities that can compete with the entropic tendency of a chain to crumple up into a random coil. 
Many studies on semiflexible polymers and their assemblies have been motivated by their importance in biology. Indeed, crosslinked networks of semiflexible polymers form a major structural component of tissue and living cells. Reconstituted networks of such biopolymers have emerged as a new class of biological soft matter systems with remarkable material properties, which have spurred many of the theoretical developments discussed here. Starting from the mechanics and dynamics of individual semiflexible polymers, we review the physics of semiflexible bundles, entangled  solutions and disordered cross-linked networks. Finally, we discuss  recent developments on marginally stable fibrous networks, which exhibit critical behavior similar to other marginal systems such as jammed soft matter. 
\end{abstract}                                                                 
\date{\today}
\maketitle
\tableofcontents

\section{Introduction}
\label{sec:intro}
\begin{figure}[t]
\centering
\vspace{.1cm}
\includegraphics[width=8cm]{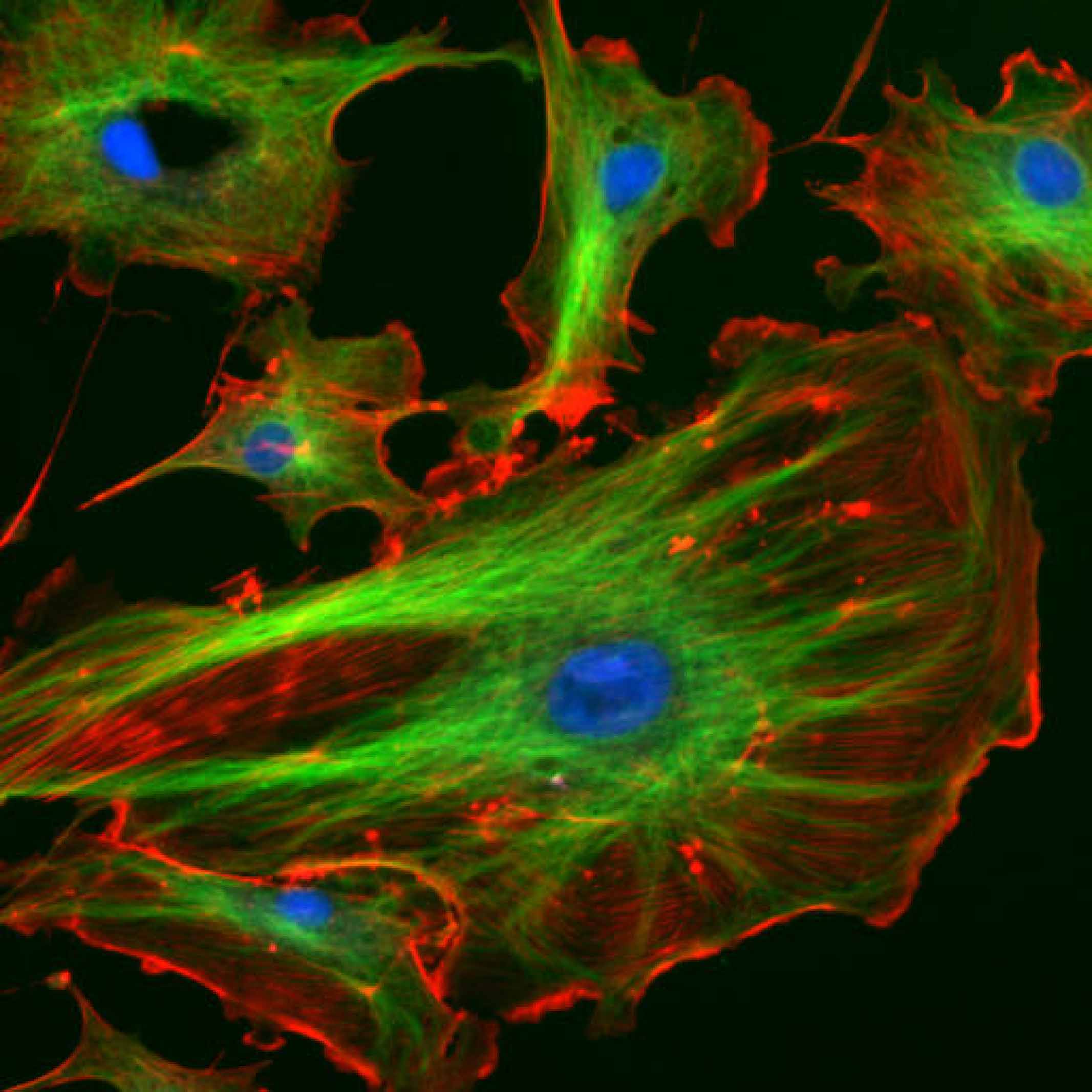}
\caption{Fluorescence microscopy image of bovine pulmonary artery endothelial cells. Nuclei are stained blue with DAPI, microtubules (green)  are labeled by an antibody bound to FITC and actin filaments (red) are labelled with phalloidin bound to TRITC. Source http://rsb.info.nih.gov/ij/images/ (example image from ImageJ (public domain)) }
\label{fig:Cliff}
\vspace{.5cm}
\includegraphics[width=8cm]{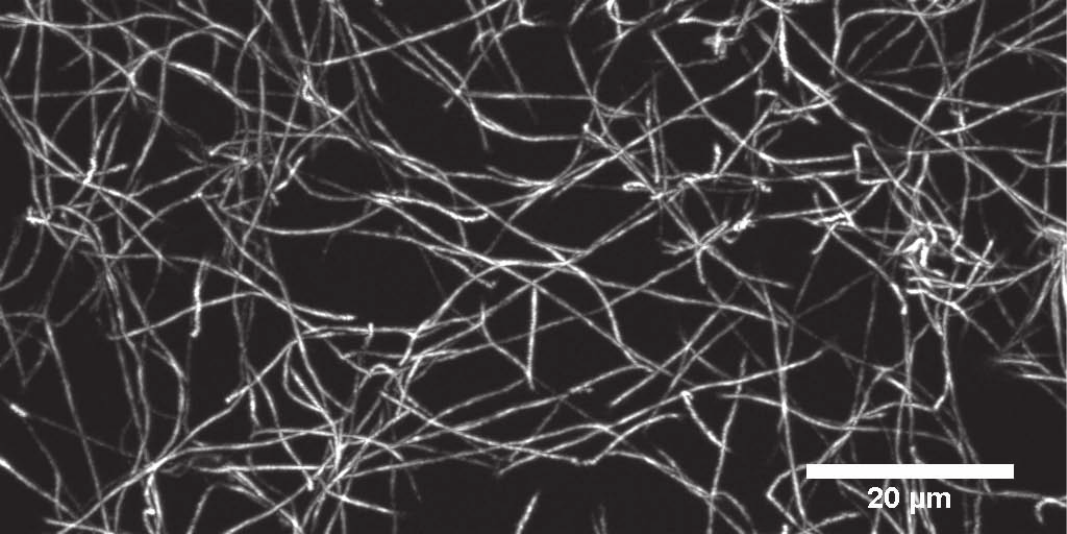}
\caption{Confocal microscopy image of a fluorescently labeled collagen network with a concentration of 0.4~mg/ml. Courtesy of Stefan M\"{u}nster (Erlangen-Nurnberg).}
\label{fig:Stefan}
\end{figure}

Over the past decades, semiflexible polymers and their assemblies in the form of solutions and networks have emerged as a distinct new class of soft condensed matter with striking properties. 
A major reason for the recent interest in semiflexible polymers has been their importance in living systems. Biopolymer assemblies form principal structural components throughout biology~\cite{Alberts1994,Bausch2006,Fletcher2010,Lieleg2010b,Kasza2007}, from the intracellular scaffold known as the \emph{cytoskeleton} to extracellular matrices of collagen, as illustrated in Figs.~\ref{fig:Cliff} and~\ref{fig:Stefan}. Cytoskelatal structures contribute to intracellular transport and organization, and ensure the structural integrity and mobility of cells~\cite{Alberts1994}. 
Thus, most of the experimental studies of semiflexible polymers have been carried out on biopolymers. From a fundamental physics perspective, a major motivation for many of the experimental and theoretical studies of biopolymers has been the diverse behavior of biopolymer systems, which are often in stark contrast to their now better understood synthetic and flexible counterparts in polymer science and materials. 

A polymer is said to be \emph{semiflexible} when its bending stiffness is large enough, such that the bending energetics---that favors a straight conformation---can just out-compete the entropic tendency of a chain to crumple up into a \emph{random coil}. Thus, semiflexiblõe polymers exhibit small, yet significant, thermal fluctuations around a straight conformation. This competition between entropic and energetic effects gives rise to many of the unique physical properties of semiflexible polymers and their assemblies~\cite{Bausch2006,Fletcher2010,Lieleg2010b,Kasza2007,MacKintosh1997}. The semiflexible nature of the polymers also has major implications for how they interact with each other to form entangled or crosslinked networks, and for the linear and nonlinear elastic and flow properties of such networks. A deep and predictive understanding of the physics of such networks has proven to be a daunting theoretical challenge,  in part due to their disordered many-body nature, and the fundamentally more peripheral role of entropy in these systems. Here, we review recent advances in modeling such systems, and highlight some of the major remaining open questions. 

Biopolymers, especially those composed of globular proteins much larger than the atomic or molecular scale, are far more rigid than most synthetic polymers, and they constitute prime examples of semiflexible polymers. Their rigidity results in conformations, both at the single polymer and network level, that are very far from the near gaussian or random coil configurations common in polymer physics~\cite{Wilhelm1996}. This difference turns out to be more than just a quantitative one: Semiflexible polymer systems exhibit qualitatively different elastic and viscoelastic properties. These properties include reversible softening under compression~\cite{Chaudhuri2007}, as well as both stiffening~\cite{Gardel2004a,Storm2005,Lieleg2007} and negative normal stress under shear~\cite{Janmey2007}. 

Because of the unusual material properties of biopolymers and their assemblies, much can be learned from them and they can serve as inspiration for new materials or new experimental systems to test fundamental physics. An example of the latter is the recent use of carbon nanotubes, which have comparable mechanical properties to many biopolymers and which can be very effectively visualized with light microscopy, to address long-standing puzzles in polymer physics \cite{Fakhri2009,Fakhri2010,Doi1988,Odijk1983}.

From a physical point of view, the main differences between various semiflexible polymers, biological or synthetic, are their dimensions and mechanical properties. One of the ways to quantify the bending stiffness of polymers is by their so-called \emph{persistence length}, which is essentially the length over which they appear straight in the presence of Brownian forces. The persistence lengths and dimensions of various semiflexible polymers are listed in Table~\ref{table:biopolymers}. An important aspect, setting biopolymers apart from most synthetic polymers, is that their persistence length is much larger than the molecular or single protein scale, and is often comparable to or larger than the relevant length scale on which the polymer is considered, such as its contour length or the cross-linking length scale of the network in which they are embedded. Thus, many biopolymers  are considered to be \emph{semiflexible}, and their dynamics is governed by a competition between entropic and energetic effects. Many theories of semiflexible polymers have been put to a test, since it became possible to study the dynamics and elastic properties of isolated biopolymers. 

Quantitative measurements of the properties of biopolymer systems in their native environment \emph{in vivo} remains a formidable experimental challenge~\cite{Fletcher2010}. However, important advances have been achieved by using a \emph{bottom-up} approach: proteins are purified and reconstituted to form simplified \emph{in vitro} models of real biopolymer systems, which can be studied quantitatively under well-controlled conditions~\cite{Kasza2007,Bausch2006,Lieleg2010b}. Most biopolymers, including filamentous actin (F-actin), intermediate filaments and microtubules, as well as associated regulatory proteins can now be purified and reconstituted into networks. These reconstituted networks have not only formed an ideal testing ground for theory, but have often lead the way for new theoretical developments. Thus, a large share of the work reviewed here was done in the context of such reconstituted biopolymer networks. 

\begin{figure}
\includegraphics[width=\columnwidth]{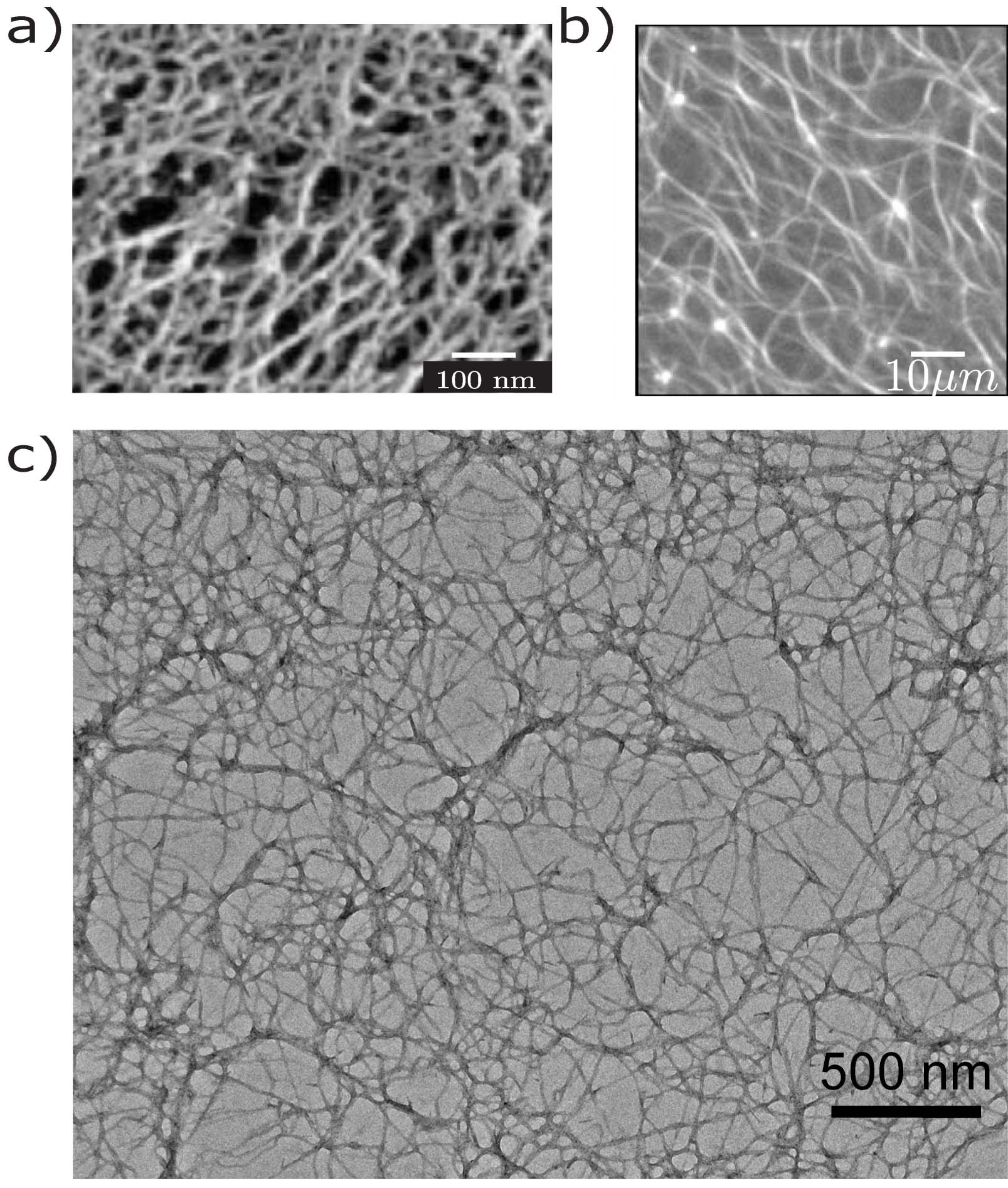}
\caption{a) Electron micrograph of a fixed and rotary-shadowed filamin-F-actin network at an actin concentration 1~mg/ml, average filament length 15~$\mu$m, and a filamin:actin molar ratio of 0.005:1. From~\cite{Kasza2009} b) Confocal microscopy image of a fluorescently labeled bundled filamin-F-actin network at high filamin concentrations~(From~\cite{Kasza2010}).
c) Electron micrograph of a fixed and rotary-shadowed Vimentin network. Courtesy of Y-C. Lin and D. Weitz (Harvard).}
\label{fig:ifintro}
\end{figure}

As an example, we show an electron micrograph and a fluorescence microscopy image of an \emph{in vitro}  F-actin network in Figs.~\ref{fig:ifintro}a,b. The microstructure and mechanics of such networks can depend sensitively on the type and concentration of polymer and crosslinking proteins~\cite{Lieleg2010b}. This represents yet another key difference with respect to flexible polymers: semiflexible polymers are fundamentally less prone to entangle with their neighbors, since they cannot readily form tight coils or knots. This renders biopolymer networks much more sensitive to cross-linking, and may be a reason why nature employs a wide variety of crosslinkning proteins. Indeed, distinguishing properties of physiological crosslinkers, such as their dynamic or transient nature~\cite{Lieleg2008,Broedersz2010b,Lieleg2010,Yao2013,Strehle2011,Ward2008,Heussinger2011} or a nonlinear elastic response~\cite{Gardel2006,Wagner2006,Broedersz2008,Kasza2009,Kasza2010}, have been found to have a major impact on the linear and nonlinear viscoelastic properties of the networks they form. The addition of motor proteins such as myosin, which can
\emph{actively} generate stochastic forces by tugging on F-actin filaments, can drive the network into a \emph{nonequilibrium} state~\cite{Mizuno2007,Koenderink2009}, with striking effects on the dynamics and mechanics of the system. 

One of the key mechanical properties of such reconstituted networks is the shear modulus. The shear modulus typically exhibits a rich frequency dependence~\cite{Koenderink2006,Hinner1998,Schnurr1997}, including frequency-independent plateau regimes at intermediate frequencies, and various power law regimes at both low and high frequencies. This variety reflects how the network can be dominated by qualitatively different dynamics on different times scales. Insights into these various frequency regimes were given by theories on the dynamics of semiflexible polymers or bundles in permanently or transiently crosslinked networks~\cite{Heussinger2007b,Heussinger2010,Gittes1998,Morse1998a,Morse1998b,Morse1998c,Broedersz2010b,Lieleg2008}. In some cases, weak power laws were observed~\cite{Semmrich2007}, suggesting soft glassy dynamics, which spurred the developments of theories on the dynamics of polymers in glassy environments~\cite{Kroy2007,Kroy2008}. These systems also exhibit a striking nonlinear response~\cite{Gardel2004a,Gardel2004b,Storm2005,Gardel2006}, in which the networks' differential stiffness can increase 10-100 fold at moderate strains between 10\%-100\%, which lead to much debate on the origins of this behavior~\cite{Onck2005,Lieleg2007,Huisman2007,Heussinger2007c,Huisman2008,Kabla2006,Wyart2008,Conti2009}.

In this review we focus largely on minimal, physical approaches.
We begin with the properties of single filaments, and then move on to the collective properties of entangled solutions and semiflexible polymer networks.

\label{sec:ind_polymers}
\begin{table*}[t]
\caption{Persistence lengths and dimensions of various biopolymers\cite{Gittes1993,Howard2001,Dogic2004, Mucke2004,Lin2010a,Lin2010b}.} 
\centering 
\begin{tabular}{l l l l} 
\hline\hline 
Type \ \  \ & approximate diameter \  \ \ &  persistence length \ \ \ & contour length\\ [0.5ex] 
\hline 
Microtubule & $25~$nm & $\sim1-5$~mm & 10s of $\mu$m\\ 
F-actin & $7~$nm & $17~\mu$m & $\lesssim20~\mu$m\\
Intermediate filament & $9~$nm & $0.2-1~\mu$m &  $2-10~\mu$m\\
DNA & $2~$nm & $50$~nm & $\lesssim1$~m \\ 
SWNTs & $<1~$nm & $\sim 10~\mu$m & $\gtrsim1~\mu$m \\ [1ex] 
\hline 
\end{tabular}
\label{table:biopolymers} 
\end{table*}

\section{Semiflexible polymers}

\begin{figure}[b]
\centering
\vspace{.2cm}
\includegraphics[width=8cm]{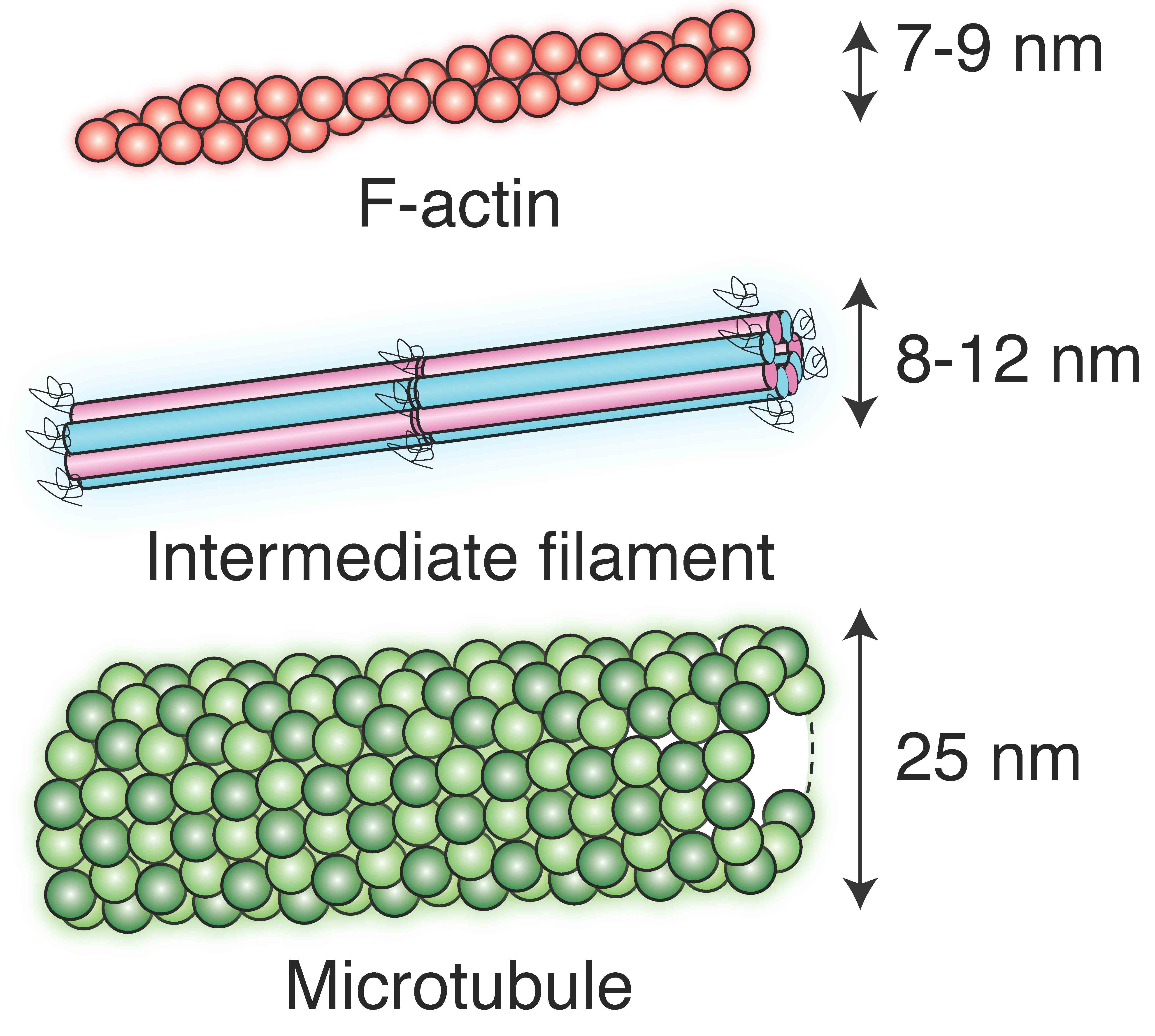}
\caption{The three families of cytoskeletal filaments, including F-actin, intermediate filaments and microtubules.}
\label{fig:biopolymers}
\end{figure}

Biopolymers such as those that make up the cytoskeleton and extracellular matrices typically consist of aggregates of large globular proteins (Fig.~\ref{fig:biopolymers}). These are usually bound together more weakly than most synthetic, covalently-bonded polymers. Biopolymers can nevertheless exhibit surprising stability and strength. Specifically, given that their diameter can be as large as tens of nanometers or more, they are far more rigid to bending than most common synthetic polymers, and it can be a good approximation in some cases to treat them as elastic fibers. Thus, their \emph{bending rigidity} is often their most important characteristic. In many cases, however, the contour length of these filaments is still long enough that they may exhibit significant thermal bending fluctuations. Thus they are said to be \emph{semiflexible} or \emph{worm-like}.

The most intuitive characterization of the stiffness of biopolymers is their \emph{persistence length} $\ell_p$, which can be regarded as the contour length at which significant bending fluctuations occur. This characterization is convent, but can be misleading. It is not correct, for instance, to think of a semiflexible polymer as rod-like and effectively athermal on length scales shorter than $\ell_p$: even for lengths much less than this, thermal fluctuations can play an important, even dominant role, e.g., in determining the axial stretching response of a semiflexible chain. Also, it is important to bear in mind that $\ell_p$ is directly related to the filament stiffness only in thermal equilibrium and only for filaments that are perfectly straight in their relaxed state. Sometimes, the term persistence length is also used merely as a way of characterizing how straight a given polymer is, for instance, in AFM experiments that measure the conformation of a polymer adsorbed on a surface. Such conformations can be far from equilibrium, and thus the shape may not directly reflect the bending rigidity of the filament. Under conditions of thermal equilibrium, the persistence length is more precisely defined in terms of the angular correlations of the local tangent along the polymer backbone, which decay exponentially with a characteristic length $\ell_p$. The persistence lengths of a few important semiflexible polymers are given in Table~\ref{table:biopolymers}, along with their approximate diameter and contour length.

\subsection{Worm-like Chain Model}
\label{sec:worm}

\begin{figure}[t]
\centering
\vspace{.3cm}
\includegraphics[width=8cm]{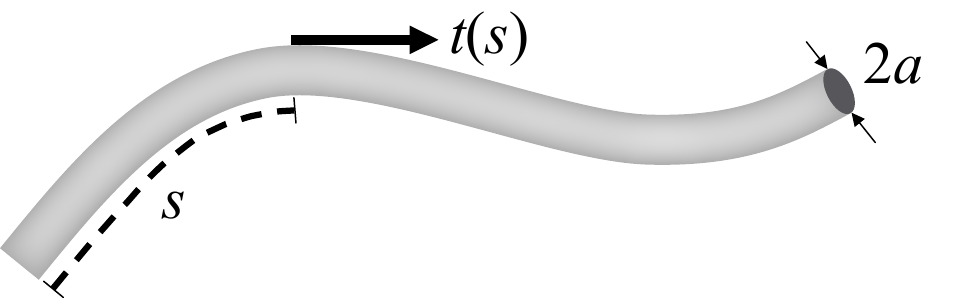}
\caption{A filamentous protein can be regarded as an elastic rod
of radius $a$. Provided the length of the rod is very long
compared with the monomeric dimension $a$, and that the rigidity
is high (specifically, the persistence length $\ell_p\gg a$), this
can be treated as an abstract line or curve, characterized by the
length $s$ along its backbone. A unit vector $\vec t$ tangent to
the filament defines the local orientation of the filament.
Curvature is present when this orientation varies with $s$. For
bending in a plane, it is sufficient to consider the the angle
$\theta(s)$ that the filament makes with respect to some fixed
axis. The curvature is then $\partial\theta/\partial s$.}
\label{fig:Filament}
\end{figure}

On the scale of several nanometers to micrometers, biopolymers are often effectively modeled as inextensible elastic rods or fibers with finite resistance to bending. This is the essence of the so-called worm-like-chain (WLC) model~\cite{Kratky1949}. This can be described by a bending energy of the form,
\be
H_{\rm bend}=\frac{\kappa}{2}\int ds\left|\frac{\partial \vec t}{\partial s}\right|^2,\label{eq:WormLikeChain}
\ee
where $\kappa$ is the \emph{bending modulus} and $\vec t$ is a (unit) tangent vector along the chain, and the integrand represents the square of the local curvature along the chain. Here, the chain position $\vec r(s)$ is described by an arc-length coordinate $s$ along the chain backbone. Hence, the tangent vector
\begin{equation}
\vec t=\frac{\partial \vec r}{\partial s}.
\end{equation}
These quantities are illustrated in Fig.\ \ref{fig:Filament}.

The bending modulus $\kappa$ has units of energy times length. A natural energy scale due to Brownian fluctuations is $\kT$, where $T$ is the temperature and $k$ is Boltzmann's constant. Thus, $\ell_p=\kappa/(\kT)$ is a length. In fact, this is precisely the \emph{persistence length} described above. For a homogeneous rod of diameter $2a$ consisting of a homogeneous material, the bending modulus should be proportional to the material's Young's modulus $E$, which has units of energy per volume. Thus, having units of energy times length, we expect that $\kappa$ to be of order $Ea^4$. In fact \cite{Landau1986},
\be
\kappa=\frac{\pi}{4}Ea^4.
\ee
This is often expressed as $\kappa=EI$, where $I$ is the moment of inertia of the cross-section. For a cylindrical fiber, the moment of inertia $I$ depends on the fourth power of the fiber radius $a$, apart from a purely geometric prefactor depending only on the cross-section. The factor $\pi a^4/4$ happens to be the right one for a cylindrical solid rod of radius $a$. For a hollow tube, the prefactor would be different, but still of order $a^4$, where $a$ is the (outer) radius. This elastic rod (or tube, as in the case of microtubules) approximation can be a good one, at least if the radius of curvature  of the filament is large compared with the molecular scale $a$. Within this approximation, the implied Young's modulus $E$ for polymers such as F-actin and microtubules can be as large as $~1$ GPa \cite{Howard2001,dePablo2003}.
\begin{table}[t]
\caption{List of main symbols used in text.} 
\centering 
\begin{tabular}{l l} 
\hline\hline 
Symbol \ \  \ & Description\\ [0.5ex] 
\hline 
$a$ & Filament radius\\
$E$ & Young's modulus\\
$\phi$ & Volume fraction\\
$G$ & Shear modulus\\
$\gamma$ & Strain\\
$\Gamma$ & Nonaffinity parameter\\
$\epsilon$ & Relative extension\\
$K$ & Differential shear modulus\\
$\kappa$ & Filament bending rigidity\\
$\ell$ & Filament length\\ 
$\ell_p$ & Persistence length\\
$\ell_c$ & Spacing between crosslinks.\\
$\mu$ & Filament stretching modulus\\
$\rho$ & Filament length density\\
$\sigma$ & Stress\\
$\tau$ & Tension\\
$\xi$ & Mesh size\\
$z$ & Network connectivity\\
$\zcf$ & Central force isostatic point\\
$z_b$ & Bending isostatic point\\
\hline 
\end{tabular}
\label{table:symbols} 
\end{table}

It is instructive to begin our analysis of the WLC model with the case of motion confined to a plane, for which there is a single transverse degree of freedom, the deflection away from a straight line. Here, the integrand above in Eq.\ (\ref{eq:WormLikeChain}) becomes $\left(\partial \theta/\partial s\right)^2$, where $\theta$ is simply the local angle that the chain axis makes relative to any fixed axis. A discrete approximation to the integral in Eq.\ (\ref{eq:WormLikeChain}) is then
$\sum_i\left(\Delta\theta_i\right)^2/\Delta s$,
where $\Delta\theta_i=\theta_i-\theta_{i-1}$ is the angle change between points separated by a small distance $\Delta s$ along the contour.
If the $\Delta\theta_i$ are independent degrees of freedom, which can be expected to be valid in the absence of long-range forces, the equipartition theorem tells us that
\begin{equation}
\langle\Delta\theta_i^2\rangle=\frac{\kT\Delta s}{\kappa}.\label{eq:Equipartition}
\end{equation}
This can be used to determine the correlations of orientations from one point along the chain to another. We note that the thermal 
average
\begin{eqnarray}
\langle\cos\left(\theta_m-\theta_n\right)\rangle
&=&\langle\cos\left(\Delta\theta_m\right)\rangle
\langle\cos\left(\theta_{m-1}-\theta_n\right)\rangle\nonumber\\
&\ldots&\\
&=&\langle\cos\left(\Delta\theta_m\right)\rangle^{m-n-1},\nonumber
\end{eqnarray}
where we have used the independence of the various $\Delta\theta_i$, and the fact that $\langle\sin\left(\Delta\theta_m\right)\rangle=0$. From this, it follows that the correlation function decays as
\be
\langle\vec t(s)\cdot\vec t(s')\rangle
= e^{-|s-s'|/\ell_p},
\ee
where $\ell_p=2\kappa/\kT$. 
Of course, this is all for motion confined to a plane. Taking into account the two independent transverse directions for thermal fluctuations in 3D, the persistence length becomes 
$\ell_p=\kappa/\kT$. This persistence length provides a \emph{geometric} measure of the \emph{mechanical} stiffness of the rod, provided that it is in equilibrium at temperature $T$.

This provides, in principle, a way to measure the persistence length, and thus the bending modulus of filaments by imaging the angular correlations along a filament. As discussed below, however, one must also be careful to account for the dynamics of filaments.

\subsection{Force-extension}\label{sec:force extension}
A single filament can respond to forces applied to it by bending, stretching or compressing. 
On length scales shorter than the persistence length, the bending can be described in mechanical terms, as for elastic rods. 
By contrast, stretching and compression may involve a purely elastic or mechanical response (as for macroscopic, elastic rods), a purely entropic response, or a combination of the two. 
For an inextensible chain, the entropic response comes from the thermal bending fluctuations of the filament. 
Perhaps surprisingly, as we shall see below, the longitudinal response can be dominated by entropy even on length scales small compared with the persistence length. 
Thus, it may be incorrect to think of a filament as truly rod-like, even on length scales shorter than $\ell_p$.

The longitudinal single filament response is often described in terms of a so-called force-extension relationship, in which 
the axial force required to extend the filament is measured or calculated in terms of the degree of extension along a line. 
At any finite temperature, there is an entropic resistance to such extension due to the presence of thermal fluctuations that make the polymer deviate from a straight conformation: since there are many more crumpled configurations of the polymer than the (unique) straight 
conformation, extending the polymer reduces the entropy and may increase the free energy. 
This entropic force-extension has been the basis of mechanical studies, for example, of long DNA \cite{Bustamante1994},
and a full, general calculation for low and high force, as well as short and long chains
is very involved, although simple approximations can be very accurate in the limit of long chains ($\ell\gg\ell_p$) or high forces \cite{Marko1995}. An interesting and universal form was recently proposed and shown to capture a broad range of polymer properties \cite{Dobrynin2010,Carrillo2013}. Here, we focus on a simple calculation appropriate for high tensile forces or for stiff polymers such as those that make up the cytoskeleton of cells, for which a nearly straight chain is most relevant \cite{MacKintosh1995}.

\begin{figure}[t]
\centering
\vspace{.2cm}
\includegraphics[width=8cm]{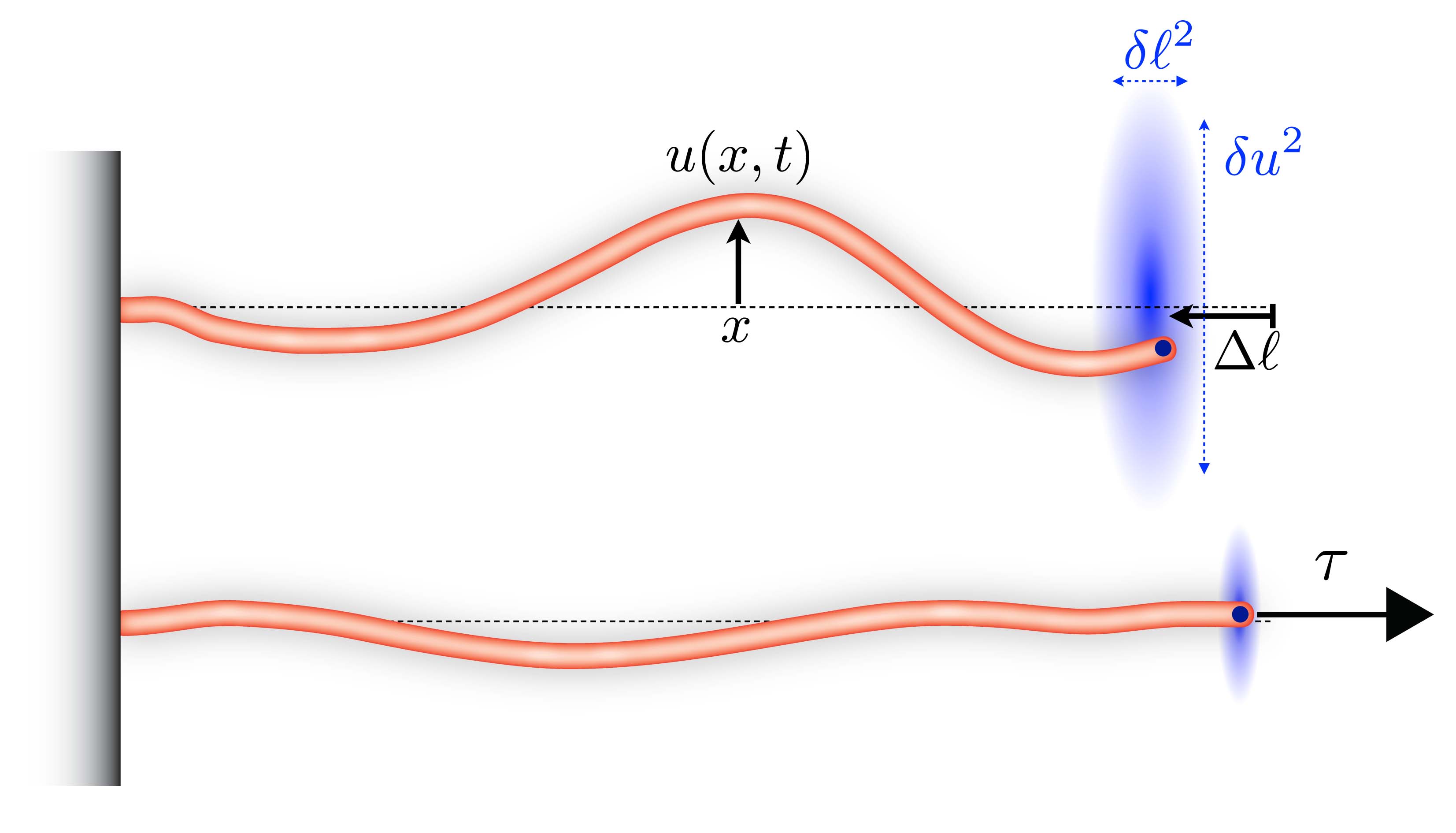}
\caption{If one end of a filament is fixed, both in position and orientation, while the other is free. The filament tends to wander in a way that can be characterized by $u(x)$, a transverse displacement field. For a fixed total arc length of the filament, thermal fluctuations result in a contraction of the end-to-end distance, which is denoted by $\Delta\ell$. In fact, this contraction is actually distributed about a thermal average value $\langle\Delta\ell\rangle$. The mean-square (longitudinal) fluctuations about this average are denoted by $\langle\delta\ell^2\rangle$, while the mean-square lateral fluctuations (\emph{i.e.}, with respect to the dashed line) are denoted by $\langle u^2\rangle$. The bottom image shows how the fluctuations are reduced and the chain is extended when a tension $\tau$ is applied. }\label{fig:Filament2}
\end{figure}

For a filament segment of length $\ell\lesssim\ell_p$ the filament is nearly
straight, with only small transverse fluctuations. We let the $x$-axis
define the average orientation of the chain segment, and let $u$
and $v$ represent the two independent transverse degrees of
freedom. These can then be thought of as functions of $x$ and time
$t$ in general. For simplicity, we first consider just a single transverse coordinate, $u(x,t)$. 
The bending energy is then
\begin{equation}
H_{\rm bend}=
\frac{\kappa}{2}\int dx\left(\frac{\partial^2 u}{\partial x^2}\right)^2
=\frac{\ell}{4}\sum_q \kappa q^4 u_q^2,
\end{equation}
where $u(x)$ is represented by a Fourier decomposition
\begin{equation}
u(x,t)=\sum_q u_q\sin(qx).\label{eq:decomp}
\end{equation}
As illustrated in Fig.\ \ref{fig:Filament2}, the local orientation
of the filament can be characterized by the slope $\partial u/\partial x$,
while the local curvature involves the second derivative
$\partial^2 u/\partial x^2$. Such a description is appropriate for
the case of a nearly straight filament with fixed boundary
conditions $u=0$ at the ends, $x=0,\ell$. Here, the wave
vectors $q=n\pi/\ell$, where $n=1,2,3,\ldots$.

If the chain is inextensible, with no compliance in its contour length,
then the end-to-end contraction of the chain in the presence of thermal fluctuations in $u$ is
\bea
\Delta\ell&=&\int dx\left(\sqrt{1+\left|\frac{\partial u}{\partial x}\right|^2}-1\right)\simeq \frac{1}{2}\int dx\left|\frac{\partial u}{\partial x}\right|^2
\nonumber\\
&=&\frac{\ell}{4}\sum_q q^2 u_q^2.\label{DeltaEll}
\eea
The integration here is actually over the projected length of the chain. But, to leading (quadratic) order in the transverse displacements, we make no distinction between projected and contour lengths here, and above in $H_{\rm bend}$.
Because the tension $\tau$ is conjugate to $\Delta\ell$, we can write the energy in terms the  applied tension as
\begin{eqnarray}
H&=&
\frac{1}{2}\int dx\left[\kappa\left(\frac{\partial^2 u}{\partial x^2}\right)^2+\tau\left(\frac{\partial u}{\partial x}\right)^2\right]\nonumber\\
&=&\frac{\ell}{4}\sum_q \left(\kappa q^4+\tau q^2\right) u_q^2.\label{EbendSum}
\end{eqnarray}
Under a constant tension $\tau$ therefore, the equilibrium amplitudes $u_q$ satisfy the equipartition theorem,
\begin{equation}
\langle\left|u_q\right|^2\rangle=\frac{2\kT}{\ell\left(\kappa q^4+\tau q^2\right)},\label{eq:Equipartition}
\end{equation}
and the contraction
\begin{equation}
\langle\Delta\ell\rangle_\tau=
\kT\sum_q \frac{1}{\left(\kappa q^2+\tau\right)}.\label{eq:amptau}
\end{equation}
There are, of course, two transverse degrees of freedom, and this final expression incorporates a factor of two appropriate for a chain fluctuating in 3D.

semiflexible filaments exhibit a strong suppression of bending fluctuations for modes of wavelength less than the persistence length $\ell_p$, as can be seen in the $q$-dependence in Eq.\ (\ref{eq:Equipartition}). This has important consequences for many of the thermal properties of such filaments. In particular, it means that the longest unconstrained wavelengths tend to be dominant in most cases of interest, provided that this length is short compared with $\ell_p$. This allows us, for instance, to anticipate the scaling form of the end-to-end contraction $\Delta\ell$ between points separated by arc length $\ell$ in the absence of an applied tension. We note that it is a length, it must vary inversely with stiffness $\kappa$, and must increase with temperature. Thus, since the dominant mode of fluctuations is that of the maximum wavelength, $\ell$, we expect the contraction to be of the form $\langle\Delta\ell\rangle_0\sim \ell^2/\ell_p$. More precisely, for $\tau=0$,
\begin{equation}
\langle\Delta\ell\rangle_0=\frac{\kT\ell^2}{\kappa\pi^2}
\sum^\infty_{n=1}\frac{1}{n^2}=\frac{\ell^2}{6\ell_p}.\label{eq:contract}
\end{equation}
Similar scaling arguments to those above lead us to expect that the typical transverse amplitude of a segment of length $\ell$ is approximately given by
\begin{equation}
\langle u^2\rangle\sim\frac{\ell^3}{\ell_p}.\label{eq:transfluct}
\end{equation}
in the absence of applied tension.  The precise coefficient for the mean-square amplitude of the midpoint between ends separated by $\ell$ (with vanishing deflection at the ends) is $1/24$. 
Apart from the prefactor, we could have anticipated the scaling form of the result in Eq.\ (\ref{eq:transfluct}) by noting that, being a thermal effect, it should increase proportional to $\kT$. It should also decrease inversely with bending rigidity $\kappa$. Thus, the expectation is that $\langle u^2\rangle\sim({\kT}/{\kappa})\times\ell^3$, where the dominating wavelength $\lambda\sim\ell$ is the longest unconstrained mode and this length enters with a third power for dimensional reasons. 

\begin{figure}[b]
\centering
\includegraphics[height=5cm]{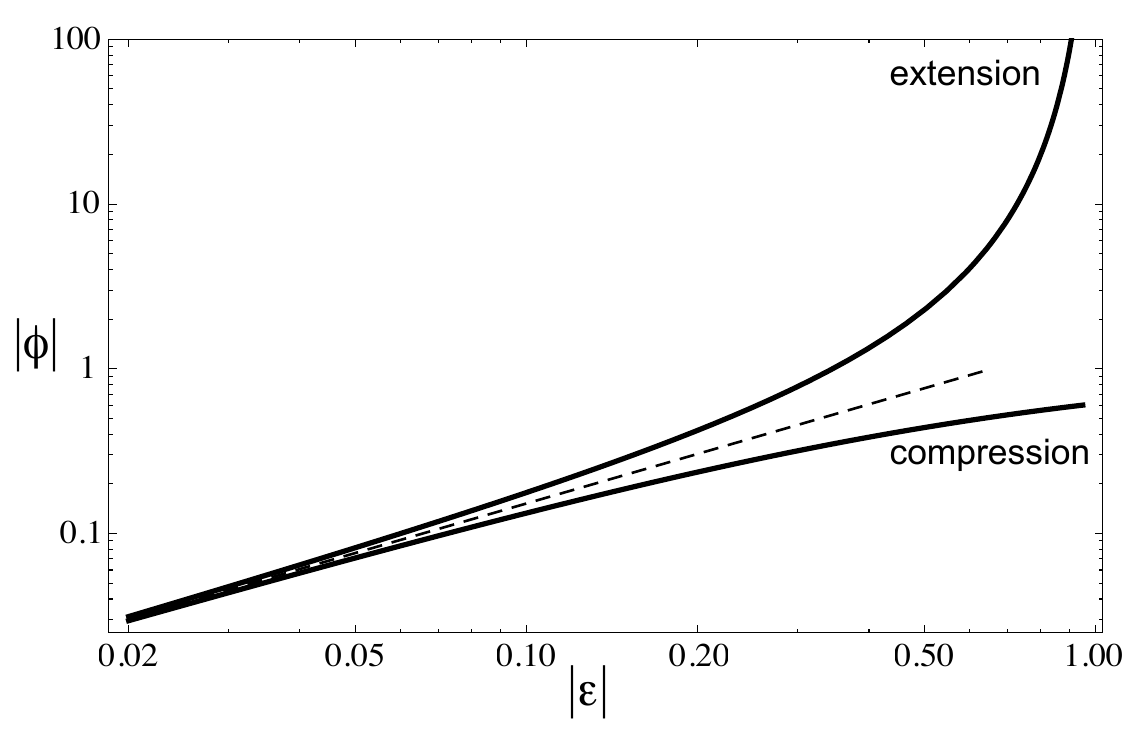}
\caption{The dimensionless force $\phi$ as a function of the relative extension $\epsilon=\delta\ell/\langle\Delta\ell\rangle$ from Eq.\ (\ref{eq:Gamma}). For small extension or compression, the response is linear. The upper curve depicts the force under extension ($\epsilon>0$), where the force is positive. The lower solid curve depicts the case of compression ($\epsilon<0$), where the force is negative. Both curves exhibit nonlinearities near the point where the extension or compression become comparable to the thermal contraction $\langle\Delta\ell\rangle_0$ in the absence of force (i.e., $|\epsilon|\simeq1$). For comparison, the linear limit is shown by the dashed line. For both extension and compression, the nonlinearities appear when the force is near the buckling threshold, $\phi=1$, indicated by the end of the dashed line. While these results are exact for inextensible chains of length $\ell\ll\ell_p$ under tension, finite-temperature buckling must be included for $-\epsilon\simeq -\phi\simeq 1$ \cite{Odijk1998,Emanuel2007,Baczynski2007}.}
\label{fig:FExt}       
\end{figure}
For a finite tension $\tau$, however, the longest unconstrained wavelength is not the only relevant length. There is also a characteristic length $\ell_t\sim\sqrt{\kappa/\tau}$ associated with the competition of bending and the tension. In short, modes of wavelength shorter than this are governed primarily by bending, while those of longer wavelength are governed by tension. Thus, the analysis above is valid provided that $\ell_t\gtrsim\ell$, i.e., for tensions $\tau$ small compared with the Euler buckling force $~\kappa/\ell^2$ of an elastic rod of length $\ell$. This also corresponds to the regime of force for which the response to tension is linear. In the other limit, $\ell_t\lesssim\ell$, nonlinearities in the response can be expected. 
In both limits, the extension of the chain (toward full extension) under tension is given by 
\begin{equation}
\delta\ell(\tau)=\langle\Delta\ell\rangle_0-\langle\Delta\ell\rangle_\tau
=\frac{\ell^2}{\pi^2\ell_p}\sum_n\frac{\phi}{n^2\left(n^2+\phi\right)},\label{eq:extension}
\end{equation}
where 
\begin{equation}
\phi=\tau\ell^2/(\kappa\pi^2)
\end{equation}
is a dimensionless force.
As suggested above, the characteristic force $\kappa\pi^2/\ell^2$ that enters here is the critical force in the classical Euler buckling problem. The summation in Eq.\ \eqref{eq:extension} can be found analytically, with the result that the relative extension
\be
\epsilon\equiv\frac{\delta\ell}{\langle\Delta\ell\rangle_0}
=1-3\frac{\pi\sqrt{\phi}\coth\left(\pi\sqrt{\phi}\right)-1}{\pi^2\phi}.
\label{eq:Gamma}\ee
Thus, the force-extension curve can be found by inverting this relationship numerically. 

In the linear regime, the extension becomes
\begin{equation}
\delta\ell=\frac{\ell^2}{\pi^2\ell_p}\phi\sum_n\frac{1}{n^4}
=\frac{\ell^4}{90\ell_p\kappa}\tau,
\label{eq:wlcstiffness}
\end{equation}
\emph{i.e.}, the effective spring constant for longitudinal extension of the chain segment is $90\kappa\ell_p/\ell^4$, which varies \emph{inversely} with $\kT$, in contrast to the freely-jointed chain and semiflexible chains in the limit $\ell\gg\ell_p$, both of which exhibit increasing stiffness with temperature $T$. The scaling form of this could also have been anticipated, based on very simple physical arguments similar to those above. In particular, given the expected dominance of the longest wavelength mode (\emph{i.e.}, $\ell$), we expect that the end-to-end contraction scales as $\delta\ell\sim\int\left(\partial u/\partial x\right)^2\sim u^2/\ell$. Thus, $\langle\delta\ell^2\rangle\sim\ell^{-2}\langle u^4\rangle\sim\ell^{-2}\langle u^2\rangle^2\sim\ell^4/\ell_p^2$, which is consistent with the effective (linear) spring constant derived above from Eq.\ \eqref{eq:wlcstiffness}, since $\langle\delta\ell^2\rangle$ should be equal to $\kT$ divided by the longitudinal spring constant.

Importantly, due in part to the asymmetry of the wormlike chain under extension and compression, the statistics of the end-to-end fluctuation of a semiflexible polymer are not described by a Gaussian distribution. The full distribution function was calculated analytically in \cite{Wilhelm1996}. The resulting distribution can be approximated by a Gaussian only near the thermal average extension, while the distribution cuts off sharply near the full extension. 

The full nonlinear force-extension curve can be calculated numerically by inversion of the expression above. This is shown in Fig.\ \ref{fig:FExt}. Here, one can see both the linear regime for small forces, with the effective spring constant given above, as well as a divergent force near full extension. In fact, the force diverges in a characteristic way, as the inverse square of the distance from full extension:  $\tau\sim|1-\epsilon|^{-2}$
\cite{Fixman1973,Odijk1995,Marko1995}.
This form of the divergence of force can be identified without preforming the full summation in Eq.\ (\ref{eq:amptau}), as follows. For $q\lesssim q_\tau=\sqrt{\tau/\kappa}\sim\ell_t^{-1}$, the tension $\tau$ governs the mode amplitudes, as noted above. Mathematically, in this range the tension term dominates the bending term in the denominator of Eqs.\ (\ref{eq:Equipartition},{\ref{eq:amptau}). In the other limit, for larger $q$, the sum rapidly converges. Thus, only a number of terms that grows as $\sqrt{\tau}$ are really needed in the sum, and these terms are themselves proportional to $1/\tau$. Thus, $\langle\Delta\ell\rangle_\tau\sim1/\sqrt{\tau}$ and 
\be\tau\rightarrow\frac{\kappa\ell^2}{4\ell_p^2|\langle\Delta\ell\rangle_0-\delta\ell|^{2}}\label{eq:DivergenceForce}\ee
for large tension.

As noted above, the force-extension relation can be found by numerically inverting Eq.\ (\ref{eq:extension}) using Eq.\ (\ref{eq:Gamma}). In practice, however, it is often preferable to use a more tractable approximation to the exact force-extension relation, as is usually done for DNA \cite{Marko1995}. However, this worm-like-chain approximation is only valid for $\ell\gg\ell_p$. In the opposite limit of $\ell<\ell_p$, the small-extension or linear response regime is characterized by a different spring constant, although the high-force asymptotic regime in Eq.\ (\ref{eq:DivergenceForce}) is the same, including prefactor. An approximate force-extension relation can be obtained from the asymptotic limit in Eq.\ (\ref{eq:DivergenceForce}), together with a constant term and a term linear in the extension, where these are chosen to yield the correct overall linear response and zero force at zero extension. The result can be expressed simply in terms of the normalized extension $\epsilon$ and force $\phi$:
\be
\phi=\frac{9}{\pi^2}\left[\frac{1}{(1-\epsilon)^2}-1-\frac{1}{3}\epsilon\right].
\ee
Under extension, this yields the correct small and large force limits. It strictly overestimates the intermediate force range between these limits, but by less than 16\%. This form has been used for efficient computation of the nonlinear elasticity of semiflexible networks \cite{Gardel2004a}. An equivalent  force-extension expression with an additional approximate term to capture buckling was derived in \cite{Huisman2008}. 

\subsubsection{Inextensible versus extensible polymers}
Before concluding our discussion of the longitudinal response of semiflexible polymers, it is worth asking about another obvious contribution to their response. This, we can think of as the \emph{zero-temperature} enthalpic or \emph{purely mechanical} response. After all, we are treating semiflexible polymers as small bendable rods. To the extent that they behave of rigid rods, we might expect them to respond to longitudinal stresses by increasing/decreasing their countour length. Based on the arguments above, it seems that the persistence length $\ell_p$ determines the length below which filaments behave like rods, and above which they behave like flexible polymers with significant thermal fluctuations. It would be tempting to expect that enthalpic stretching dominates for any $\ell\lesssim\ell_p$. Perhaps surprisingly, however, even for segments of semiflexible polymers of length much less than the persistence length, their longitudinal response can be dominated by the entropic force-extension described above.

To examine this, we consider a simple model of a semiflexible polymer as a homogeneous elastic rod of radius $a$. We have already seen that the bending modulus is $\kappa\sim Ea^4$. Likewise, the (linear) stretching/compression of such an elastic rod is described by the Hamiltonian
\begin{equation}
{H}_{\rm stretch}
=
\frac{1}{2}
\mu
\int{\rm d}s
\left(
\frac{{\rm d}\ell(s)}{{\rm d}s}
\right)^{2}
\label{e:H_stretch}
\end{equation}
\noindent{}where ${\rm d}\ell/{\rm d}s$ gives the relative change in length (strain) along the filament. The stretch modulus $\mu\sim Ea^2$. The effective (mechanical) spring constant of a segment of length $\ell$ is thus $k_{\rm M}\sim\mu/\ell\sim Ea^2/\ell$, compared with the effective (thermal) spring constant $k_{\rm T}\sim\kappa\ell_p/\ell^4\sim k_{\rm M}a^2\ell_p/\ell^3$, since $\kappa\sim Ea^4$. Since the system will respond primarily according to the softer effective spring constant, the dominant response will be thermal if $\ell^3\gtrsim a^2\ell_p$, and will be mechanical only if $\ell^3\lesssim a^2\ell_p$. Thus, even segments of length much less than $\ell_p$ can still respond according to the thermal response described above. For F-actin, for example, even filament segments as short as 100-200 nm in length may be dominated in their longitudinal compliance by the thermal response arising from bending fluctuations. The extent of this compliance, however, will be quite limited. Thus, there is expected to be a crossover from a nonlinear thermal compliance to an enthalpic stretching regime. This crossover can be characterized by two mechanical springs in series, in which one adds a purely enthalpic compliance $\delta\ell^{(e)}=\ell\tau/\mu$ to the compliance in Eq.\ (\ref{eq:extension}) \cite{Odijk1995}. However, the entropic stiffness is nonlinear, and nonlinear compliances do not simply add. There is an additional higher-order correction that corresponds to a renormalization of the force in the nonlinear entropic force extension curve~\cite{Storm2005}. The resulting extension in Eq.\ \eqref{eq:extension} for stiff chains, where $\ell\gg\langle\Delta\ell\rangle_0$, is given by 
\be
\delta\ell(\tau)=\ell\tau/\mu+\delta\ell\left(\tau\left[1+\tau/\mu\right]\right).
\ee

\subsubsection{Euler buckling}\label{sec:buckling}

In addition to the mechanical or enthalpic response to stretching, there is also another purely mechanical response under compression: Euler buckling \cite{Landau1986}. When a straight elastic rod is subject to a compressive load, it initially responds by compressing longitudinally along its axis. Above a well-defined force threshold, however, it undergoes a buckling instability. To a good approximation, the rod simply cannot bear any additional compressive load beyond this threshold. Thus, the rod is simply unstable and any additional load will cause the rod to completely collapse. This threshold compressive force can be calculated as follows. 

When a rod of length $\ell$ undergoes an oscillatory transverse deflection of amplitude $u(x)=\sum_q u_q\sin(qx)$, it contracts longitudinally by an amount given by Eq.\ (\ref{DeltaEll}). The energy of this deflection is given by Eq.\ (\ref{EbendSum}), where $\tau$ is the (tensile) load. For $\tau>0$, each term in the series contributes a positive energy, so that the system is (harmonically) stable against increasing amplitude $u_q$, for each $q$. For compressive loads, however, where $\tau<0$, this is no longer the case if $\tau<-\kappa q^2$. In that case, the energy as function of $u_q^2$, transitions from being concave to being convex. Thus, for compressive tensions exceeding this $q$-dependent threshold, the corresponding transfer deflection modes become unstable. For such modes, the system is unstable to transverse displacements. As the compressive load $-\tau$ increases from zero, this instability first occurs for the smallest $q$ possible, which is determined by the length of the rod: $q=q_1=\pi/\ell$. Thus, the buckling instability occurs for compressive force given by 
\be
f_c=\kappa\left(\frac{\pi}{\ell}\right)^2,
\ee
and this instability corresponds to the fundamental oscillatory mode for the rod, where the wavelength of the instability is twice the length of the rod, as illustrated in the upper panel of Fig.\ \ref{fig:Buckling}. One important aspect of this buckling is its threshold nature, which has been used effectively in numerous biopolymer experiments to measure axial forces precisely \cite{Dogterom1997,Footer2007}.

Thermal fluctuations modify this classical, Euler buckling result. As might be expected for thermal fluctuations, the sharp force threshold no longer applies for filaments at finite temperature. Rather, the smooth force-compression curve is found, and for even small forces there is a finite compression. This can be seen in the linear regime for compression in Fig.\ \ref{fig:FExt}. Qualitatively, thermal fluctuations can be said to enhance the compliance under compression, resulting in reduced forces, relative to Euler buckling, for the same degree of compression \cite{Odijk1998,Emanuel2007}. Interestingly, however, this only holds for buckling in three dimensions, and there are qualitative differences for buckling in two dimensions: for large compressive forces ($f\agt f_c$), finite temperature filaments actually exhibit \emph{reduced} compressive compliance, relative to zero-temperature buckling \cite{Emanuel2007,Baczynski2007}. As discussed by Emanuel et al., the enhanced compliance arises from fluctuations out of the plane of buckling.

\begin{figure}[]
\centering
\vspace{0.4cm}
\includegraphics[width=8cm]{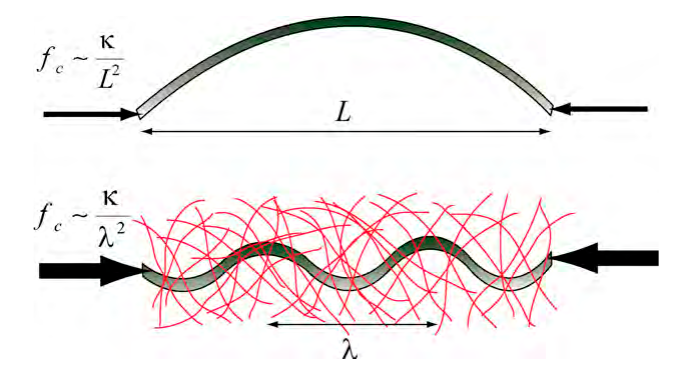}
\caption{Schematic of classical Euler buckling (above), showing the expected shape for an elastic rod that is free to bend between its ends. Under compression defined by a fixed end-to-end distance, beyond a well-defined threshold force, the elastic energy can be lowered by relieving the compression at the cost of a bend to accommodate the reduced end-to-end distance \cite{Landau1986}. When the lateral deflection of the rod is suppressed by a surrounding elastic matrix (indicated in red), the buckling exhibits a shorter wavelength bend at a correspondingly higher force threshold \cite{Landau1986}. Microtubules in vivo exhibit such a constrained buckling under compression \cite{Brangwynne2006}. \copyright Brangwynne et al., 2006.\ Originally published in \emph{Journal of Cell Biology.}\ doi:10.1083/jcb.200601060.}
\label{fig:Buckling}
\end{figure}

Interestingly, if an athermal elastic rod is embedded in a surrounding elastic material, as can be the case for microtubules embedded in the matrix formed by the rest of the cytoskeleton \cite{Brangwynne2006}, then this classical Euler buckling problem and corresponding threshold is altered \cite{Landau1986}. We can see this from Eg.\ (\ref{DeltaEll}), since a given amount of longitudinal compression $\Delta\ell$ requires a larger amplitude $u_q$ for smaller $q$. Such large amplitude lateral deflections are suppressed by the surrounding matrix. More precisely, for a transverse deflection $u$, the elastic energy per unit length along the rod is approximately given by
\be
\frac{1}{2}\alpha_\perp u^2\;,
\ee
where 
\be
\alpha_\perp\simeq\frac{4\pi G}{\ln(\lambda/b)}.
\ee
Here, $G$ is the shear modulus of the surrounding elastic medium, $\lambda$ is the bending wavelength and $b$ is a microscopic length of order the radius of the rod. 
This result is the elastic analogue of a possibly more familiar hydrodynamics result for the viscous drag per unit length on a thin rod of length $\sim\lambda$ moving transverse to its axis \cite{Lamb1945}. The hydrodynamics case will be discussed in the next section on dynamics. There are two important points to note in deriving either result: (1) on dimensional grounds, the elastic spring constant $\alpha_\perp$ for transverse displacement cannot involve an additional factor of length beyond the obvious term $\propto G$, and (2) the elastic as well as viscous stress balance equations involve the Laplacian, for which the solutions in the two transverse dimension naturally lead to logarithms. 
Thus, using the mode decomposition above, we obtain the elastic energy 
\begin{eqnarray}
H&=&\frac{1}{2}\int dx
\left[\kappa\left(\frac{\partial^2 u}{\partial x^2}\right)^2\kern-4pt+
\tau\left(\frac{\partial u}{\partial x}\right)^2\kern-4pt+\alpha_\perp u^2\right]\nonumber\\
&=&\frac{\ell}{4}\sum_q \left(\kappa q^4+\tau q^2+\alpha_\perp\right) u_q^2.\label{EbendSumBuckling}
\end{eqnarray}
The elastic energy of the matrix has the effect of both increasing the threshold force and shifting the instability to a shorter wavelength. The first unstable mode corresponds to 
\begin{equation}
q_*=\sqrt[4]{\alpha_\perp/\kappa}\;,
\end{equation}
and the critical buckling force is now 
\begin{equation}
f_c=2\kappa q_*^2\;,
\end{equation}
which can be much larger than for unconstrained buckling. Such constrained buckling has been reported for microtubules under compression in cells, where the surrounding cytoskeletal meshwork can greatly enhance the compressive load bearing capability of microtubules by up to 100 times \cite{Brangwynne2006,Das2008,Shan2013,Liu2012}.

\subsection{Dynamics}
\label{sec:dynam}
In the above, we have considered only static properties of individual polymer chains. The dynamics of single chains exhibit rich behavior that can have important consequences even at the level of bulk solutions and networks. The principal dynamic modes come from the transverse motion, \emph{i.e.}, the degrees of freedom $u$ and $v$ above. The equation of motion of these modes can be found from $H_{\rm bend}$ above, together with the hydrodynamic drag of the filaments through the solvent. In the presence of only thermal forces, this is done via a Langevin equation describing the net force per unit length on the chain at position $x$, 
\begin{equation}
0=-\zeta\frac{\partial}{\partial t}u(x,t)
-\kappa\frac{\partial^4}{\partial x^4}u(x,t)+\xi_\perp(x,t),
\label{eq:Langevin}
\end{equation}
which is, of course, zero within linearized, inertia-free (low Reynolds number) hydrodynamics that we assume here. 

The first term represents the hydrodynamic drag per unit length of the filament. Here, we have assumed a constant transverse drag coefficient that is independent of wavelength. In fact, given that the actual (low Reynolds number) drag per unit length on a rod of length $\ell$ is 
\be
\zeta=\frac{4\pi\eta}{\ln\left(A \ell /a\right)},
\ee
where $\ell/a$ is the aspect ratio of the rod, and $A$ is a constant of order unity that depends on the precise geometry of the rod \cite{Lamb1945}. As noted above, the logarithm is a natural consequence of the 2D nature of the transverse plane in which the motion occurs. For a filament undergoing free bending fluctuations in solution, the relevant length $\ell$ is the wavelength $\lambda$ of the bending mode. Thus, a week logarithmic dependence of the relaxation rate on the wavelength is expected. However, this hydrodynamic effect is weak and has not been directly observed. In practice, the presence of other chains in solution gives rise to an effective screening of the long-range hydrodynamics beyond a length of order the typical separation $\xi$ between chains, which can then be taken in place of $\ell$ above. 

The second term in the Langevin equation above is the restoring force per unit length due to bending, which is obtained by the functional derivative
\begin{equation}
-\frac{\delta}{\delta u}H_{\rm bend}=-\kappa\frac{\partial^4}{\partial x^4}u(x,t).
\label{eq:GenForce}
\end{equation}
Finally, we include a random force $\xi_\perp$, which can be taken to be uncorrelated white-noise in the case of a purely viscous solvent. Equation (\ref{eq:Langevin}) represents an example of model $A$ dynamics, in which the dissipation (here, hydrodynamic drag) is local and the field $u(x,t)$ is non-conserved \cite{Chaikin,Hohenberg1977}.

A simple force balance in the Langevin equation above after thermal averaging (i.e., without the noise term) leads us to conclude that the characteristic relaxation rate of a mode of wavevector $q$ is \cite{Farge1993}
\begin{equation}
\omega(q)=\kappa q^4/\zeta.\label{eq:omega}
\end{equation}
This is valid provided end-effects are not important, i.e., provided that the wavelength is short compared to the contour length of the chain.
The fourth-order dependence of this rate on $q$ is to be expected from the appearance of a single time derivative along with four spatial derivatives in Eq.\ (\ref{eq:Langevin}).
This relaxation rate determines, among other things, the correlation time for the fluctuating bending modes. Specifically, in the absence of an applied tension,
\begin{equation}
\langle u_q(t)u_q(0)\rangle=\frac{2\kT}{\ell\kappa q^4}e^{-\omega(q) t}.\label{eq:amp}
\end{equation}
That the relaxation rate varies as the fourth power of the wavevector $q$ has important consequences. For example, while the time it takes for an actin filament bending mode of wavelength 1~$\mu$m to relax is of order 10~ms, it takes about 100~s for a mode of wavelength 10~$\mu$m.

The very strong wavelength dependence of the relaxation rates in Eq.\ (\ref{eq:omega}) has important consequences, for instance, for imaging of the thermal fluctuations of filaments, as is done in order to measure $\ell_p$ and the filament stiffness (Gittes 1993). This is the basis, in fact, of most measurements to date of the stiffness of DNA, F-actin, and other biopolymers. Using Eq.\ (\ref{eq:amp}), for instance, one can both confirm thermal equilibrium and determine $\ell_p$ by measuring the mean-square amplitude of the thermal modes of various wavelengths. However, in order to both resolve the various modes, as well as establish that they behave according to the thermal distribution, one must sample over times long compared with $1/\omega(q)$ for the longest wavelengths $\lambda\sim 1/q$. At the same time, one must be able to resolve fast motion on times of order $1/\omega(q)$ for the shortest wavelengths. Given the strong dependence of these relaxation times on the corresponding wavelengths, for instance, a range of order a factor of 10 in the wavelengths of the modes requires a range of a factor $10^4$ in observation times.

Another way to look at the result of Eq.\ (\ref{eq:omega}) is that a bending mode of wavelength $\lambda$ relaxes (\emph{i.e.}, fully explores its equilibrium conformations) in a time of order $\zeta\lambda^4/\kappa$. Since it is also true that the longest (unconstrained) wavelength bending mode has by far the largest amplitude, and thus dominates the typical conformations of any filament (see Eqs.\ (\ref{eq:amptau}) and (\ref{eq:amp})), we can see that in a time $t$, the \emph{typical} or dominant mode that relaxes is one of wavelength $\ell_\perp(t)\sim\left(\kappa t/\zeta\right)^{1/4}$. As we have seen above in Eq.\ (\ref{eq:transfluct}), the mean-square amplitude of transverse fluctuations increases with filament length $\ell$ as $\langle u^2\rangle\sim\ell^3/\ell_p$. Thus, in a time $t$, the expected mean-square transverse motion is given by (Farge and Maggs 1993; Amblard \emph{et al.}\ 1996)
\begin{equation}
\langle u^2(t)\rangle\sim \left(\ell_\perp(t)\right)^{3}\kern -0.3em/\ell_p\sim t^{3/4},
\end{equation}
because the typical and dominant mode contributing to the motion at time $t$ is of wavelength $\ell_\perp(t)$.

The dynamics of longitudinal motion can be calculated similarly. Here, however, we must account for the fact that the mean-square longitudinal fluctuations $\langle\delta\ell^2(t)\rangle$ of a long filament involve the sum (in quadrature) of independently fluctuating segments along a full filament of length $\ell$. The typical size of such independently fluctuating segments at time $t$ is $\ell_\perp(t)$, of which there are $\ell/\ell_\perp(t)$. As shown above, the mean-square amplitude of longitudinal fluctuations of a fully relaxed segment of length $\ell_\perp(t)$ is of order $\ell_\perp(t)^4/\ell_p^2$. Thus, the longitudinal motion is given by (Granek 1997; Gittes 1998)
\begin{equation}
\langle\delta\ell(t)^2\rangle\sim\frac{\ell}{\ell_\perp(t)}\times\frac{\ell_\perp(t)^4}{\ell_p^2}\sim t^{3/4},
\label{eq:longmotion}
\end{equation}
where the mean-square amplitude is smaller than for the transverse motion by a factor or order $\ell/\ell_p$. Thus, for both the short-time fluctuations as well as the static fluctuations of a filament segment of length $\ell$, a point on the filament explores a disk-like region with longitudinal motion smaller than perpendicular motion by a factor of order $\ell/\ell_p$, which is assumed here to be small. This is illustrated in Fig.\ \ref{fig:Filament2}. From the end-to-end fluctuations in Eq.\ \eqref{eq:longmotion}, it is also possible to determine the time- or frequency-dependent compliance of such a filament to a longitudinal force applied at one end, using the Fluctuation-Dissipation Theorem. The result can be expressed in terms of a (complex) spring constant $K_{\rm eff}(\omega)$:
\begin{equation}
K_{\rm eff}(\omega)=\kappa\ell_p\left(-2i\zeta/\kappa\right)^{3/4}\omega^{3/4}\;.\label{eq:Keff}
\end{equation}
This is valid in the limit of high frequency and high molecular weight. As discussed in Sec.\ \ref{sec:solutions}, additional relaxations are expected for finite length polymers.

For the problem as stated above, \emph{i.e.}, for an isolated fluctuating filament in a quiescent solvent, there is a potential problem with the analysis above, which includes only the effect of drag for motion perpendicular to the filament (Everaers 1999). In fact, there is a finite propagation of tension along a semiflexible filament, expressed by yet another length (Morse 1998b)
\begin{equation}
\ell_\parallel(t)\sim\sqrt{\ell_\perp(t)\ell_p}\sim t^{1/8}.\label{Everears}
\end{equation}
This represents, for instance, the range along the filament over which the tension has spread from a point of disturbance. At very short times, it is possible to observe a $t^{7/8}$ motion of ends of a freely fluctuating filament in a quiescent solvent, rather than the $t^{3/4}$ in Eq.\ (\ref{eq:longmotion}) (Everaers 1999). For the high-frequency rheology of semiflexible polymer networks, however, only a dynamical regime corresponding to Eq.\ (\ref{eq:longmotion}) is observed (Gittes 1997; Koenderink 2006) and expected (Morse 1998a; Gittes 1998). This will be examined in greater detail in Sec.\ \ref{sec:solutions}.  

There are two important extensions to the dynamic analysis above. First, when a filament is subject to tension $\tau$, the equation of motion in Eq.\ (\ref{eq:Langevin}) must be modified to include $\tau\frac{\partial^2}{\partial x^2}u(x,t)$. Then, the relaxation rate in Eq.\ (\ref{eq:omega}) becomes
\begin{equation}
\omega(q)=\frac{1}{\zeta}\left(\kappa q^4+\tau q^2\right).\label{eq:omega-tau}
\end{equation}
As noted above, tension becomes dominant for wavelengths longer than $\ell_t\sim\sqrt{\kappa/\tau}$. Here, the relaxation rate reduces to
$\tau q^2/\zeta$
This has important consequences,e.g., for the end-to-end fluctuations \cite{Granek1997}:
\begin{equation}
\langle\delta\ell(t)^2\rangle\sim t^{1/2}.
\label{eq:longmotion-tau}
\end{equation}
Tension also leads to a corresponding $G(\omega)\sim\omega^{1/2}$ regime in the rheology of networks \cite{Mizuno2007}.
The analysis based on Eq.\ (\ref{eq:omega-tau}) fails for large applied tensions. As the linearized theory above appears to be sufficient for many purposes, including a quantitative understanding of network and solution rheology, this is what we have focused on here. It is interesting to note, however, that there are rich nonlinear aspects of the full dynamics. These have been addressed in part by scaling analyses \cite{Seifert1996} and more rigorous multi-scale perturbative approaches \cite{Hallatschek2007a,Hallatschek2007b}.

The second important extension of the analysis above is required for quantitative dynamics of finite length filaments, such as any practical situation of isolated or dilute chains freely fluctuating in a solvent. The important thing to note here is that, while the mode decomposition and amplitudes in Eqs.\ (\ref{eq:decomp} and \ref{eq:Equipartition}) are valid when considering static fluctuations, they do not correspond to proper normal modes, i.e., modes that exhibit single-exponential relaxation, as in Eq.\ (\ref{eq:amp}). Put differently, the simple Fourier modes mix in their dynamics. For a discussion of the proper analysis of this situation, we point the interested reader to excellent discussions and derivations in Ref.\ \cite{Aragon1985,Wiggins1998}, as well as specific discussion of the dynamics of isolated F-actin and microtubule filaments in Refs.\ \cite{Gittes1993,Brangwynne2007}.

\subsection{Wormlike bundles}
\begin{figure}[b]
\includegraphics[width=\columnwidth]{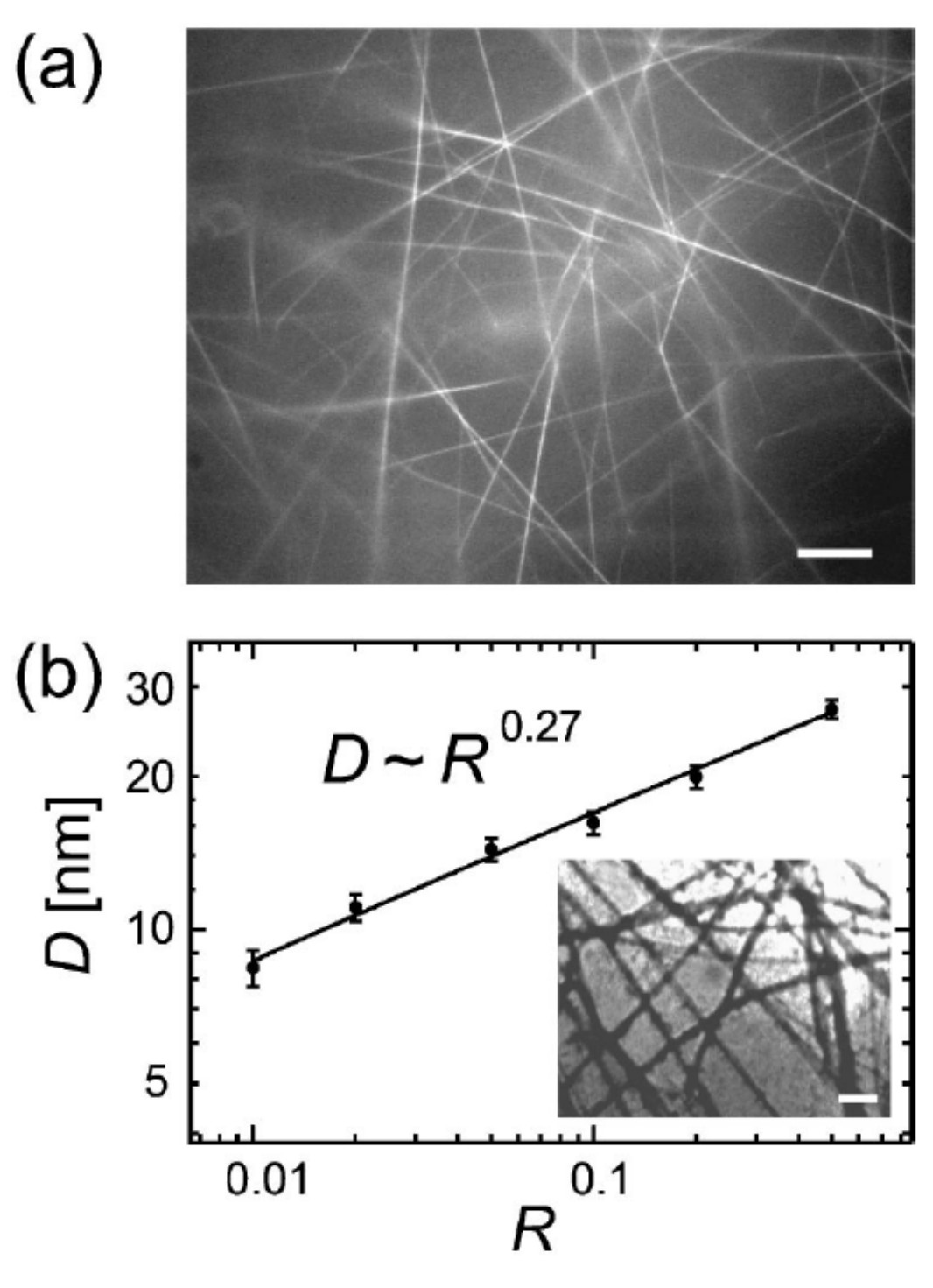}
\caption{ a) Fluorescence microscopy image of a bundled F-actin network
(0.1 mg/ml actin) crosslinked by  fascin proteins (scale bar is 10 micron). (b) From transmission electron micrographs
 (inset, scale bar is 0.2 micron) a scaling relation for the
average bundle diameter $D$ is obtained. Image and data adapted from~\cite{Lieleg2007}.}
\label{fig:lielegbundle}
\end{figure}

In many biological systems individual filaments can be crosslinked or ligated together to form hierarchical bundles, which in some cases combine to form networks. For instance, actin is known to form networks of thick bundles when polymerized in the presence of certain actin crosslinker proteins~\cite{Hirst2005,Pelletier2003,Gardel2004a,Schmoller2009,Lieleg2010b,Kasza2010,Claessens2008}, and such bundles constitute important cytoskeletal components of live cells, including  filopodia, sensory hair cells and microvilli~\cite{Claessens2006,Bathe2008,Fletcher2010}. An illustrative example of a network of long, straight semiflexible bundles obtained by polymerizing actin in the presence of fascin crosslinking proteins is shown in Fig.~\ref{fig:lielegbundle}~\cite{Lieleg2007}. In this review we focus on work that addressed the mechanical properties of such bundles, and for more detail on the structural properties of bundles we refer the reader to~\cite{Kierfeld2005,Claessens2008,Grason2007,Grason2009,Shin2009}.

Can such bundles be described  as an inextensible chain with the standard worm like chain model with some effective, renormalized bending stiffness? We may already suspect various problems. For instance, the WLC model would not account for internal deformation modes of the bundle, i.e twisting or relative sliding (shear)  of the constituent filaments in the bundle. To address this, a new theory was proposed termed the worm-like bundle (WLB) model~\cite{Bathe2008, Heussinger2007b,Heussinger2010}, which explicitly accounts for the discrete character of the internal  architecture of the bundle, and its internal deformation modes. 
  
The WLB model describes a bundle of length $\ell$ as a collection of  $N$ filaments oriented in parallel, which are connected by crosslinks with a stiffness $k_x$ and spacing $\delta$ (see Figs.~\ref{fig: WLB} and ~\ref{fig: WLB2}). The individual fibers run the full length of the bundle, have a bending rigidity $\kappa$, and a stiffness $k_s$ on the scale of the crosslinking distance $\delta$. An important dimensionless parameter in this model is~\cite{Bathe2008, Heussinger2007b,Heussinger2010}
\begin{equation}
\alpha=\frac{k_x \ell^2}{k_s \delta^2},
\label{eq:bundlealpha}
\end{equation}
which is a measure of the competition between crosslink shearing and filament stretching. The continuum limit of this model is obtained by taking $N \rightarrow \infty$ at fixed bundle diameter, at which the WLB model describes a Timoshenko beam.

Importantly, it was found that the bundle can not be described by a single bending rigidity. Instead, the bending stiffness is state-dependent
and, in particular, depends on the wavenumber $q_n$ of a  bending mode of the bundle~\cite{Heussinger2007b}
\begin{equation}
\kappa_n=N \kappa \left[1+\left(\frac{12 \hat{\kappa}}{N-1}+(q_n \lambda)^2\right)^{-1}\right],
\end{equation}
with a dimensionless bending stiffness $\hat{\kappa}=\kappa/k_s \delta b^2$ and length scale $\lambda=(\ell/\sqrt{\alpha})\sqrt{M \hat{\kappa}/(M-1/2)}$, where $b$ is the inter filament distance and $M=\sqrt{N}/2$.  
Note, however, that $\lambda$ itself does not depend on bundle length $\ell$.

For a fixed wavenumber $q_n\sim n/\ell$, three elastic regimes can be identified: 1) If the shear stiffness of the bundle is large, $\alpha \gg N$, the system is in the ``tightly coupled" limit and the fibers do not slide (shear) relative to each other to accommodate bundle bending, and $\kappa_n\sim N^2 k_s$. 2) If the shear stiffness is small, $\alpha\ll1$, the system is  ``decoupled" and the fibers contribute to bundle bending independently, and  $\kappa_n\sim N \kappa$. 3) For intermediate shear stiffness, $1\ll \alpha <<N$, the bundles' bending stiffness is dominated by internal shearing, and thus has a strong wavelength dependence, $\kappa_n\sim N k_{x} q_n^{-2}$. 

The various regimes for the dependence of the bundles bending rigidity on $N$ have been observed for single F-actin bundles formed with variety of actin binding proteins~\cite{Claessens2006}. This was done by inferring the scaling dependence of the bundles persistence length, $\ell_p$, on $N$. For actin bundles crosslinked by plastin, evidence of a decoupled regime was found ($\ell_p\sim N$), while for actin crosslinked with fascin and  $\alpha$-actinin, the results were consistent with tightly-coupled behavior ($\ell_p\sim N^2 $) for lower values of $N$, which then crossed over to decoupled behavior for larger values of $N$. By contrast, actin bundles formed by the depletion agent PEG exhibited tightly-coupled behavior for a broad range of $N$. Thus, the molecular details of the crosslinker can have an important impact on the mechanical (and dynamical) behavior of bundles. Clearly, this will impact the mechanics of a bundle network on the macroscopic level. Indeed, various studies have found evidence for decoupled or coupled regimes in the macroscopic rheology of bundle networks~\cite{Lieleg2007}. Moreover, evidence for such behavior has also been found for networks of fibrinogen, which forms complex, hierarchical fibers consisting of many protofibrils. The resulting fibers can be modeled as bundles, although the protofibrils are less distinct than, e.g., actin in bundles. Interestingly, fibrin networks have also shown evidence of a nonlinear thermal compliance that goes beyond the mechanical models above \cite{Piechocka2010}.

It has also been argued that the WLB model is a more appropriate description than the WLC model for microtubules, owing to their anisotropic molecular architecture~\cite{Taute2008}: as shown in Fig.\ \ref{fig:biopolymers}, microtubules are cylindrical and tubular in shape, but the axial binding of tubulin proteins into \emph{protofilaments} is stronger than the lateral binding of these filaments. Interestingly, it has been reported that microtubules have a length- or wavelength-dependent stiffness~\cite{Pampaloni2006,Taute2008}, and the elastic anisotropy of their structure has been implicated as the cause of this~\cite{Heussinger2010}. However, it was also noted that the degree of anisotropic elasticity required to account for the reported length dependence is extremely large, and much larger than recent detailed simulations found~\cite{Sept2010}. Moreover, AFM experiments probing the response of microtubules to radial forces have not shown evidence for significant anisotropy~\cite{dePablo2003}. Thus, this remains and interesting and unresolved puzzle \cite{Liu2012}.

The remarkable wavelength dependence of the bundles bending rigidity has a number of important implications.
The entropic stiffness of a bundle in the parameter range $\lambda \sqrt{N} \gg \ell \gg \lambda$ is given by~\cite{Heussinger2007b}
\begin{equation}
k_{\rm entr} \sim \frac{\left(  N \kappa\right)^2}{\ell \lambda^3 k_B T}
\end{equation}
Importantly, the strong $1/\ell^4$ dependence of length in the WLC model (see Eq.~\eqref{eq:wlcstiffness}), is replaced here by $1/\ell \lambda^3$ with implications for the modulus of a network and its scaling with concentration~\cite{Heussinger2007b}. Under a compressive force, the bundle  buckles (see Sec.~\ref{sec:buckling}) at a force threshold $f_c\sim \kappa_n/\ell^2$ (with $n=1$) . However, in the intermediate regime, where internal shear dominates, $\kappa_n\sim \ell^2$, and thus the strong length-dependence of the buckling force threshold drops out $f_c\sim N \kappa/\lambda^2$. It has been argued that this mechanism may stabilize biopolymer assemblies under compression, with possible implications for microtubules, and actin bundles in growing filliopodia~\cite{Bathe2008,Heussinger2007b}.

\begin{figure}
\includegraphics[width=\columnwidth]{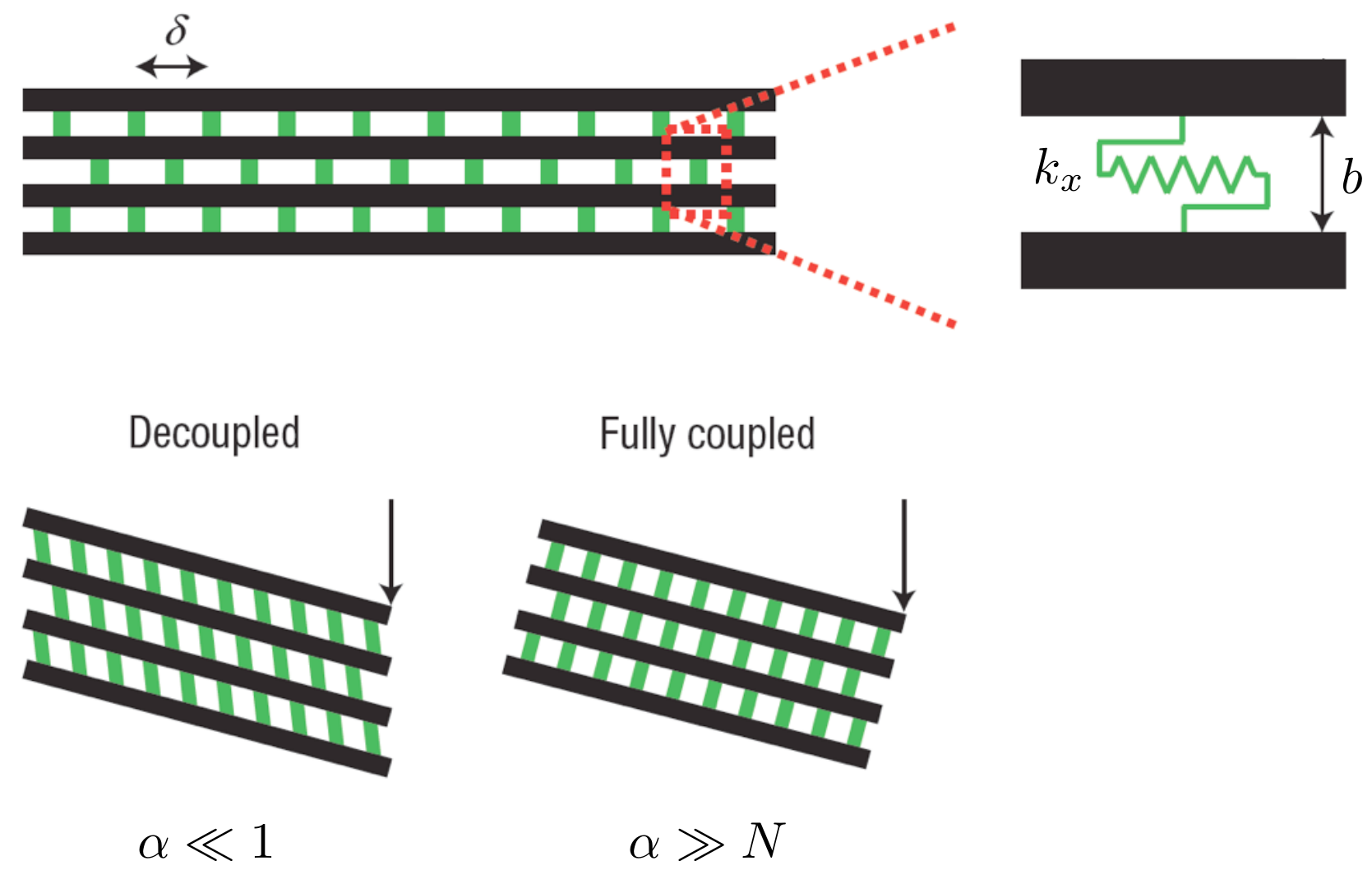}
\caption{Semiflexible filaments (black) are coupled to nearest-neighbor filaments by crosslinks (green) with axial spacing, $\delta$, and stiffness, $k_x$.
The inter-filament spacing, $b$, is fixed by the length of the intervening ABPs and remains constant in tightly crosslinked bundles. The ratio $\alpha$ (See Eq.~\eqref{eq:bundlealpha}), represents the competition between crosslink shearing and filament extension or compression during bundle bending. This ratio determines the degree of coupling in the bundle.  Image adapted from~\cite{Claessens2006}.}
\label{fig: WLB}
\end{figure}

\begin{figure}
\includegraphics[width=\columnwidth]{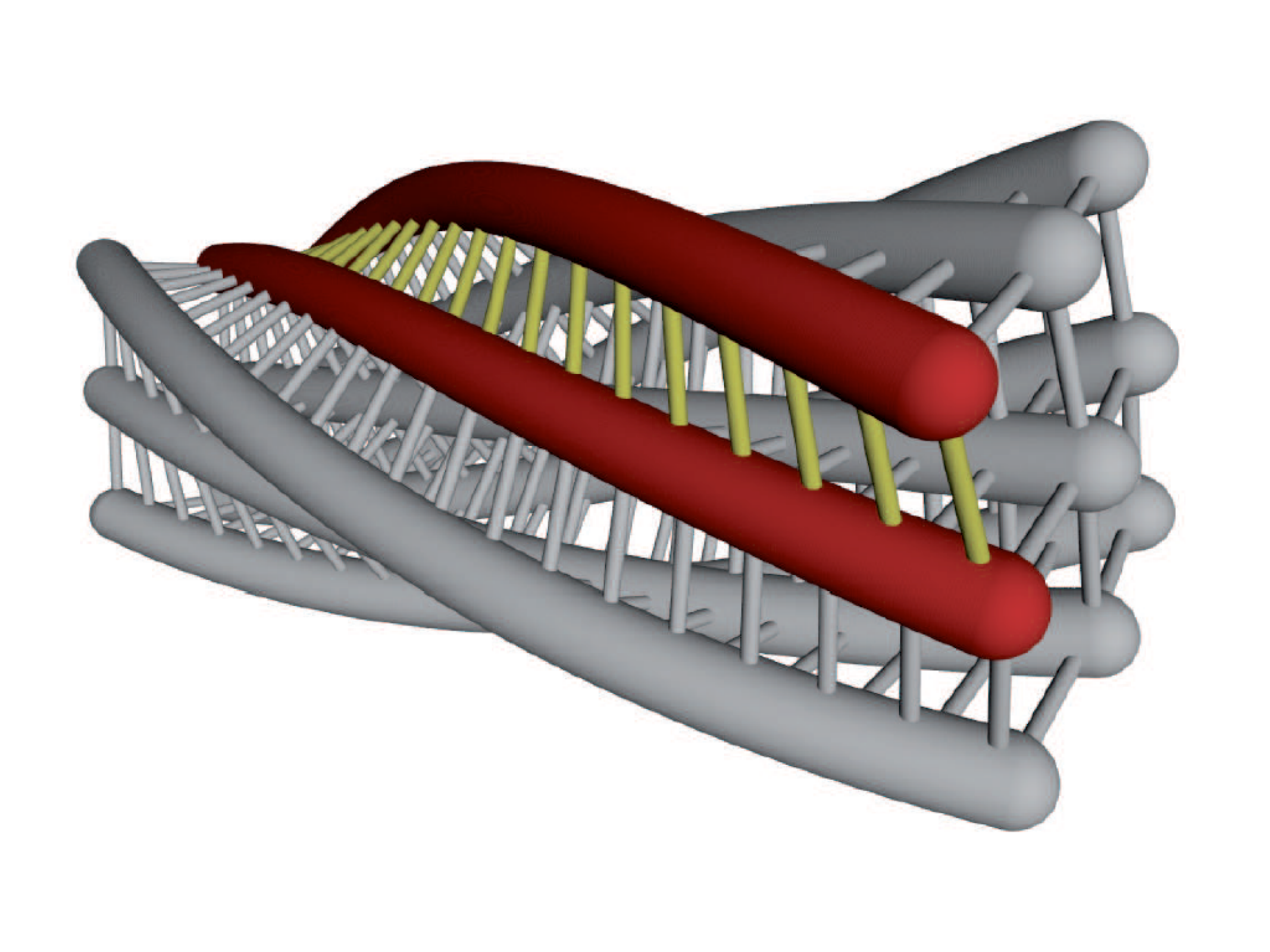}
\caption{Schematic illustrating the wormlike bundle model.
Bundles  consist of regular arrangements of filaments held together by equally spaced crosslinking proteins. The bundle
can bends and twists in space, which can result in internal filament sliding/shear. Image adapted from~\cite{Heussinger2010}.}
\label{fig: WLB2}
\end{figure}

\section{Entangled solutions of semiflexible polymers}
\label{sec:solutions}
\subsection{Rheology of entangled networks}
Given the rigidity of semiflexible polymers at scales shorter than their contour length, it is not surprising that in solutions they interact with each other in very different ways than flexible polymers would, \emph{e.g.}, at the same concentration. In addition to the important characteristic lengths of the molecular dimension (say, the filament diameter $2a$), the material parameter $\ell_p$, and the contour length of the chains, there is another important new length scale in a solution, the \emph{mesh size}, or typical spacing between polymers in solution, $\xi$. A simple estimate~\cite{Schmidt1989} shows how $\xi$ depends on the molecular size $a$ and the polymer volume fraction $\phi$. In the limit that the persistence length $\ell_p$ is large compared with $\xi$, we can approximate the solution on the scale of the mesh as one of rigid rods. Hence, within a cubical volume of size $\xi$, there is of order one polymer segment of length $\xi$ and cross-section $a^2$, which corresponds to a volume fraction $\phi$ of order $(a^2\xi)/\xi^3$. Thus,
\begin{equation}
\xi\sim a/\sqrt{\phi}.
\end{equation}

While the mesh size characterizes the typical spacing between polymers within a solution, it does not entirely determine the way in which they interact sterically with each other. For instance, for a random static arrangement of rigid rods, it is not hard to see that polymers will not touch each other on average except on a much larger length: imagine threading a random configuration of rods at small volume fraction with a thin needle. An estimate of the distance between typical interactions (entanglements) of semiflexible polymers must account for their thermal fluctuations~\cite{Odijk1983}. As we have seen above, the transverse range of fluctuations $\delta u$ a distance $\ell$ away from a fixed point grows according to $\delta u^2\sim\ell^3/\ell_p$. Along this length, such a fluctuating filament explores a narrow cone-like volume of order $\ell\delta u^2$. An entanglement that leads to a constraint of the fluctuations of such a filament occurs when, with probability of order unity, another filament crosses through this volume, in which case it will occupy a volume of order $a^2\delta u$, since $\delta u\ll\ell$. Thus, the volume fraction and the contour length $\ell$ between constraints is of order $\phi\sim a^2/(\ell\delta u)$. Taking the corresponding length as an entanglement length
\begin{equation}
\ell_e\sim (a^4\ell_p)^{1/5}\phi^{-2/5},
\label{eq:meshsizeconcen}
\end{equation}
which is larger than the mesh size $\xi$ in the semiflexible limit $\ell_p\gg\xi$.

These transverse entanglements, separated by a typical length $\ell_e$, govern the elastic response of solutions, in a way first outlined by~\cite{Isambert1996}. A more complete discussion of the rheology of such solutions can be found in \cite{Morse1998b,Morse1998c}, along with experimental evidence in~\cite{Hinner1998}. The basic result for the rubber-like plateau shear modulus $G_0$ for such solutions can be obtained by noting that the number density of entropic constraints (entanglements) is $n\sim 1/(\xi^2\ell_e)$, where $\rho\sim 1/\xi^2$ is the concentration in total chain length per volume. In the absence of other energetic contributions to the modulus, the reduction in entropy associated with these constraints results in a shear modulus proportional to $k_{\rm\small B}T$ per entanglement:
\be
G\sim \frac{\kT}{\xi^2\ell_e}\sim\phi^{7/5}.\label{eq:Isambert}
\ee 
This is analogous to the case of flexible polymers, where $G\sim k_{\rm\small B}T/\xi^3$. It is also interesting to note that the semiflexible result in Eq.\ (\ref{eq:Isambert}) is strictly smaller than the corresponding flexible polymer result for the same mesh size, since stiff polymers are fundamentally less entangled than flexible polymers, and $\ell_e>\xi$. The modulus in Eq.\ (\ref{eq:Isambert}) has been well-established in experiments, such as those of~\cite{Hinner1998}.

With increasing frequency, or for short times, the macroscopic shear response of solutions is expected to show the underlying dynamics of individual filaments. One of the main signatures of the frequency response of polymer solutions in general is an increase in the shear modulus with increasing frequency. In practice, for high molecular weight F-actin solutions of approximately 1 mg/ml, this is seen for frequencies above a few Hertz. Initial experiments measuring this response by imaging the dynamics of small probe particles have shown that the shear modulus increases as $G(\omega)\sim\omega^{3/4}$~\cite{Gittes1997,Schnurr1997}, which has since been confirmed in other experiments and by other techniques \cite{Koenderink2006,Hoffman2006,Deng2006,Gardel2004b,Gisler1999}.

This behavior can be understood in terms of the dynamic longitudinal fluctuations of single filaments, as shown above \cite{Gittes1998,Morse1998a}. Much as the static longitudinal fluctuations $\langle\delta\ell^2\rangle \sim \ell^4/\ell_p^2$ correspond to an effective longitudinal spring constant $\sim \kT\ell_p^2/\ell^4$, the time-dependent longitudinal fluctuations shown above in Eq. (\ref{eq:longmotion}) correspond to a time- or frequency-dependent compliance or stiffness, in which the effective spring constant increases with increasing frequency, as shown in Eq.\ (\ref{eq:Keff}). This is because, on shorter time scales, fewer bending modes can relax, which makes the filament less compliant and stiffer. Accounting for the random orientations of filaments in solution results in a frequency-dependent shear modulus
\begin{equation}
G(\omega)=\frac{1}{15}\rho\kappa\ell_p\left(-2i\zeta/\kappa\right)^{3/4}
\omega^{3/4}-i\omega\eta,
\label{eq:Gomega}
\end{equation}
where $\rho$ is the polymer concentration measured in length per unit volume. As shown in Fig.\ \ref{fig:Koenderinkhighfreq}, this frequency dependence of the shear modulus has been quantitatively confirmed in measurements on \emph{in vitro} actin networks~\cite{Koenderink2006}, and has also been reported for the high-frequency response of living cells~\cite{Deng2006,Hoffman2006}. Moreover, the experiments of Koenderink et al., also showed evidence of an additional relaxation predicted by Pasquali et al.\ \cite{Pasquali2001}. These authors identified a macroscopic relaxation associated with tension propagation along the polymer chains, which was predicted to occur at frequencies intermediate between the high-frequency $\omega^{3/4}$ regime and an entangled plateau in the rheology. The experiments in in the inset of Fig.\ \ref{fig:Koenderinkhighfreq} are consistent with the predicted relaxation, and are also consistent with the fact that the relaxation mechanism is predicted to be absent in cross-linked networks. 

\begin{figure}
\includegraphics[width=\columnwidth]{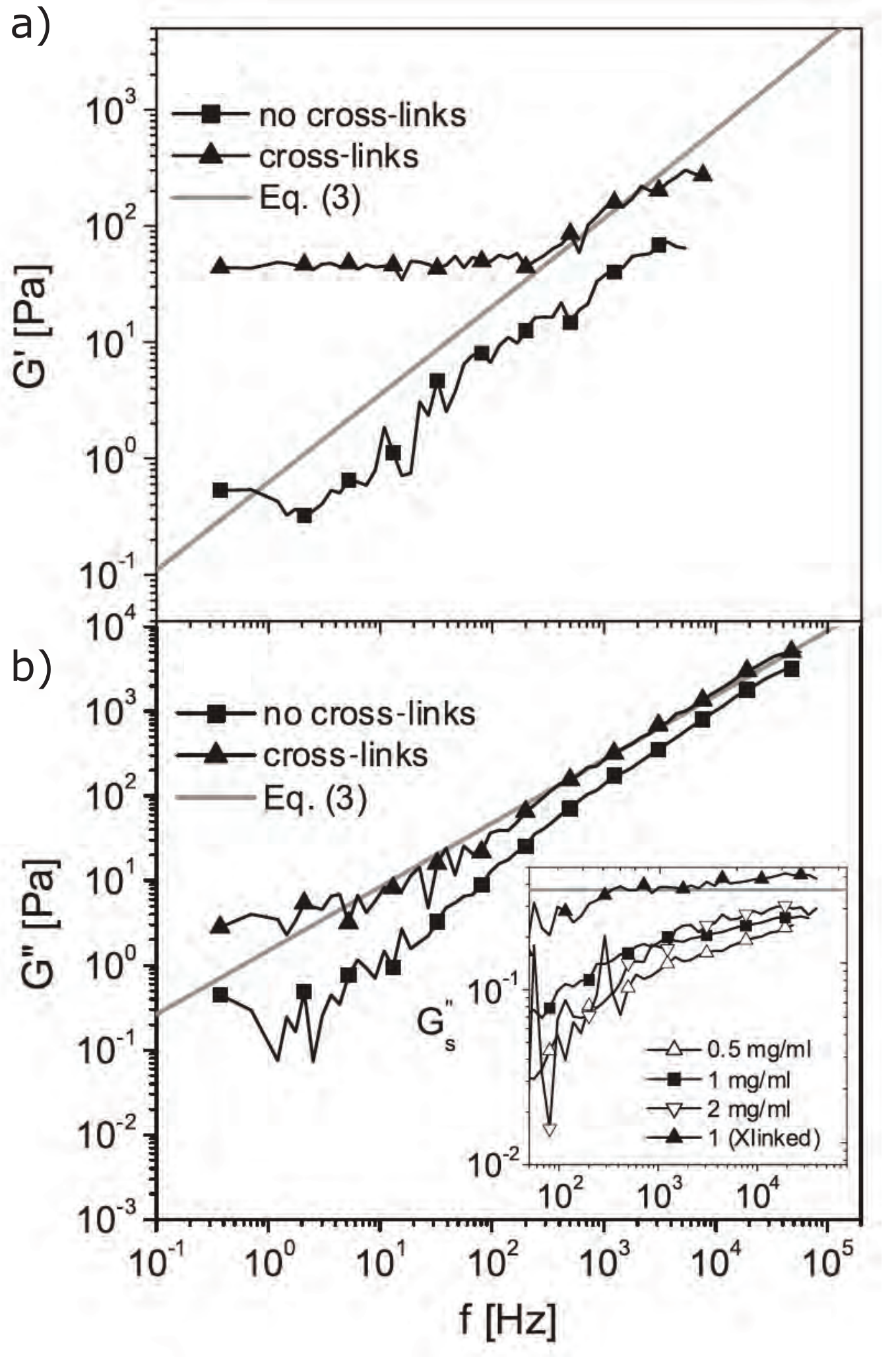}
\caption{(a) Storage modulus $G'(\omega)$ and b) loss modulus $G''(\omega)$
of 1mg/ml solutions of F-actin filaments with (triangles) and without (squares) crosslinks plotted against frequency.
The solid lines indicate theoretical predictions from Eq.~\eqref{eq:Gomega} Inset: scaled loss modulus $G_s''(\omega)=-\left[G''(\omega)+i \omega \eta\right]/(c_a \omega^{3/4})$. From \cite{Koenderink2006}.}
\label{fig:Koenderinkhighfreq}
\end{figure}

\subsection{Glassy wormlike chain model}

\begin{figure}
\includegraphics[width=\columnwidth]{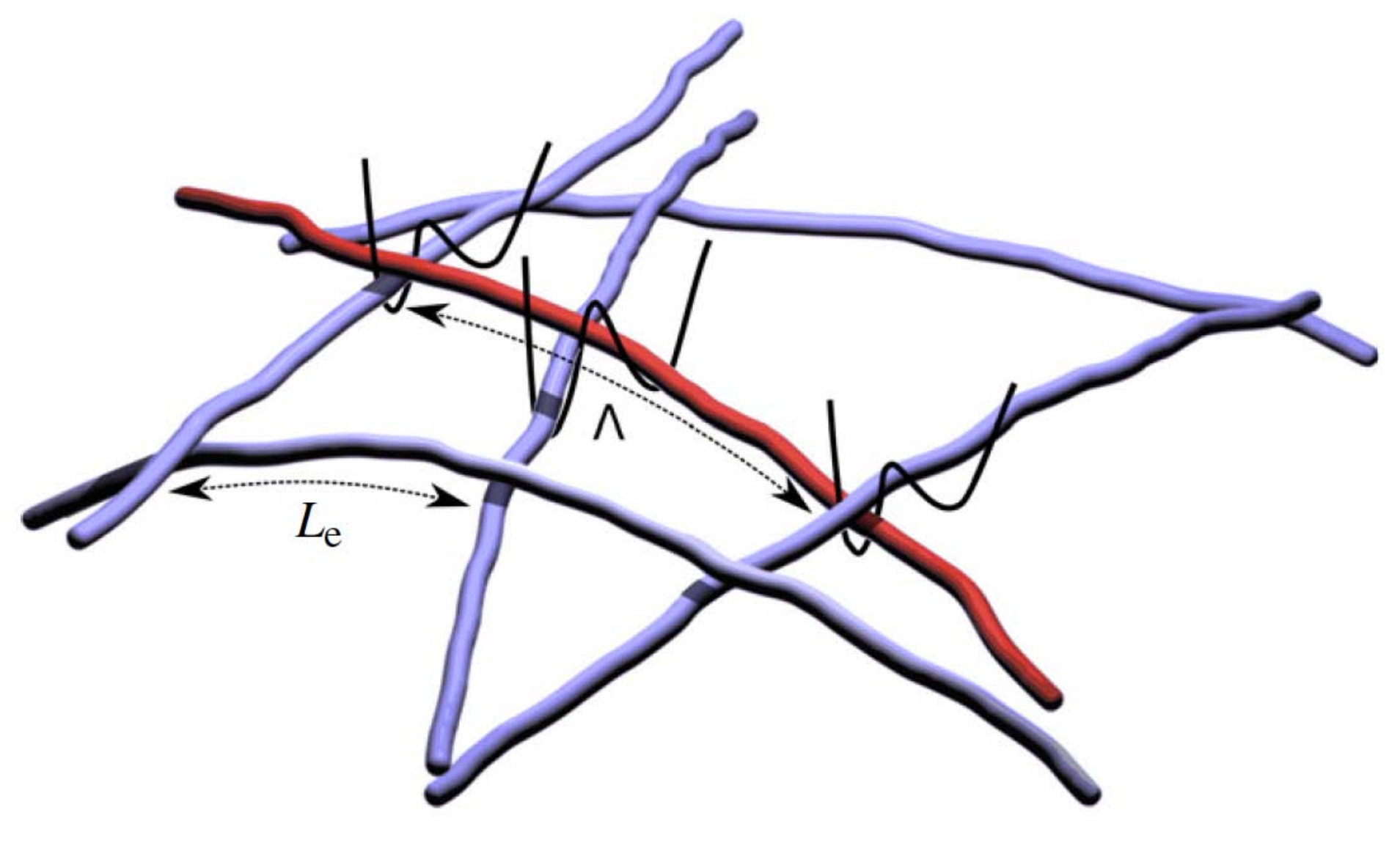}
\caption{Schematic to illustrate the Glassy worm-like chain model. The (red) test polymer can be trapped through effective interactions with the surrounding polymer solution. These interaction are indicated by the potential wells at sticky entanglement
points, which are on average separated by the entanglement length $\ell_e$. The test polymer
can bind or unbind by overcoming an energy barrier of height $\epsilon$. The average
distance length between the ``closed" bonds is represented by $\Lambda$. Image from~\cite{Wolff2010a}.}
\label{fig:GWLC}
\end{figure}

Microrheology experiments on live cells have provided evidence that cells may behave as soft glassy materials---existing close to a glass transition~\cite{Fabry2001,Deng2006}.
The rheology of cells was found to exhibit weak power law behavior $G\sim\omega^{0.17}$ over five decades in frequency. This weak frequency dependence can not be understood within existing theories for semiflexible polymer networks or solutions, and indeed appears to be reminiscent to the glassy rheology of other soft matter systems. This suggest that the cytoskeleton and its polymer constituents, which are  thought to provide the dominant contribution to the rheological measured by Fabry et al., may be surrounded by a ``glassy" environment. However, the affect of such a glassy environment of the dynamic rheology of a semiflexible polymer network remains poorly understood.

Kroy and Glaser addressed this issue by considering the affect of a glassy environment on the dynamics of a semiflexible polymer~\cite{Kroy2007,Kroy2008}. Their approach starts from the relaxation spectrum of a polymer under tension using the ordinary worm like chain model. From Eq.\ (\ref{eq:omega-tau}), one obtains a relaxation time for the $n$-th mode of wavevector $q_n=n\pi/\ell$
give by
\begin{equation}
t_n=\frac{1}{\omega(q_n)}=\frac{\zeta \ell^4}{\kappa \pi^4}\frac{1}{n^4+n^2 \tau/\tau_\ell}
\end{equation}
where $\tau_\ell=\kappa \pi^2/\ell^2$ and $\tau$ is the backbone tension. This spectrum describes the relaxation time of the $n$-th mode of the transverse fluctuations (see Sec.~\ref{sec:dynam}) for a filament under tension.

From the equilibrium mode amplitudes and the relaxation spectrum one can calculate various quantities of interest, such as  the dynamic structure factor and the rheological response. The question, asked by Kroy and Glaser, was how a glassy environment might affect the dynamics of the wormlike chain. Such a glassy environment may be thought of as a collection of traps distributed in space, with a broad power law distribution in strength (set by the height of a free energy barrier), locally pinning the polymer---along the lines of the trap models underlying soft glassy rheology~\cite{Sollich1997}

The trapping interactions will impact the relaxation timescales of the transverse fluctuations, and longer wavelength modes are expected to be slowed down more substantially, since there will be more trapping interactions (see Fig.~\ref{fig:GWLC}). To obtain a description of the glassy worm like chain (GWLC), 
the relaxation spectrum of the ordinary WLC is ``stretched" as follows
\begin{equation}
\tilde{t}_n=t_n \exp(N_n \epsilon)
\end{equation}
for wavelengths longer than $\lambda_n=2 \pi/q_n>\Lambda$. Here, $N_n=\lambda_n/\Lambda-1$ represents the number of trapping interactions per wavelength $\lambda_n$ of a given mode. The so-called stretching parameter $\epsilon$ controls how modes get slowed down by the trapping interaction through an Arrhenius factor; $\epsilon$ may be thought of as the characteristic height of the free energy barriers associated to the traps in units of $k_{\rm B} T$, while $\Lambda$ represent the typical distance between traps. In principle, one  expects that the rough energy landscape of a glassy environment might also affect the mode amplitudes, but this presents a very daunting calculation, and has so far remained elusive. However, it was argued that the slowing down of the relaxation times could capture the most relevant aspects of the glassy wormlike chain. Thus, it was assumed that the mode amplitudes of the ordinary WLC, but with a stretched relaxation spectrum. 

The predictions of the GWLC model include logarithmic tails in the long-time behavior of the dynamic structure factor, which can account for experiments on F-actin solutions~\cite{Semmrich2007}. In addition, within an affine framework the rheology of a collection of GWLC's can be calculated, also in the presence of a prestress on the network~\cite{Kroy2007}. The role of prestress was  included in the model by its affect on the parameter $
\epsilon$, since the free energy landscape gets tilted by the presence of a force directed along the reaction coordinate. It is interesting to contrast the GWLC predictions for the dynamic rheology with that predicted for a crosslinked network, which exhibits a frequency-independent plateau at low frequencies. The GWLC still allows relaxation beyond the interaction length $\Lambda$, in contrast to a permanently crosslinked network for which the transverse filament fluctuations can not relax for modes with wavelengths beyond the crosslinking scale $\ell_c$. Thus, in the GWLC model, the ``plateau" regime is no longer flat but appears to increase with frequency as a weak power law, consistent with experiments on live cells~\cite{Fabry2001,Deng2006}. In the GWLC model, the exponent of this power law depends on  the level of prestress and the interaction strength $\epsilon$, which is a phenomenological parameter that is hard to predict from first principles. Nonetheless, various predictions of the glassy wormlike chain agree favorably with the rheology of actin solutions, as well as live cells, and this was taken as evidence that these systems operate near a glass transition~\cite{Semmrich2007}. 

\subsection{Transient linkers and cross-link governed dynamics}
\begin{figure}
\includegraphics[width=\columnwidth]{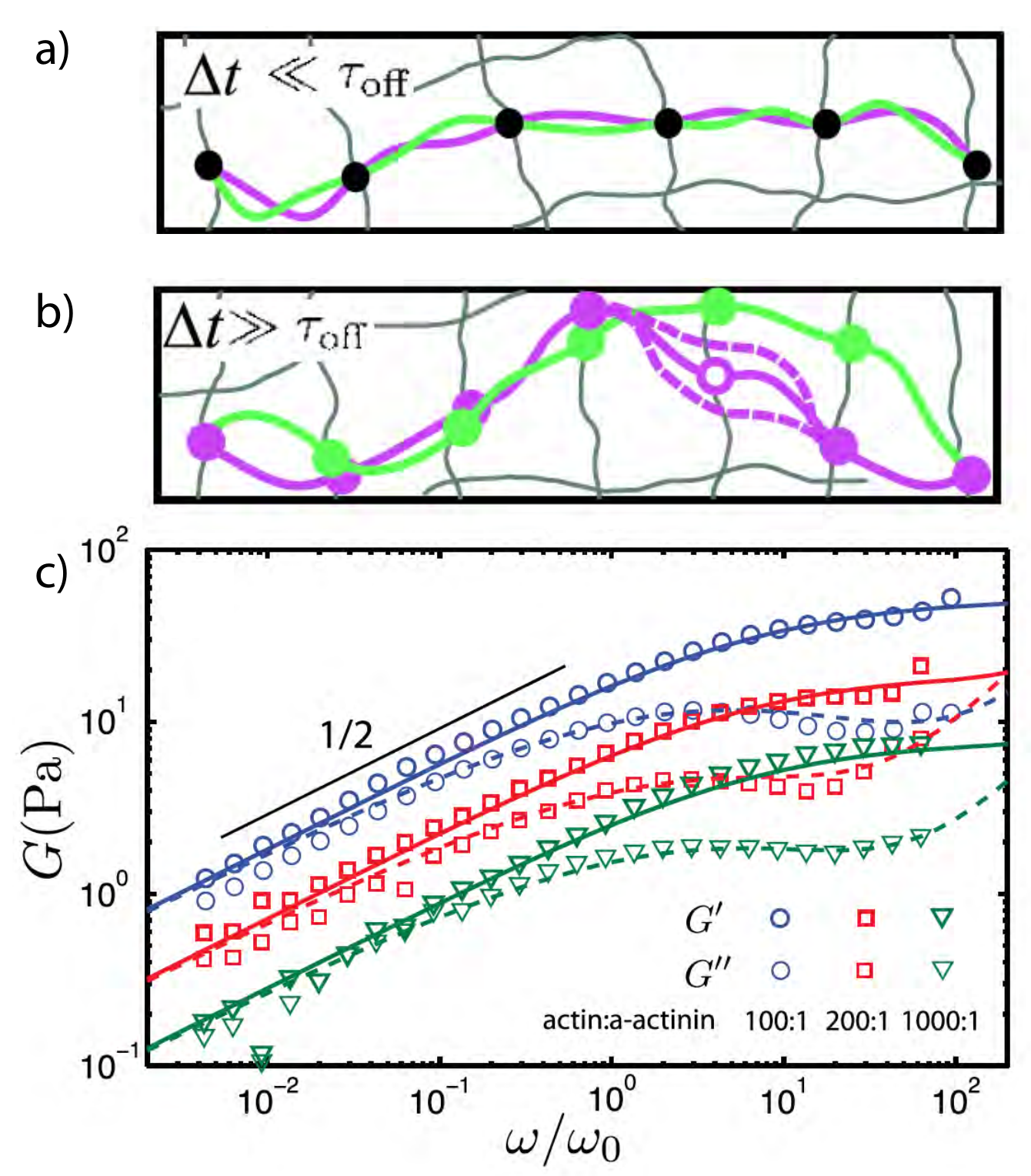}
\caption{a) For times shorter than the
unbinding time $\tau_{\rm off}$, only small scale
bend fluctuations between the effectively permanent cross-inks can relax, resulting
in a plateau in $G_0$ for frequencies $>1/\tau_{\rm off}$.
b) For longer times, large scale conformational relaxation can
occur via linker unbinding (open circle) and subsequent rebinding
at a new location. c) Measured linear rheology of a $23.8 \mu$M
actin network cross-linked with various concentrations of $\alpha$-actinin-4. The low-frequency behavior is consistent with
$G\sim\omega^{1/2}$. The solid and dashed lines are global fits using
the mean-field CGD model for the low-frequency regime
together with the known high frequency response. Image adapted from~\cite{Broedersz2010b}.}
\label{fig:CGD}
\end{figure}

One essential feature of many physiological biopolymer networks is the intrinsically dynamic nature of their cross-links. For instance, many actin-binding crosslinking proteins only form transient bonds between filaments.
Such systems represent a distinct class of polymeric materials whose long-time dynamics are not governed by viscosity or reptation~\cite{Doi1988}, but rather by the transient nature of their cross-links~\cite{Ward2008,Lieleg2008,Broedersz2010b,Lieleg2010,Lieleg2011,Yao2011,Yao2013,Heussinger2011,Strehle2011}.
This can give rise to a complex mechanical response, particularly at long times, where the network is expected to flow. In the next section we will discuss permanently crosslinked networks in detail. Here, we briefly discuss some recent work on transient networks, which forms a natural segue between solutions and crosslinked networks.

The simplest possible description of a material that is elastic on short time scales while flowing on long time scales is that of a Maxwell fluid; this exhibits a single relaxation time. Indeed, some experiments on transient networks have been modeled with a single relaxation time~\cite{Lieleg2008,Lieleg2010}; however, those experiments and others~\cite{Ward2008,Broedersz2010b,Yao2011,Yao2013}---probing longer relative time scales compared to the linker unbinding time---show amore complex low-frequency viscoelastic behavior, indicative of multiple relaxation times. 

To address these experimental observations, a microscopic model was developed for long-time network relaxation that is controlled by cross-link dynamics~\cite{Broedersz2010b}. This cross-link-governed dynamics (CGD) model describes the structural relaxation that results from many independent unbinding and rebinding events (FIg.~\ref{fig:CGD}), leading to very slow relaxation of stress. Using a combination of Monte Carlo simulations and an analytic approach, it was shown that this type of cross-link dynamics yields power-law rheology 
\be G\sim \omega^{1/2}\label{eq:HalfPower}\ee 
at frequencies below the crosslink unbinding rate, arising from a broad spectrum of relaxation rates. An important difference with the GWLC model for entangled solutions described in the previous section, is that in the CGD model the relaxation spectrum of the bending modes is derived from the microscopic unbinding of individual crosslinks at a constant rate $1/\tau_{\rm off}$, and the exponent of the power law rheology is fixed at a value of $1/2$. The predictions from the CGD model are in  quantitative agreement with experiments on F-actin networks using the transient linker protein $\alpha$-actinin-4~\cite{Broedersz2010b,Yao2011,Yao2013}, as shown in Fig.\ \ref{fig:CGD}c.

The CGD model can be qualitatively understood in simple physical terms, as follows. Each filament is assumed to be cross-linked into the network, with an average spacing $\ell_c$. Only filament bending modes between cross-links can relax (Fig.~\ref{fig:CGD}a), and the thermalization of these results in an entropic, spring-like response (\emph{e.g.}, that of Eq.\ \eqref{eq:wlcstiffness}). To account for transient crosslinking, the linkers may unbind at a rate $1/\tau_{\rm off}$. This initiates the relaxation of long-wavelength ($\lambda> \ell_c$) modes (Fig.~\ref{fig:CGD}b), giving rise to a reduced macroscopic modulus. However, the relaxation of successively longer wavelength modes becomes slower, as an increasing number of unbinding events are needed for such a relaxation. Interestingly, this suppression of longer wavelength relaxation cannot be accounted for, even phenomenologically, by an effectively larger viscosity for polymer motion. This would lead to a terminal relaxation and viscous-like response. Rather, this simple physical picture of multiple, uncoordinated unbinding events suggests a broad spectrum of relaxation times for the different mode wavelengths $\lambda\gtrsim\ell_c$, leading to a power-law rheology. Specifically, this model predicts the scaling in Eq.\ \eqref{eq:HalfPower} below the characteristic frequency  $1/\tau_{\rm off}$. 

This model does not account for the steric entanglements of the surrounding chains that can hinder the relaxation of bending modes. These constraints will begin to affect the relaxation of modes with wavelength $\lambda\gtrsim\ell_e$. Thus, the $\omega^{1/2}$ regime above can be expected for networks with $\ell_c$ substantially smaller than $\ell_e$, and this power-law regime is expected to give way to a solution-like plateau in Eq.\ \eqref{eq:Isambert}, albeit at much lower frequencies due to the transient binding. Finally, at the lowest frequencies, a dramatically slowed-down reptation, and correspondingly enhanced viscous behavior is expected.

The unbinding kinetics of the transient crosslinks may also depend on the level of an imposed macroscopic stress. One possibility is that stress leads to forced unbinding~\cite{Lieleg2007b,Kasza2010}, which has been studied within the GWLC framework~\cite{Wolff2010b}. Interestingly, various experiments with actin and myosins~\cite{Lieleg2010,Norstrom2011} or $\alpha$-actinin-4 crosslinks~\cite{Yao2013} appear to show a counter intuitive response to stress; macroscopical rheological experiments indicate that the upper bound of the cross-link governed dynamical regime shifts to lower frequencies with increasing stress, suggesting that the unbinding rate slows down with increasing stress. Such behavior may be caused by a ``catch-bond" mechanism at the molecular level, causing an enhances gel-like range in the macroscopic rheology (the fluid like
regime is shifted to lower frequency with increasing stress levels). 

Another framework to describe networks with transient crosslinks was proposed in~\cite{Heussinger2012}. This model went beyond the assumption of affine deformations (See Sec.~\ref{sec:affinemodel}) on the scale of a filament, and assigned an elastic stiffness to the transient crosslinks. Using a self-consistent effective medium approach, this model treated a test-filament connected by reversible crosslinks to a ``tube", representing the surrounding network. When a finite strain is applied, this confining tube deforms, stretching  the crosslinks and bending the test filament. Thus, using this approach, it is possible to self-consistently calculate the nonlinear mechanical properties of the network. When the network is deformed, crosslink unbinding processes lead to stress relaxation, resulting in a reduction of the network modulus with increasing strain. However, in the current model, both the crosslinks and filaments were treated as linear elements, and it will be interesting to investigate how the network softening caused by linker unbinding competes with stiffening contributions from the filaments and cross links~\cite{Kasza2009,Kasza2010,Wagner2006,Broedersz2008,Didonna2006,Gardel2006}.

\section{Cross-linked networks}
\label{sec:crosslinked_networks}

A major challenge in this field is to construct a formulation to bridge the gap between the mechanical properties of an individual polymer and the collective response of a network of such polymers. Here, among other things, the disordered nature of these networks complicates such a description because it can lead to nonuniform deformations, which may depend sensitively on the local details of network inhomogeneities. In some cases, such nonuniform strain fields can have a major qualitative impact on the network's elastic response. Nonetheless, we shall start by ignoring these spatial strain fluctuations. This constitutes the central assumption of the affine model~\cite{MacKintosh1995,Morse1998b,Storm2005}. This is inspired, in large part, by the success of such an approach in flexible polymer systems \cite{Doi1988,Rubinstein2003}. Then, after discussing various experimental studies of reconstituted biopolymer gels that have made direct comparisons with the affine model, we will resume with theoretical approaches that have gone beyond the affine assumption. 

In the following, we treat the crosslinks as freely hinging but otherwise noncompliant. Thus, we shall not cover various interesting aspects of physiological crosslinking proteins that are themselves compliant, or that tend to impose specific bond angles. Recent reviews covering these topics can be found in~\cite{Lieleg2010b,Fletcher2010}. Moreover, we focus here on isotropic networks. We point the interested reader to several recent studies addressing phase behavior and possible anisotropic networks in \cite{Zilman2003,Borukhov2005,Benetatos2007,Cyron2013}.

\subsection{Affine  model}\label{sec:affinemodel}
The affine model assumes that all crosslinks in the network deform according to the externally imposed (uniform) strain, $\gamma$. A polymer strand of length $\ell$ between two such crosslinks will deform through stretching or compressing by an amount that scales with its length and depends on both its orientation and the macroscopic strain $\gamma$. In the limit of small $\gamma$, this extension/compression is simply proportional to strain, $\delta\ell\sim\ell\gamma$, with a simple trigonometric prefactor related to the polymer orientation relative to the shear direction \cite{MacKintosh1995,Morse1998a}. Thus, this affine assumption completely specifies how each polymer strand in the network deforms, making the calculation of all the stresses in the system straightforward. The stress tensor is found by adding contributions from polymer strands over all orientations. These aspects form the essence of affine models of both the semiflexible networks reviewed here, as well as much earlier approaches to rubber elasticity \cite{James1943,Wall1951,Rubinstein2003,Doi1988} that have been the inspiration for much of what follows in this section. 

We begin with the approach by~\cite{Storm2005}, which extends the small-strain approach of Refs.\ \cite{MacKintosh1995,Morse1998a} to larger strains. Here, we consider a semiflexible polymer strand with an orientation $\hat{n}$ in the initially undeformed network. The deformation of such a strand is described by the \emph{uniform} Cauchy deformation tensor $\Lambda_{ij}$. For example, for a simple shear of the $xy$-plane in the $x$-direction we have
\begin{equation}
{\bf \Lambda}=\left(
\begin{array}{ccc}
 1 & 0 & \gamma \\
 0 & 1 & 0 \\ 
 0 & 0 & 1
\end{array}
\right).\label{eq:Lambda}
\end{equation}
The total polymer length per unit volume in the undeformed network, $\rho$, is not conserved under this deformation. While the volume is conserved, the total length of polymer will increase when the network is deformed. 

To calculate all components of the stress tensor, we need to decompose the various contributions to the stress. It is helpful to recall that the stress is a rank-2 tensor because it contains both information about the direction of the forces in a given plane, as well as the orientation of this plane. The length density of strands per unit volume crossing a plane oriented perpendicular to the $j$-direction transforms as $\frac{\rho}{\text{det}\Lambda}\Lambda_{jk} n_k$, where the determinant, $\text{det}\Lambda$, accounts for the volume change associated with the deformation. For a simple shear, as considered here, the volume is conserved, and thus, $\text{det}\Lambda=1$. The tension in a strand between the two affinely deforming crosslinks is denoted as $\tau(|\Lambda\hat{n}|-1)$, where $|\Lambda\hat{n}|-1$ represents the axial strain of the polymer. 
The $i$-component of this tension is given by $\tau(|\Lambda \hat{n}|-1)\Lambda_{il}n_l/|\Lambda\hat{n}|$. Adding all these contributions, weighed by the amount of polymer length crossing the $j$-plane, results in the $ij$-component of the symmetric stress tensor~\cite{Storm2005,Morse1999},
\begin{equation}\label{EQ:NL_stresstensor}
\sigma_{ij}=\frac{\rho}{\text{det}\Lambda}\left\langle\tau(|\Lambda
\hat{n}|-1)\frac{\Lambda_{il}n_l \Lambda_{jk}n_k}{|\Lambda
\hat{n}|}\right\rangle,
\end{equation}
where summation of repeated indices is implied.
The angular brackets indicate an average over the distribution of chain orientations in the initial, undeformed network.
For a small shear deformation $\gamma\ll1$, the stress simplifies to~\cite{Gittes1998,Morse1998b}
\begin{equation}
\sigma_{ij}=\rho
\left<\tau(\gamma n_x n_z)\hat{n}_i\hat{n}_j\right>+\mathcal O(\gamma^3),
\label{eq:small strain affine}
\end{equation}
where $\gamma n_x n_z$ represents the axial strain of a polymer with orientation $\hat n$ due to the strain tensor in Eq.\ (\ref{eq:Lambda}).

\subsubsection{Affinely deforming semiflexible polymer networks}\label{sec:AffThermalSmallStrain}
So far we have not specified the force-extension behavior of the network's polymer constituents, which is implied by the $\tau(\gamma n_x n_z)$ term in Eq.\ (\ref{eq:small strain affine}).
Here, we consider the case of a semiflexible polymer networks. To model an inextensible, semiflexible polymer of length
$\ell_c$ between two point-like, freely-hinging cross-links in the network, we can use
the WLC model~\cite{Kratky1949,Marko1995} in the semiflexible limit $\ell_p \gtrsim \ell_c$~\cite{MacKintosh1995}, which was discussed
in Sec.\ref{sec:force extension}.

Taking the linearized force-extension relation for a semiflexible polymer in Eq.\ (\ref{eq:wlcstiffness}), the tension above in the small strain limit becomes
\be
\tau(\gamma n_x n_z)=\frac{90\kappa\ell_p}{\ell_c^3}\gamma n_x n_z 
\ee
and the shear stress becomes 
\be
\sigma_{xz}=\rho\frac{90\kappa\ell_p}{\ell_c^3}\gamma \langle n_x n_z n_x n_z\rangle.
\ee
For an assumed isotropic distribution of orientations $\hat n$, we obtain the linear shear modulus of a semiflexible polymer network,
\begin{equation}
G_0=6 \rho \frac{\kappa^2}{k_B T \ell_c^3}.
\label{eq: linear thermal modulus}
\end{equation}
Thus, the network stiffness depends sensitively on the crosslinking density.

\begin{figure*}
\includegraphics[width=2 \columnwidth]{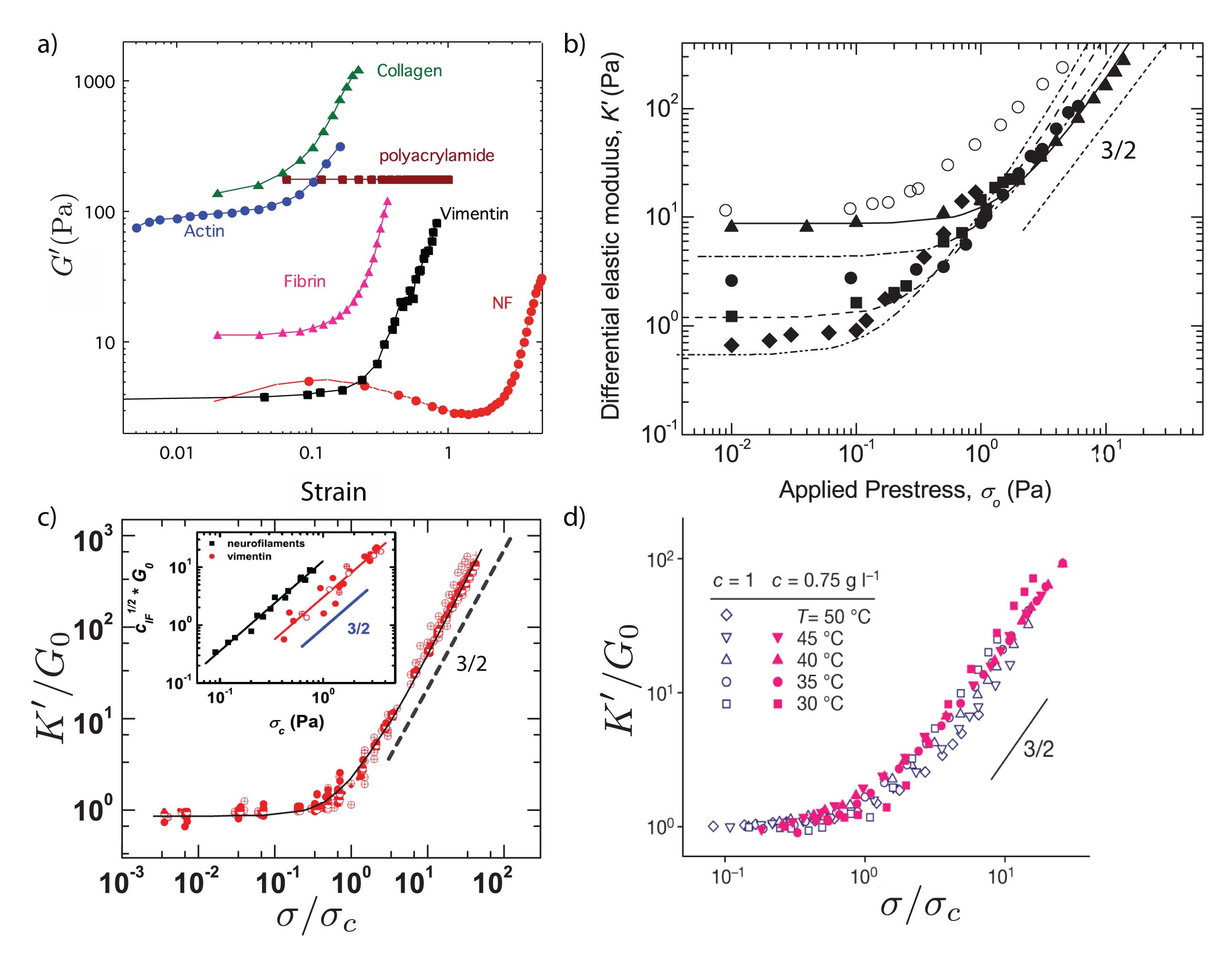}
\caption{a) The shear modulus $G'=\sigma/\gamma$ as a function of strain for various reconstituted biopolymer networks and polyacrylamide (from~\cite{Storm2005}). b) The differential shear modulus $K=d\sigma/d\gamma$ as a function of applied external stress $\sigma_0$ for reconstituted actin networks croslinked by scruin \cite{Gardel2004a}. The lines indicate the theoretically predicted form of stiffening for small strains, as outlined in Sec.\ \ref{sec:AffThermalSmallStrain}. 
c) Neurofilament network moduli ($K=d\sigma/d\gamma$) normalized by the linear modulus $G_0$ vs applied stress normalized by the characteristic stress $\sigma_c$ for the onset of nonlinearity (adapted from \cite{Lin2010b}). These data are compared with the predicted behavior for the small-strain approximation (solid line) introduced in \cite{Gardel2004a}, as well as the asymptotic 3/2 scaling (dashed line). The inset also shows a comparison with the behavior predicted in Eq.\ (\ref{eq:PredGvsSigma-c}) for both Neurofilaments and Vimentin intermediate filaments \cite{Lin2010b}. The difference between these two filament types is consistent with a difference between their persistence lengths. 
d) Biomimetic polyisocyanopeptide hydrogels also show nonlinear rheology consistent with the affine thermal model in Sec.\ \ref{sec:AffThermalSmallStrain} \cite{Kouwer2013}.}
\label{fig: fiber network rheology}
\end{figure*}

Using the small strain approximation, $\gamma\ll1$, for the stress tensor (Eq.~(\ref{eq:small strain affine})), we can also cast the nonlinear network response in a universal form~\cite{Gardel2004a}
\begin{equation}
 \tilde{\sigma}_{ij}=\left<\phi(\tilde{\gamma} n_x n_z)\hat{n}_i\hat{n}_j\right>
\label{eq: AFM universal form}
\end{equation}
where $\tilde{\sigma} =\sigma/ \sigma_c$, $\tilde{\gamma}=\gamma/\gamma_c$, and $\phi$ was defined in Eq.~\eqref{eq:extension}. Here, the characteristic strain and stress for the onset of nonlinearity are defined as
\begin{equation}
\gamma_c=\frac{1}{6}\frac{\ell_c}{\ell_p} \ \ \ \ \text{and} \ \ \ \ \sigma_c=\rho \frac{\kappa}{\ell_c^2}.
\label{eq: AFM critical strain and stress}
\end{equation} 
Beyond these characteristic values, the differential shear modulus, $K=d\sigma/d\gamma$, asymptotically approaches a scaling regime where $K\sim\sigma^{3/2}$. This can be seen by the high tension limit of the force extension relation in Eq.\ (\ref{eq:DivergenceForce}), since $\sigma\sim\tau$ and 
\be
\frac{d\sigma}{d\gamma}\sim\frac{d\tau}{d(\delta\ell)}\sim\frac{1}{|\delta\ell-\Delta\ell|^{-3}}\sim\tau^{3/2}.
\ee
This scaling form is not exact, as it does not account for the angular distribution of filaments, but this does not significantly affect the asymptotic behavior \cite{Gardel2004a}.

\subsubsection{Comparison of the affine model to experiments on reconstituted biopolymer networks}\label{sec:affinethermalmodel}
A comparison of the functional form of the nonlinear elastic response of a range of biopolymer networks reveales a remarkable qualitative similarity, even between intracellular and extracellular biopolymers, as shown in Fig.~\ref{fig: fiber network rheology} \cite{Storm2005}. All these systems stiffen under applied strain. These data suggested that the nonlinear elasticity across systems may have the same biophysical origins, despite large differences in architectural details and mechanical properties at both the filament and network level. Indeed, it has been shown, for systems ranging from actin \cite{Gardel2004a,Gardel2004b,Koenderink2006,Tharmann2007} and intermediate filaments \cite{Lin2010a,Lin2010b,Yao2010} to synthetic stiff polymers \cite{Kouwer2013}, that aspects of both linear and nonlinear rheological response can be accounted for by simple affine thermal models. In this section we discuss a few of these experimental studies.

First, we would like to relate some of the quantities introduced in the previous section, such as the crosslinking length scale $\ell_c$, to experimental control parameters. For example, consider a reconstituted F-actin network. Two important experimental control parameters are the concentration of monomeric actin, $c$, and the concentration of crosslinking protein, $c_{\times}$. It will also turn out to be useful to quantify the degree of crosslinking by the ratio of crosslinkers to polymer, which is often most conveniently expressed in terms of the molar ratio ratio $R=c_{\times}/c$. 

We will limit this discussion to the affine thermal model for homogenous, isotropic networks in which the crosslinks do not lead to bundling of filaments. For such cases, we expect that varying the polymer concentration will not only affect the polymer length density $\rho\sim c$ but also the crosslinking lenghtscale $\ell_c$, since the number of potential physical bonds between crosslinks increases with more polymer (see Eqs.\eqref{eq: linear thermal modulus} and \eqref{eq: AFM critical strain and stress}). The precise dependence of the crosslinking distance $\ell_c$ on parameters is subtle. The mostly likely binding sites on the polymer for effective crosslinks are sites where two polymer strands interact sterically, i.e., the entanglement points~\cite{Odijk1983}. Thus, we expect that crosslinking will occur on the entanglement length scale $\ell_c\sim\ell_e\sim (a^4\ell_p)^{1/5}\phi^{-2/5}$ (see Eq.~\eqref{eq:meshsizeconcen}), where $a$ is the polymer's diameter, and the polymer volume fraction $\phi \sim c$. However, if we fix the actin concentration to hold the entanglement length constant, while varying $R$, we expect that $\ell_c$ will also vary. It is often postulated that $\ell_c \sim R^{-x}$, where $x$ is a phenomenological exponent that may depend on the type of crosslinker. Taking the scaling of $\ell_c$ with the entanglement length scale and the degree of crosslinking together yields,
\begin{equation}
\ell_c~\sim (a^4\ell_p)^{1/5}c^{-2/5} R^{-x}\label{eq:lcAff}
\end{equation}
If $R$ is held fixed, then it is expected that $G_0\sim c^{11/5}$, which  is a stronger dependence than the prediction for an entangled solution, $G\sim c^{7/5}$ (see section \ref{sec:solutions}). Various measurements of the linear shear modulus of F-actin and intermediate filament networks have been found to be consistent with the $11/5$ scaling predicted by the affine model~\cite{MacKintosh1995,Gardel2004a,Gardel2004b,Tharmann2007,Lin2010a,Lin2010b,Yao2010}.

It is often found that the network response becomes nonlinear for stresses (or strains) beyond a characteristic or \emph{critical} stress $\sigma_c$ (or strain $\gamma_c$). From Eqs.\ (\ref{eq: linear thermal modulus}) and (\ref{eq: AFM critical strain and stress}), the critical stress can be expressed in terms of the linear modulus according to 
\begin{equation}
G_0=6\sqrt{\frac{\kappa}{\rho}}\sigma_c^{3/2}\sim c^{-1/2}(\kT\ell_p)^{1/2}\sigma_c^{3/2}.\label{eq:PredGvsSigma-c}
\end{equation}
Note that $\ell_c$ drops out of this equation, and hence, this relationship should only depend on directly measured rheological quantities and material parameters such as $\ell_p$ that are usually known by independent means. The predicted relationship in Eq.\ (\ref{eq:PredGvsSigma-c}) is in good agreement with recent experiments on intermediate filaments \cite{Lin2010b} (see Fig.\ \ref{sec:AffThermalSmallStrain}) and has also been tested in synthetic hydrogels \cite{Kouwer2013}. In particular, this relation should be insensitive to the details of crosslinking. Moreover, from the measured rheological quantities $G_0$ and $\sigma_c$, one can the infer microscopic quantities 
\begin{equation}\label{eq:IF_eq2}
\ell_p = \frac{1}{36}\rho k_B T \frac{G_0^2}{\sigma_c^3},
\end{equation}
and
\begin{equation}\label{eq:IF_eq3}
\ell_c = 6 \ell_p \frac{\sigma_c}{G_0}.
\end{equation}
Since the persistence length is often known independently, the first of these represents an additional test of the model. These relations have been tested recently in intermediate filament gels and in synthetic hydrogels \cite{Lin2010a,Lin2010b,Yao2010,Kouwer2013}. 

Another important prediction of this model is the universality of the stiffening response to applied stress. Scaling the differential shear modulus $K=\frac{d \sigma}{d \gamma}$ by $G_0$ and the stress by $\sigma_c$ should result in a collapse of all data on a universal curve (see Eq.~\eqref{eq: AFM universal form}) that exhibits a high stress scaling regime in which $K\sim \sigma^{3/2}$ \cite{Gardel2004a}. Such behavior has been observed for actin, intermediate filaments and for synthetic hydrogels \cite{Gardel2004b,Lin2010a,Yao2010,Kouwer2013} (See Fig.~\ref{fig: fiber network rheology}). However, this universal response is only valid if the polymers are truly inextensible. Real polymers will have some purely energetic/enthalpic (as apposed to entropic) mode of extension~\cite{ Odijk1995}, which could start playing a role at high stresses, resulting in a departure from the universal stiffening curve. Evidence of this has been seen for fibrin gels and intermediate filament gels \cite{Storm2005,Lin2010a,Lin2010b,Piechocka2010}. 

\subsection{Contractility and motor-generated stiffening in affine thermal networks}
Given the ubiquitous nonlinear elastic response of biopolymer networks to applied stress, it is natural to ask whether internal stresses in living systems might also couple to such nonlinearities. These internal stresses might arise, for instance, due to molecular motors that are known to induce motion and exert forces within cytoskeletal networks in the cytoplasm of living cells. While the evidence for this \emph{in vivo} remains indirect, reconstituted systems \emph{in vitro} with added motor activity have observed mechanical stiffening, qualitatively consistent with the elastic stiffening due to externally applied stress \cite{Mizuno2007,Koenderink2009}. Specifically, when myosin motor proteins are added to actin networks, together with adenosine triphosphate (ATP) that acts as a fuel for the activity, an approximate 100-fold increase in the modulus was observed, both by local micromechanical measurements \cite{Mizuno2007} and by macroscopic rheological measurements \cite{Koenderink2009}. While such a stiffening can also arise from an increase of mechanical crosslinking, e.g., due to dead or inactive myosins, Mizuno et al., were able to confirm that there was a significant increase in network tension that was coincident with the mechanical stiffening, consistent with a mechanisms due to network nonlinearity. These authors were also able to demonstrate directly the non-equilibirum nature of both reconstituted and living systems \cite{Mizuno2007,Mizuno2009,Mizuno2008}. Interestingly, analogous behavior is beginning to be studied in synthetic systems \cite{Bertrand2012}.

These observations are consistent with mean-field theories of stiffening due to network tension induced by motor activity \cite{MacKintosh2008,Liverpool2009}, as well as simulations of networks of stiff fibers activated by motors \cite{Broedersz2011b}. Both of these approaches model the motors by \emph{force dipoles} of pairs of equal and opposite forces within the network. This was done to ensure the necessary force balance within the network. Subsequent theory has begun to account for the nonlinear nature of the networks in determining the local response of networks to internal motor forces \cite{Shokef2012}. Here, we shall only briefly return to motor-activated systems in Sec.~\ref{sec:contractilenonaffine}, but we would point the interested reader to recent reviews on the subject \cite{MacKintosh2010,Brangwynne2008,Marchetti2012}. 

\subsection{Nonaffine approaches for disordered fiber networks}\label{sec:NAsection}
Considering the success of the affine model in describing rubber elasticity \cite{Doi1988}, its application to semiflexible polymer networks may seem reasonable. Indeed, in many cases, the affine model for semiflexible polymers agrees well with \emph{in vitro} experiments for a range of biopolymer systems, including intracellular actin and intermediate filament gels as well as extracellular fibrin networks~\cite{MacKintosh1995,Storm2005,Gardel2004a,Gardel2004b,Tharmann2007,Yao2010,Lin2010a,Lin2010b,Piechocka2010}. There are reasons, however, to expect the affine approximation to fail. In detail, of course, even rubber or flexible polymer networks are not strictly affine in their response~\cite{Basu2011,Wen2012}. But, in such cases, nonaffinity does not result in a significant deviation from the affine limit \cite{Carrillo2013}. The more interesting question for semiflexible networks is whether nonaffinity leads to a qualitatively different behavior. This can happen, for instance, if the response becomes dominated by chain/fiber bending, rather than stretching. Naively, affine shear cannot lead to bending of straight filaments. If bending dominates, one can expect, for instance, a softer network response than for purely affine deformations, as well as a different scaling of $G$ with the concentration $c$~\cite{Kroy1996,Satcher1996,Broedersz2012}. Recent evidence seems to point in this direction for some systems \cite{Lieleg2007,Stein2011,Piechocka2011}, and the affine state may even be considered to be unstable~\cite{Heussinger2006a}. This leaves us now with the question: when should we expect this model to break down and what are the signatures of a network response that is dominated by nonaffine deformations?

An important aspect of affine models is that the polymer strands only deform through stretching-modes. In contrast to flexible polymers, however, the force-extension behavior of a semiflexible polymer is highly anisotropic: semiflexible polymers are typically much softer to bending than to stretching. In particular, the ratio of the restoring force for a transverse displacement  (bending) to that for an axial deformation (stretching) is $k_{\perp}/k_{\parallel}\sim\ell_c/\ell_p$. Naively, this may suggest that in semiflexible polymer networks, for which $\ell_p\gtrsim \ell_c$, it may be considerably more favorable to avoid costly affine stretching-modes by, instead, favoring deformation through the (presumed) softer bending modes. Interestingly, however, it turns out to be not that simple. The energy of a deformation mode is not just set by the associated elastic rigidity, but also by the amplitude of the deformation; even though the fibers themselves can be softer to bending, the bending deformations required to accommodate the macroscopically imposed strain can still be large compared to the stretching deformations, which may render the nonaffine scenario energetically less favorable than the affine alternative. Clearly, there is a tradeoff, which may depend sensitively on the system properties. Moreover, given the connectivity of the network, it might simply be impossible to construct modes of deformation that avoid stretching of bonds. Thus, we seek to determine a ``phase" or regime diagram of some sort for fiber networks, describing which deformation modes dominate the macroscopic response, given certain network and fiber parameters. To set the stage for this, we start by discussing how the nonaffine deformation field can be characterized and quantified.

 \begin{figure}
\includegraphics[width=\columnwidth]{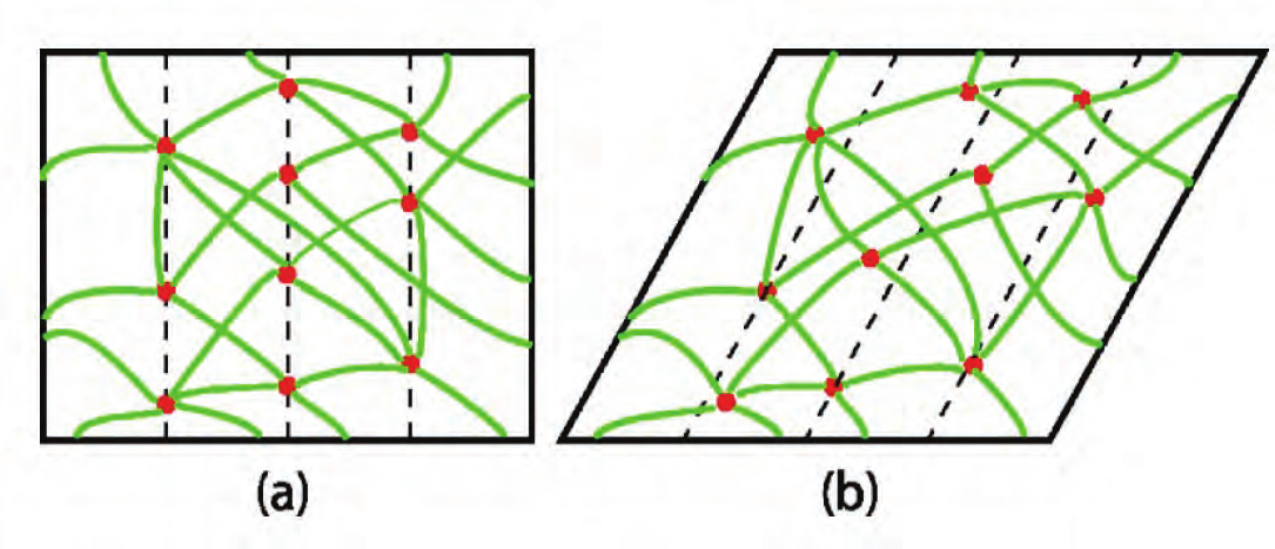}
\caption{Schematic to illustrate nonaffine deformations in networks~\cite{DiDonna2005}. a) Undeformed reference state. b) Sheared state with
nonaffine displacements. Under affine deformation, points on the vertical dotted lines in a) would map to points on the slanted dotted
lines. The schematic in b) illustrates a nonaffinely deformed network where they do not.
}\label{fig:nonaffinescheme}
\end{figure}\

\subsubsection{Characterizing nonaffinity in disordered elastic media}\label{sec:char_nonaffinity}

A detailed discussion of nonaffine correlation functions  of inhomogeneous elastic media was provided in Ref.~\cite{DiDonna2005}, and here we summarize some of their results.

The most general and straightforward way of characterizing the nonaffine deformation field (Figs.~\ref{fig:nonaffinescheme} and \ref{fig:nonaffinescheme2}) is by the correlation function
\begin{equation}\label{eq:NAcorrelations}
G_{ij}({\bf x},{\bf x'})=
\langle \delta u_i({\bf x})\delta u_j({\bf x'})\rangle,
\end{equation}
or, perhaps, the related form,
\begin{equation}
\mathcal{G}({\bf x})=\langle \left[{\bf \delta u}({\bf x})-{\bf \delta u}({\bf 0})\right]^2\rangle.
\end{equation}
Here, $\delta u_i$ represents the nonaffine component of the displacement $\delta u=u-u_{\rm aff}$ in the $i$ direction.
Since $\delta u$ will scale with strain, both these nonaffinity measures scale with $\gamma^2$.

The scaling of this function with distance may be anticipated from continuum elasticity theory. Considering the nonaffine displacement, ${\bf \delta u}({\bf 0})$, multiplied by the local modulus, as a force applied at the origin we should expect a scaling ${\bf \delta u}({\bf x})\sim{\bf \delta u}({\bf 0})|{\bf x}|^{-(d-2)}$. Indeed it was shown by Didonna and Lubensky that for disordered media with spatially varying elastic properties that $\mathcal{G}$ scales logarithmically with distance in 2D and inversely with distance in 3D. The 3D results makes intuitive sense as we expect the network to become more affine on larger scales~\cite{DiDonna2005,Liu2007, Hatami2008,Head2003b}. Furthermore, Didonna and Lubensky reported that, within this continuum approach, the nonaffinity parameter is expected to be proportional to the variance of the spatial fluctuations of the elastic modulus. 

\begin{figure}
\includegraphics[width=\columnwidth]{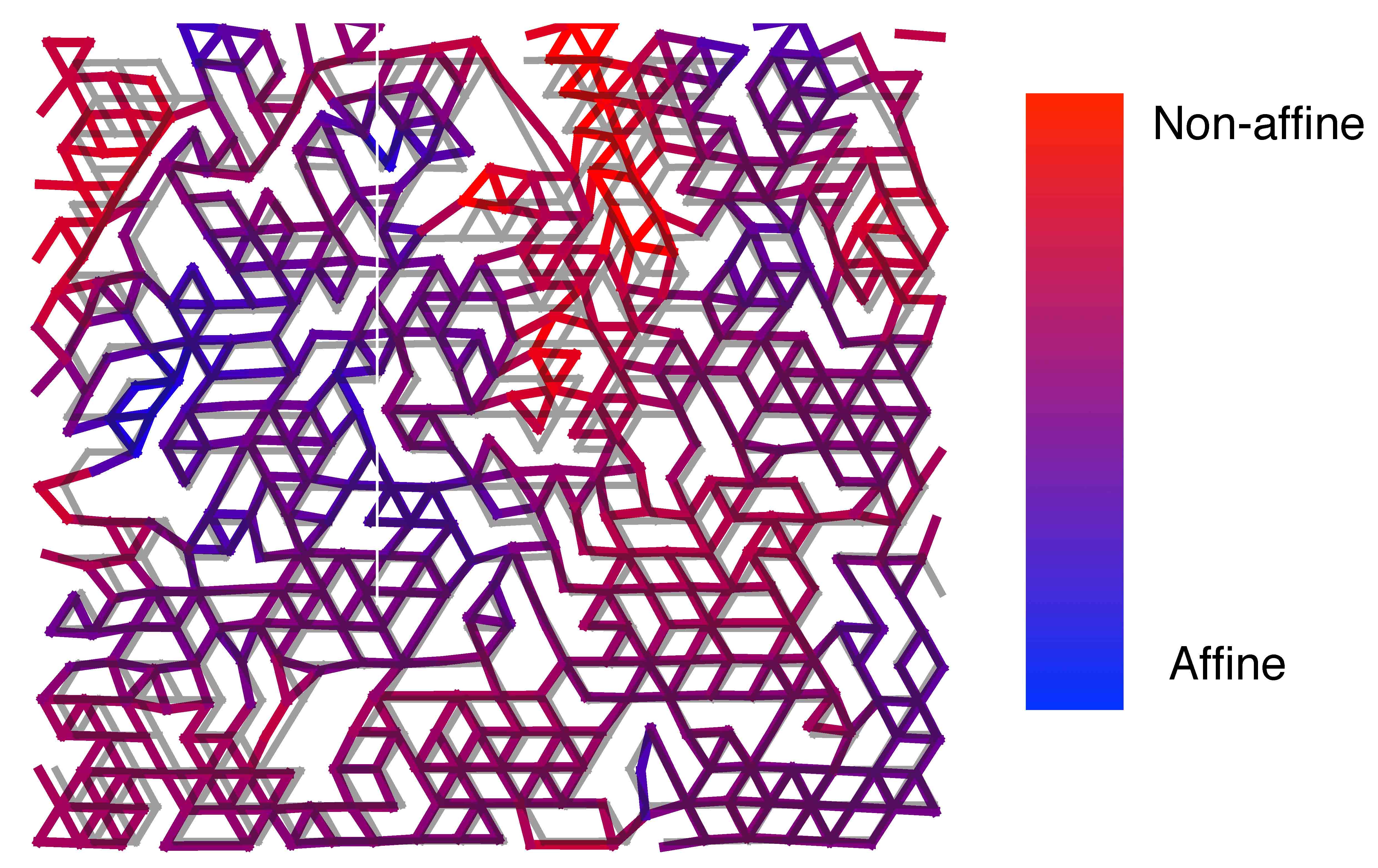}
\caption{Simulated fiber network on a diluted triangular lattice (see Sec.\ref{sec:lattice}). The gray network in the background represents the affinely sheared state. Segments in the simulated network are colored red for large nonaffine deformations and blue for deformations closer to the affine state.}\label{fig:nonaffinescheme2}
\end{figure}

There are numerous other nonaffinity parameters in the literature. However, it has been shown in~\cite{DiDonna2005} that many of these nonafinity parameters are closely related and can be expressed in terms of the nonaffinity correlation function in Eq.~\eqref{eq:NAcorrelations}. Perhaps the simplest measure of nonaffinity is the variance of local nonaffine fluctuations 
\begin{equation}
\Gamma=\frac{1}{\gamma^2}\langle \left[{\bf \delta u}({\bf x}))\right]^2\rangle
\label{eq:1pointNApar}
\end{equation}
Note that in this measure the strain dependence is scaled out, and it thus represents an intrinsic network property, equal to the trace $\gamma^{-2} G_{ii}({\bf x},{\bf x})$. The advantage of this measure is its simplicity and convenience in experiments (an accurate estimate of a one-point measure requires less data than a two-point measure). However, there are drawbacks: in 2D this measure has a logarithmic dependence on system size, and this measure may be more sensitive to spurious long-wavelength inhomogeneities compared to a two-point measure.
 
Finally, another example of a two-point measure is one that characterizes the change in orientation of a vector between two nodes separated by a distance x under deformation, relative to the affine prediction~\cite{Liu2007,Head2003a,Head2003b}. The variance in the nonaffine component of this angular deformation is a measure of nonaffinity, and depends on the distance between two points. This measure can be shown to scale as $\sim\mathcal{G}(x)/x^2$, and is thus related to the other measure discussed above. 

We are still left with the question: What can we learn from these nonaffinity measures? The message we hope to convey in the next few sections is that although these nonaffinity parameters can be extremely insightful, they need to be interpreted with caution. For instance, paradoxically, a system with higher values of the nonaffinity parameter may still have a mechanical response that is dominated by affine deformation modes, compared to a system with lower nonaffine fluctuations that is governed by a nonaffine mechanical response. Again, it must be remembered that all disordered networks can be expected to exhibit some level of nonaffinity, so that simply measuring a non-zero value of any of the above measures of nonaffine deformation does not imply an essential break-down of the affine limit, e.g., in the form of qualitative changes to the elastic response. We will discuss these issues in Sec.\ref{sec:phantom}.

\subsubsection{Unit cell approaches}\label{sec:unitcellapproaches}

Various unit cell approaches have been developed for networks such as rubber, especially in the mechanics literature \cite{Arruda1993}. In such approaches, rather than assuming that all network strands deform affinely, a small \emph{cell} consisting of a few strands of different orientations is repeated to form a 2D or 3D structure. The resulting networks are studied as mechanical and athermal structures. Such approaches have been adapted go beyond the affine formalism to calculate the viscoelastic or viscoplastic properties of biopolymer solutions~\cite{Fernandez2009,James1943,JerryQi2006,Brown2009,Palmer2008,Cioroianu2013,Satcher1996}. 

Among the earlier theoretical studies of nonaffine behavior of crosslinked nonaffine, Refs.\ \cite{Kroy1996,Satcher1996} borrowed from the field of cellular solids. The assumption of this model is that fibers only deform through bending and that such network deformations can be characterized using a cubic unit cell with sides equal to the mesh size, $\xi$. Assuming that the fiber strands deform by an amount $\delta u\sim\gamma \xi$, perpendicular to their orientation, the bending energy becomes, $E_b\sim \kappa (\frac{\delta u}{\xi^2})^2 \xi=\kappa (\frac{\gamma^2}{\xi})$. Thus, the energy density amounts to $\kappa (\frac{\gamma^2}{\xi^4})$. Using this unit cell picture, we can relate the mesh size to the polymer concentration $\xi\sim \sqrt{c}$ (see Eq.~\eqref{eq:meshsizeconcen}), suggesting a scaling for the shear modulus
\begin{equation}
G\sim \kappa c^2.
 \end{equation}
This concentration dependence should be contrasted with the affine thermal model for which $G\sim c^{11/5}$ in Sec.~\ref{sec:affinethermalmodel}, although from this scaling alone, it is difficult to distinguish the two models in practice.
 
Unit cell models have the advantage that an affine response does not need to be enforced at the crosslink level, and aspects such as averaging over different fiber orientations are included naturally. Recently, a variation on this theme was introduced in Ref.\ \cite{Carrillo2013}, in which a diamond lattice unit cell was used to create and study fully thermalized networks of semiflexible filaments spanning a wide range of mechanical properties. These authors found good agreement with several experimental observations. 

Such unit-cell approaches, whether thermal or athermal, should be appropriate, at least for networks of short stubby filaments with lengths comparable to the network's mesh size, such as in a foam like architecture~\cite{Heussinger2006a}. However, in many cases filaments have lengths greatly exceeding the networks' mesh size. At the very least, this introduces additional correlations since the filaments exhibit mechanical integrity beyond the scale of the unit cell. A unit cell-based approach cannot account for this, and one may need to deal with network of linear dimension at least as large as the fibers themselves. In addition, one might challenge the assumption that deformations of the fibers scale with mesh size and the implicit assumption made here that nonaffine deformations are uncorrelated. These assumptions clearly fail in some cases, such as networks that are close to marginal stability (see Sec.\ref{sec:isostaticiy}), where network deformations may be correlated over large length scales, and where nonaffine deformations may be controlled by other network parameters such as connectivity or filament length. To address these issues, we now continue with a discussion of whole-network models that have been studied numerically.

\subsubsection{A minimal model for disordered, athermal fiber networks in 2D: The Mikado model}\label{sec:Mikado}

\begin{figure}
\includegraphics[width=\columnwidth]{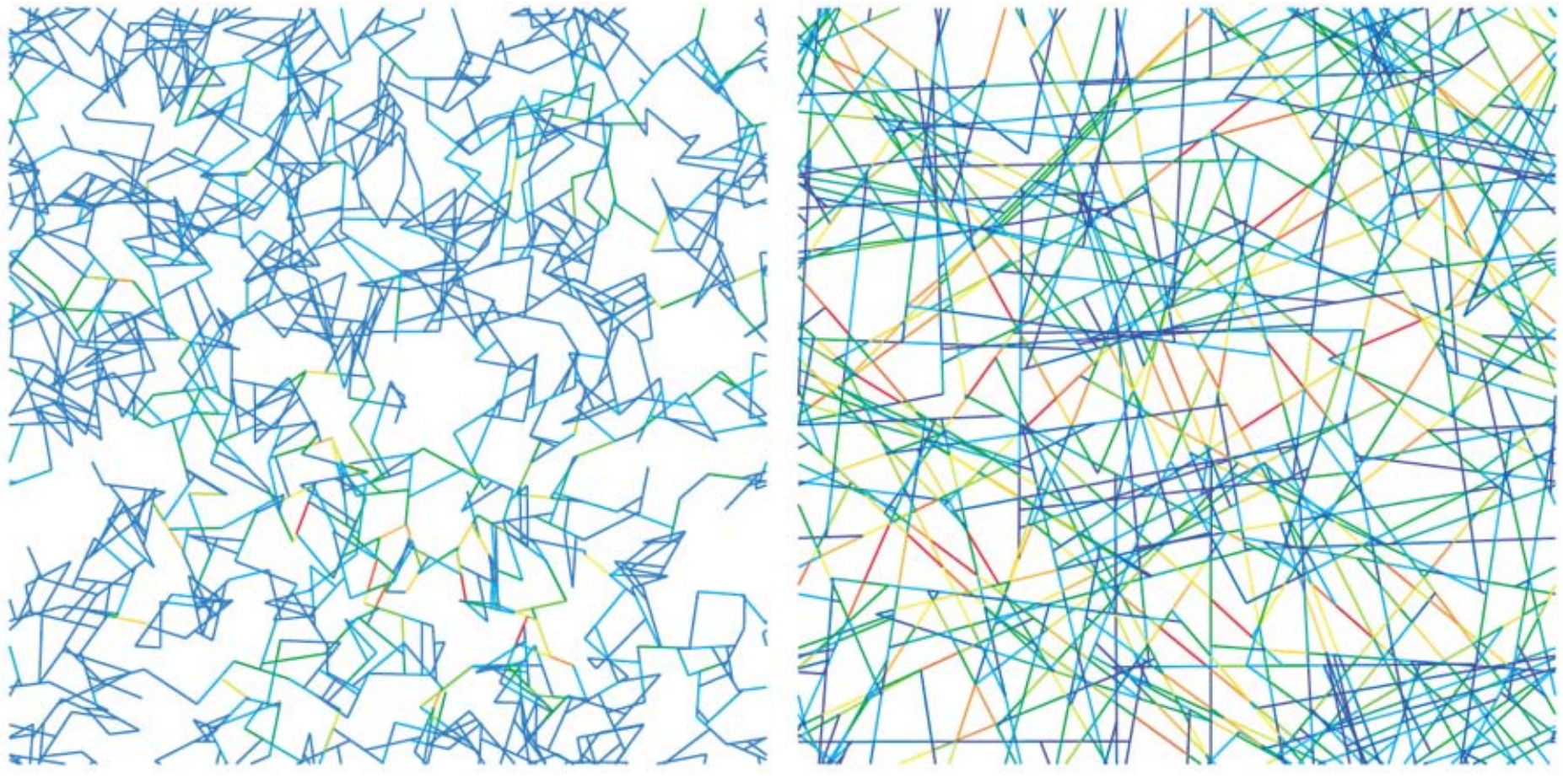}
\caption{2D Mikado networks at
low and high density under a small shear. The colors indicate distribution of tensions on a filament; the load on a filament
increases from blue to red. Adapted from~\cite{Wilhelm2003}.}
\label{fig:Mikadonetwork}
\end{figure}

One of the simplest whole-network models for crosslinked filamentous networks is the \emph{Mikado} model~\cite{Wilhelm2003,Head2003a,Head2003b}. Mikado networks are constructed by randomly depositing monodisperse filaments of length $\ell$ onto a two dimensional square of size $W\times W$, as shown in Fig.~\ref{fig:Mikadonetwork}. The intersections between filaments are identified as point-like, freely-hinging crosslinks. The energy of this system can be expressed as
\begin{equation}\label{NLrigidlinker_linearmodulus}
H=\frac{\mu}{2} \sum_i \frac{\delta \ell_i^2}{\ell_i}+\frac{\kappa}{2} \sum_{\langle ij \rangle} \frac{\delta \theta_{ij}^2}{\ell_{ij}}
\end{equation}
Here, $\ell_i$ indicates the length of segment $i$, $\ell_{ij}$ is the average length of segments $i$ and $j$, and $\delta\theta_{ij}$ is the angular deflection between segments $i$ and $j$. For fibers, the second sum runs only over neighboring segments along the same fiber. This is a purely mechanical model, and the thermal properties of semiflexible polymers can be captured, at best in a coarse-grained sense, by setting the modulus $\mu=90 \kappa \ell_p/\ell_c^3$---the entropic elasticity of a semiflexible polymer \cite{Head2003b}. Moreover, the model treats the filaments as linearly elastic elements with respect to both bending and stretching. Thus, it does not capture filament nonlinearities such as the entropic stiffening behavior. This does not mean, however, that networks of these simple elastic filaments are necessarily linear in their macroscopic response: networks of purely linear elements can have a nonlinear response, as we shall see below. In fact, even at the level of single fibers with linear bending and stretching elasticity, nonlinearity can appear due to buckling under compression. 

The parameter space of the Mikado model can be expressed in terms of a line density $\rho=\pi/\langle \ell_c \rangle $ and the dimensionless parameter $\ell_b/\ell_c$. Here, the material length scale $\ell_b=\sqrt{\kappa/\mu}$ characterizes the ratio of the bending and stretching rigidities. For simple elastic beams of length $\ell_c$, $\ell_b/\ell_c$ is a measure of their aspect ratio, while for thermal semiflexible polymers $\ell_b/\ell_c=\sqrt{\ell_c/90 \ell_p}$. 

Simulations of the Mikado model reveal various qualitatively distinct mechanical regimes, including a stretching dominated regime where the shear modulus is close to the affine limit $G\approx G_{\text{aff}}$ (large $\ell/\ell_c$ or $\ell_b/\ell_c$). The affine shear modulus of the Mikado model can easily be calculated and is given by 
\begin{equation}
G_{\text{aff}}=\frac{\pi}{16}\frac{\mu}{\ell_c}
\end{equation}
in 2D. This affine value forms a strict upper bound on the shear modulus of such networks: networks always have the affine deformation mode available to them, and any deviations from this deformation will only occur if they lower the elastic energy. In addition to the affine regime, there is a \emph{nonaffine} bending regime $G\sim \kappa$, and a rigidity percolation point $\rho_c$ at low network densities at which the shear modulus vanishes continuously. The crossover between bending dominated elastic behavior $G\sim \kappa$ and stretching dominated behavior $G\sim\mu$ is shown in Fig.~\ref{fig:HeadMikadoG} for a simulated Mikado network response. 

\begin{figure}
\includegraphics[width=\columnwidth]{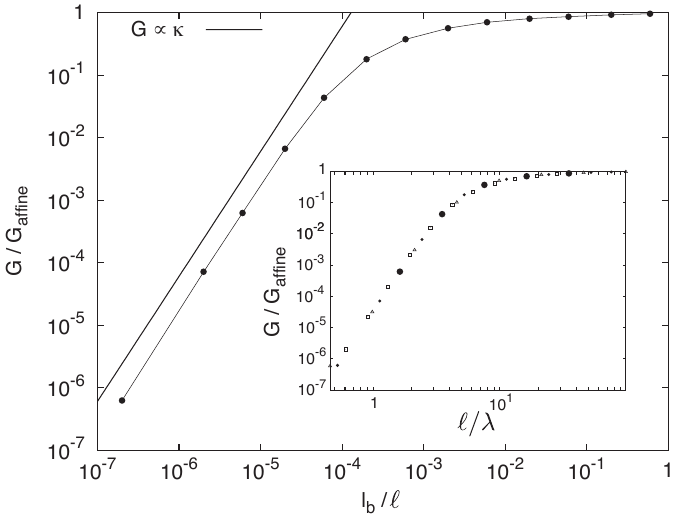}
\caption{ Simulations of the Mikado model indicate a transition between a nonaffine bending regime and an affine stretching regime (adapted from~\cite{Head2003a}).
Shear modulus $G$ as a function of filament rigidity $\ell_b=\sqrt{\kappa/\mu}$ for $\ell/\ell_c=$
29.09, where $G$ has been scaled to the affine prediction for this
density and $\ell_b$ is shown in units of $\ell$. The straight line corresponds to the bending-dominated regime
with $G \sim \kappa$, which gives a line of slope $2$ when plotted on
these axes. The inset depicts the shear modulus $G$, normalized by the affine value, for different densities as function of filament length scaled by the non affinity length scale $\lambda$ (See Eq.~\eqref{eq:lambda}). This rescaling results in a good data collapse, as described by Eq.~\eqref{eq:HeadScaling}.}
\label{fig:HeadMikadoG}
\end{figure}

An understanding of the bending regime poses one of the main theoretical challenges in this model. Owing to the disordered nature of the network, bending deformations may be correlated and could extend over large length scales. Initially, two basic approaches where offered to provide insight into the origins of this nonaffine bending regime and the crossover to affine network behavior. One argument approaches the problem from the rigidity percolation point~\cite{Wilhelm2003}, which we will discuss first, while the other approach starts from the affine limit~\cite{Head2003a,Head2003b}. In Sec.\ref{sec:floppy mode theory} we discuss a more recent approach, based on constructing the networks' floppy modes~\cite{Heussinger2007c,Heussinger2006b}.

For densities beyond the regime controlled by the rigidity percolation point, Wilhelm and Frey proposed the following form for the shear modulus~\cite{Wilhelm2003}
\begin{equation}
G=\frac{\kappa}{\ell^3} |\delta \hat{\rho}|^\mu g\left(\frac{\ell_b}{\ell} |\delta \hat{\rho}|^{\nu'}\right),\label{eq:FreyScaling}
\end{equation}
where $\delta \hat{\rho}=\ell (\rho-\rho_c)$ is dimensionless, and $g(x)$ is a universal scaling function. To capture the affine limit $G\rightarrow G_{\text{aff}}$, the universal function should scale as  $g(x)\rightarrow x^{-2}$ for $x\gg1$, and the exponents must satisfy $\mu-2\nu'=1$. By contrast, in the bending-regime $x\ll1$, the function $g(x)$ should be constant such that $G\sim\frac{\kappa}{\ell^3}|\delta \hat{\rho}|^\mu$. Wilhelm and Frey found an excellent collapse of their numerical data with this scaling function with the values $\hat{\rho}_c=5.71$, $\mu=6.67$, and thus $\nu'=2.83$.

One interesting implication of this scaling law is a length scale $\xi'=\ell |\delta \hat{\rho}|^{-\nu'}$---distinct from the length scale of the incipient percolation cluster---controlling the crossover between the bending and stretching regimes. In particular, the crossover is expected when $\xi'\simeq\ell_b$, yielding the crossover line density $ \delta \rho_{\text{cross}}=\ell^{-1} (\ell_b/\ell)^{-1/\nu}$ (reinstating units of length). Alternatively, this crossover line density can be understood from the limiting expressions of the shear moduli: In the bending regime the shear modulus scales more strongly with $\delta \hat{\rho}$ than in the affine limit. Thus, at high density the modulus of the bending-dominated regime would surpass the affine ``ceiling", implying a crossover to affine network behavior beyond $\delta \rho_{\text{cross}}$.

An alternative approach to understanding the rich mechanical behavior of the Mikado model uses the affine limit as a bench mark, together with a self-consistent scheme to find the crossover to the nonaffine bending regime~\cite{Head2003a,Head2003b}. The implicit assumption is that the bending regime near the nonaffine-affine crossover is governed by different physics than the rigidity percolation point. Conceptually, the main idea of this argument is to estimate when the total energy can be reduced by relaxing the axial strain of a filament of length $\ell$ over a scale $\lambda$ near the ends at the cost of bending other filaments in the surrounding network. This argument leads to the following scaling prediction
\begin{equation}
G=\frac{\mu}{\ell_c} f\left(\frac{\ell}{\lambda}\right),\label{eq:HeadScaling}
\end{equation}
where
\begin{equation}
 \lambda=\ell_c \left(\frac{\ell_c}{\ell_b}\right)^{z'}\label{eq:lambda}
 \end{equation}
is a length scale controlling the crossover (at $\ell \simeq\lambda$) between the bending and the affine stretching regime. This argument predicts a crossover exponent $z'=2/5$. For the system to crossover to a bending dominated regime, given here by $G\sim\frac{\kappa}{\ell_c^3}(\ell/\ell_c)^{2/z'}$, the universal scaling function 
$f(x)\sim x^{2/z'}$ for $x \ll 1$, while for large arguments $f(x)$ is constant. Refs.~\cite{Head2003a,Head2003b} report a good  collapse of numerical data using this scaling form for $z=1/3$ for low network density data, while a better collapse is obtained using $z=2/5$ at higher densities (see inset to Fig.\ \ref{fig:HeadMikadoG}). However, this scaling form does not capture the continuous vanishing of the shear modulus at the rigidity threshold, which may explain the different scaling at low density. An effective medium description for diluted Kagome lattices was offered by Mao et al., as a model for 2D filamentous networks. This effective medium approach provided an analytical calculation of the crossover function that captures  the bend-stretch transition~\cite{Mao2013b}. Interestingly, the  exponent that governs the crossover appears to be different in lattice based networks than in Mikado networks.} Other studies have discussed the impact of orientational order and length polydispersity of the filaments on this scaling~\cite{Bai2011,Missel2010}.

Although the physical reasoning in the two approaches is different and the two scaling forms differ in detail, some reconciliation is obtained by identifying the length scales $\xi'$ and $\lambda$.
Far from the rigidity threshold $\delta \rho\sim\rho\sim 1/\ell_c$ and the two scaling forms become similar, implying a correspondence between the two lengthscales if $z'=1/(\nu'-1)$ or,  equivalently, $\mu=2/z'+3$. However, the numerical values for these exponents reported in the two studies are inconsistent with these equations, still leaving the puzzle partially open.

\begin{figure*}
\includegraphics[width=2 \columnwidth]{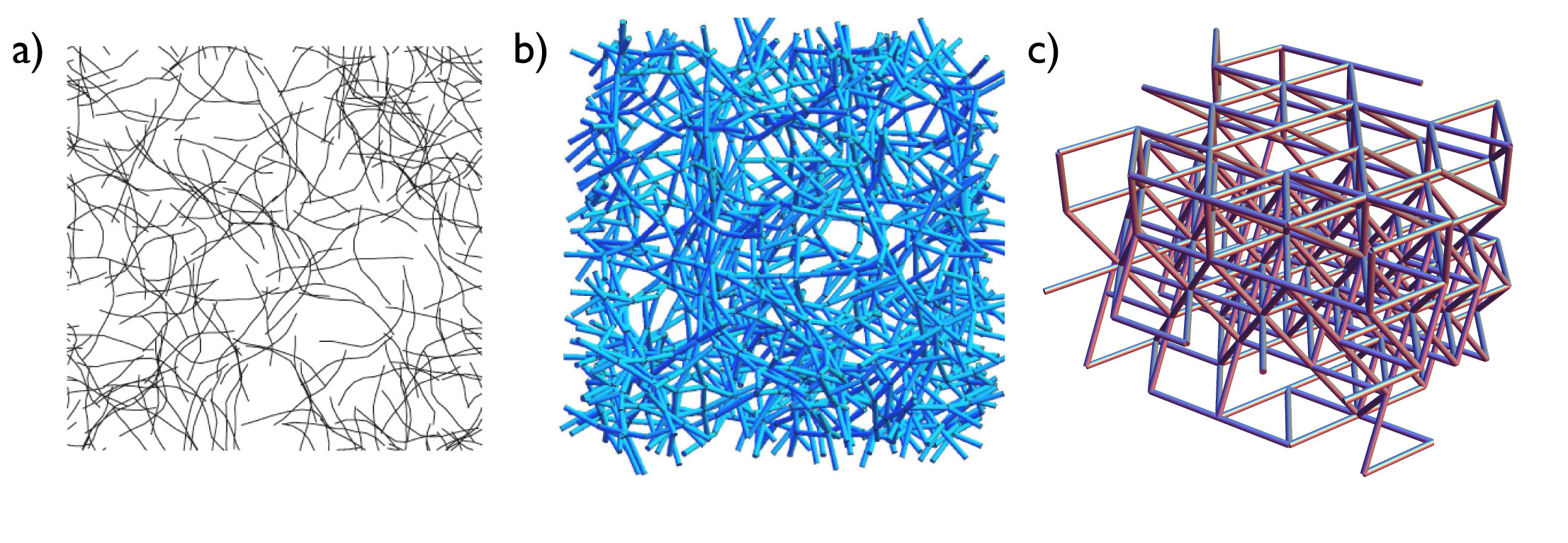}
\caption{Comparison between various fiber network models. a) Mikado model with curved filaments~\cite{Onck2005}  b) Thermalized semiflexible polymer network model~\cite{Huisman2008,Huisman2011}. c)  Lattice-based dilution model of a 3D fiber network on an fcc lattice~\cite{Broedersz2011a}.}
\label{fig: fiber network models}
\end{figure*}

\subsubsection{Floppy mode theory}\label{sec:floppy mode theory}

Heussinger and Frey proposed a theoretical framework to calculate the properties of the nonaffine bending regime of semiflexible polymer networks~\cite{Heussinger2007c,Heussinger2006b}, although the general framework
is not limited to fibrous networks and could find applications in other soft matter systems. The main premise of this model is that the low-frequency, soft 
deformation modes in the system derive from a set of zero-energy, \emph{floppy modes}~\cite{Liu2010}. In the $\ell_b\rightarrow0$ limit, the network can deform through these floppy modes with no contribution to the
mechanical energy up to harmonic order. However, at a finite, yet small bending rigidity these modes are no longer floppy, but are still considered to be the softest modes, and may thus be used to calculate the network's properties. 

By using a self-consistent effective medium approach in which the floppy modes for the Mikado model are constructed explicitly, the elastic properties of the nonaffine bending regime can be calculated~\cite{Heussinger2007c,Heussinger2006b}.
This calculation yields the prediction $G\sim\kappa\rho^{\mu}$, with $\mu=6.75$, in good agreement with simulations~\cite{Wilhelm2003}. 

The high value of this exponent ($\mu$) may be understood from a simple scaling argument, which conveys the main idea behind the Floppy mode theory. 
The typical bending  energy of a segment of length $\ell_s$ is $w_b\sim\kappa  \delta u_{\rm na}^2/\ell_s^3$, where $ \delta_{\text na}$ is the amplitude of a bend on the scale $\ell_s$. This amplitude represent the nonaffine deformation, and is independent of $\ell_s$, as apposed to an affine model; by assuming that individual fiber centers follow the affine deformation field, it was argued that  $ \delta u _{\text na}\sim \ell$. Thus, the average bending energy stored in a fiber is
\begin{equation}
\langle W_b\rangle=\rho \int_{\ell_{\rm min}}^\infty {\text d}\ell_s P(\ell_s)  \frac{\kappa \delta u_{\rm na}^2}{\ell_s^3},
\end{equation}
where $P(\ell_s)$ represents the distribution of segment lengths and $\ell_{\rm min}$ is a cutoff. Below this cutoff the segments become too stiff to contribute to the bending energy. 
A localized bend on this scale can be relaxed by exciting a ``floppy mode" in the fiber to which it is connected, with a corresponding typical energy $\langle W_b \rangle $. Hence, this cutoff length scale can be found self-consistently from the condition $w_b(\ell_{\rm min})=\langle W_b \rangle $.
Using a Poissonian distribution for $P(\ell_s)$, as in a Mikado network, the shear modulus is predicted to scale as $G\sim\kappa\rho^{\mu}$, with $\mu=7$. This is remarkably close to the result of the more elaborate effective medium calculation and the numerical result. This model has provided insight in the mechanics of bundled actin networks~\cite{Lieleg2007} and in simulations of composite networks~\cite{Huisman2010}.

\subsubsection{2D versus 3D networks}\label{sec:implications for 3D networks}

Biological filamentous networks are usually three-dimensional. To what extent should one expect the mechanical behavior of such networks to be captured by the simple 2D models described above? Put differently, are there essential \emph{qualitative} differences in the mechanics of semiflexible polymer networks in 2D and 3D.

Various computational approaches to address 3D networks of semiflexible polymers or elastic fibers have been developed recently, including Brownian dynamics models~\cite{Kim2009,Huisman2010b}, Monte Carlo simulations~\cite{Blundell2011}, energy minimization schemes for minimal mechanical models \cite{Buxton2007,Broedersz2012,Lubensky2011} and  coarse grained approaches~\cite{Huisman2007,Huisman2008,Huisman2011,Stein2011}. These have shown significant nonaffine effects. 

On very general grounds, nonaffine deformations could be expected to have a greater impact on the network's elastic response in 3D systems. More specifically, the case of binary crosslinks between fibers can be expected to be more bend-dominated than corresponding 2D systems. This can be seen as a consequence of arguments going back to Maxwell \cite{Maxwell1865}, showing that the critical coordination number for mechanically stable networks of springs (i.e., with stretching and no bending resistance) is greater in 3D than than the local coordination of binary crosslinked networks. Moreover, this critical coordination or connectivity depends on dimensionality, and is close to that of filament networks in 2D, while it is far from that of such networks in 3D. Thus, the mechanical stability of 3D fibrous networks with binary crosslinks is expected to rely on the bending elasticity of the fibers, while a 2D network is (marginally) stable without fiber bending elasticity. This suggests that nonaffine bending deformations could play a more dominant role in 3D fibrous networks with binary crosslinks~\cite{Broedersz2012,Lubensky2011,Huisman2007,Huisman2011}. However, the scaling arguments in \cite{Head2003b} appear to suggest that the behavior in 2D and 3D, at least for high molecular weight, should not be qualitatively different. 

Thus, there are fundamental questions regarding the behavior of 3D networks that makes their study more important than simply the need to examine more realistic systems. However, addressing these questions in 3D proved to be a significant computational challenge, partly because, by analogy with the Mikado work in 2D, the transition from bending to stretching may only occur for long fibers, compared to the spacing of the crosslinks, which would suggest the need for large networks and, thus, computationally slow models in 3D.

To develop a three-dimensional network model that reflects architectural characteristics of an actual biopolymer gel, Huisman et al. used a Monte Carlo scheme to generate thermalized networks~\cite{Huisman2007,Huisman2008,Huisman2011} using the worm-like chain model for semiflexible polymers. Starting from a random, isotropic network, Monte Carlo moves that alter the topology of the network are performed to minimize the free energy of the network.  Subsequently, segments are cut until an average filament length $\ell$ is obtained. This procedure results in a disordered network of curved filaments (see Fig.~\ref{fig: fiber network models}b). Though these filaments have disordered intrinsic curvatures, they maintain directionality over their persistence length. Once the network is generated, the filament segments are described by a bending rigidity, and the nonlinear force-extension curve for semiflexible filaments (Eq.~\eqref{eq:extension}). Finally, an energy minimization scheme is used to simulate the network under shear.

Simulations of this model found a nonaffine bending regime that covered the range of network parameters studied~\cite{Huisman2011}. Importantly, however, computational limits did not permit the authors to exceed system sized about an order of magnitude larger than the network mesh size. Also, the persistence length was held constant, while increasing the molecular weight. Networks in the high molecular weight limit constructed in this way consist of filaments that are much longer than their persistence length. Thus, the question of what happens in networks of long stiff filaments that are approximately straight over their full contour length remained. 

An obvious practical problem associated with this high molecular weight limit is that networks with large $\ell/\ell_c$ have a large number of degrees of freedom, which may not be computationally tractable. To overcome this problem, fiber networks with underlying lattice geometries were developed for which such large networks are computationally feasible, although the architectures were obviously simplified. We will discuss these lattice-based networks in the next section, as well as in Sec.~\ref{sec:lattice}.

\subsubsection{3D Phantom and generalized Kagome networks }\label{sec:phantom}
Three dimensional lattice-based fiber network models with binary crosslinks have been developed~\cite{Broedersz2012,Lubensky2011}. Because of the computational efficiency of lattice-based networks, these models have been able to address the outstandinc question of whether 3D networks exhibit a bend-to-stretch crossover analogous to 2D networks. In particular, these approaches have been able to address the high molecular weight limit. Stennul and Lubensky generated a 3D generalization of the Kagome lattice by appropriately combining  2D Kagome lattices. The result was a large unit cell with 54 nodes. In \cite{Broedersz2012} a network based on a face centered cubic (fcc) lattice ((see Fig.~\ref{fig: fiber network models}c)) was constructed, and disorder was introduced in such a way as to reduce the maximum coordination number to 4 while maintaining individual fibers of arbitrary length. Although an fcc lattice has local 12-fold coordination, a simple trick can be used to achieve the desired network structure in which the maximum coordination number at each vertex can be reduced: Three \emph{independent} pairs of crosslinked fibers are formed out of the six fibers crossing at a vertex. Thus, this results in 3 binary crosslinks that may overlap in space, but do not interact with or constrain each other; these 3 pairs of fibers move through each other as phantom chains. This lattice is termed the 3D Phantom network. 

In both the Kagome-based lattice and the 3D Phantom lattice networks, the fiber length can be tuned $\ell=\ell_0 /(1-p)$ by cutting bonds with a probability $1-p$, where $\ell_0$ is the distance between vertices. In the Phantom model at least one bond is removed along every fiber to avoid filaments that span the system; such spanning filaments will deform more affinely. Thus, this model can only approach $z=4$ asymptotically from below. Although this may seem like a technical detail, some of the most subtle and interesting behavior in this model occurs in the limit where filaments are long, and spanning filaments can completely overshadow the macroscopic elastic response of the network. 

The perfect, undiluted lattice is mechanically rigid when $\kappa=0$ in both models, and there is a first-order jump in the shear modulus to zero when $p$ is less than 1. Surprisingly, however, for diluted networks it was found that for a finite bending rigidity, no matter how small, the network shear modulus approaches its affine value in the high molecular weight limit, which becomes insensitive to the fiber bending stiffness. This is similar to what was observed in 2D Mikado networks (Sec.\ref{sec:Mikado}). However, the reason this is particularly surprising in 3D is that in this case the network connectivity is still well below the Maxwell isostatic threshold, which governs the stability of networks with $\kappa=0$; only beyond a higher local coordination number do stretching constraints imposed by the connectedness of the network force the system to be stretch-dominated and nearly affine (see discussion on isostaticity in Sec.\ref{sec:lattice}). Put differently, these are networks that are strictly mechanically unstable ($G=0$) when $\kappa=0$, and yet stretch-dominated and approximately affine ($G\approx G_{\rm affine}$) for any $\kappa>0$, provided $\ell/\ell_0$ is chosen to be sufficiently large. Another interesting finding in the 3D 4-fold networks is that, in the limit of floppy filaments with weak bending rigidity or infinite molecular weight, the elastic response of the system  becomes intrinsically nonlinear with a vanishing linear response regime~\cite{Broedersz2012}. We will discuss nonlinear properties of filamentous networks in more detail in Sec.\ref{sec:nonaffinenonlinear}.

The linear response of these systems can be understood within an effective medium framework developed for 2D Kagome networks~\cite{Mao2013b}, as discussed in~\cite{Lubensky2011}. An alternative approach~\cite{Broedersz2012} builds on some of the ideas of the floppy mode theory~\cite{Heussinger2007c,Heussinger2006b} (see Sec.\ref{sec:floppy mode theory}), as well as ideas presented in~\cite{Head2003a,Head2003b}. We start by considering a deformed network in which the fibers are softer to bending than to stretching. Network nodes along a fiber are assumed to undergo independent nonaffine deformations scaling as $\delta u_{\mbox{\rm \tiny NA}} \sim \gamma \ell$ to avoid stretching of the other fibers to which they are connected. As a result, we would anticipate a scaling for the nonaffine fluctuations of the form $\Gamma \sim \ell^2$ independent of $\kappa$, which is indeed observed numerically~\cite{Broedersz2012} ((see Sec.\ref{sec:char_nonaffinity} and Eq.~\eqref{eq:1pointNApar} for more detail on the definition of the nonaffinity parameter).

Nonaffine fluctuations of this form have interesting implications for the bending energy in the system. Such length-controlled nonaffine deformations store an amount of elastic energy that scales as $\kappa(\delta u_{\mbox{\rm \tiny NA}}/\ell_0^2)^2\ell_0$ per segment of length $\ell_0$, which at the macroscopic level results in a shear modulus for the bending regime given by
\begin{equation}
G_{\rm LC}\sim \frac{\kappa}{\ell_0^2} \left(\frac{\delta u_{\mbox{\rm \tiny NA}}}{\ell_0^2}\right)^2\frac{1}{\gamma^2} \sim\frac{\kappa}{\ell_0^6} \ell^{2}.
\label{eq:GLC}
\end{equation}
We can relate this to the behavior discussed for the 2D Mikado model (Sec.~\ref{sec:Mikado}). Thus, for 3D lattice networks we expect a similar scaling, but with the exponent $\mu=5$ in Eq.\ \eqref{eq:FreyScaling}.
This type of bending elasticity implies that the energetic cost of nonaffine bending deformations grows with increasing  $\ell$. But, the affine shear modulus $G_{\rm A}\sim\mu/\ell_0^2$ represents an upper bound. Thus, with increasing $\ell$, at some point the nonaffine modulus in Eq.~\eqref{eq:GLC} exceeds the affine upper bound, and thus becomes unphysical. This suggests a crossover from bend-dominated to stretch-dominated elasticity, as the nonaffine bending deformations become less favorable than the $\ell$-independent affine stretching deformations. This crossover is expected to occur for an average molecular weight comparable to $\lambda_{\mbox{\rm \tiny NA }} $, which can be identified as a nonaffine length scale. This length can be estimated by comparing $G_{\rm LC}$ with the affine stretching shear modulus $G_{\rm A}\sim\mu/\ell_0^2$, which become comparable for
\begin{equation}
\ell\sim\lambda_{\mbox{\rm \tiny NA }} = \ell_0^2/\ell_b,
\end{equation}
where $\ell_b=\sqrt{\kappa/\mu}$. Consistent with this expected crossover, numerical simulations of both the 2D andy 3D Kagome-based and 3D Phantom models show a length-controlled crossover in $G$ to the affine prediction for large $\ell$ \cite{Mao2013b,Lubensky2011,Broedersz2012}. Moreover, $G/G_{\rm A}$ is a universal function of $\ell/\lambda_{\mbox{\rm \tiny NA }}$, for which $G/G_{\rm A}\simeq 1$ when $\ell/\lambda_{\mbox{\rm \tiny NA }}\gtrsim1$. 

These results are qualitatively consistent with the earlier Mikado model in 2D, which also showed a length-controlled crossover from non-affine to affine elasticity with increasing fiber length, indicating that dimensionality does not play a qualitatively important role, in spite of the Maxwell argument \cite{Maxwell1865}. In detail, however, the prior 2D work showed a different nonaffine length scale: $\lambda_{\mbox{\rm \tiny NA }}\sim\ell_b^{-\alpha}$, with $\alpha\approx 0.3-0.4$ \cite{Head2003a,Head2003b,Wilhelm2003} (see Sec.\ref{sec:Mikado}). However, for such 2D Mikado networks it is difficult to unambiguously identify the origin of the crossover as the same length-controlled mechanism in 3D, since the high molecular weight limit also corresponds to the CF isostatic point for the Mikado model, which also leads to a bend-stretch crossover.\cite{Broedersz2011a,Heussinger2007c,Heussinger2006b, Buxton2007} . Head et al.\ argued for a length-controlled mechanism that was independent of dimensionality \cite{Head2003b}, and it may be that the difference in the exponent $\alpha$ is due primarily to the difference in local network structure: the Mikado model exhibits inherently larger polydispersity of fiber segment lengths than  lattice-based networks \cite{Heussinger2007c,Heussinger2006b}. By contrast, 2D diluted Kagoma lattice networks~\cite{Mao2013b}, which do not exhibit a large polydispersity in fiber segment length, exhibit crossover behavior quantitatively more similar to the 3D networks with binary crosslinks~\cite{Lubensky2011,Broedersz2012}.

\begin{figure*}
\includegraphics[width=2 \columnwidth]{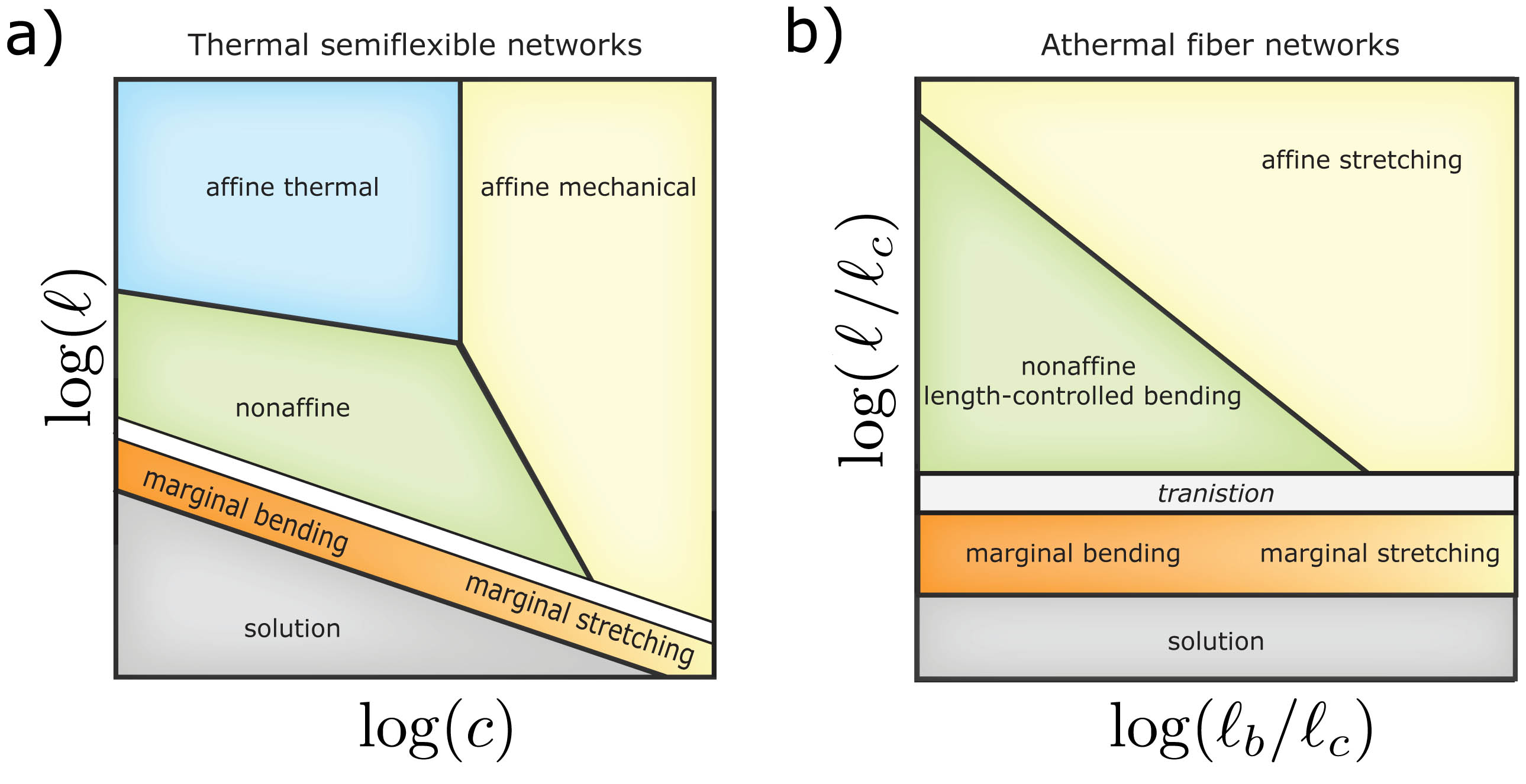}
\caption{a) Schematic for the elastic regimes (a) for a thermal semiflexible polymer network with binary crosslinks as a function of two important control parameters: Filament length $\ell$ and polymer concentration $c$, and (b)  for an athermal fiber network in 3D with binary crosslinks as a function of $\ell/\ell_c$ and $\ell_b/\ell_c$. Here $\ell_c$ is the distance between crosslinks measured along a filament, and $\ell_b^2$ is set by the ratio between the bending and stretching rigidity of a fiber.
}
\label{fig: phase diagrams}
\end{figure*}

The scaling argument discussed in this section also provides some insight into the amplitude of the nonaffine fluctuations at the nonaffine-affine transition. Intriguingly, at this crossover the nonaffine fluctuations reach a maximum $\Gamma_{\lambda}=\lambda_{\mbox{\rm \tiny NA }}^2/\ell_0^2=\ell_0^2/\ell_b^2$~\cite{Broedersz2012}.  This counter-intuitive result shows that the amplitude of the nonaffinity parameter can actually be large (or maximal), even if the networks shear modulus is very close to the affine value. This analysis suggest that we should be cautious interpreting the nonaffine fluctuations in an absolute sense; these nonaffinity parameters may only be meaningful when considered in the context of the elastic properties of the relevant modes of deformation.

We can  summarize the main results from these studies in a simple diagram, as shown in Fig.~\ref{fig: phase diagrams}. In panel a) we show the expectation for a thermal semiflexible polymer network with binary crosslinks as a function of two important control parameters: Filament length $\ell$ and polymer concentration $c$.
When the connectivity is too low, the network is mechanically unstable, and thus is best described as a solution. However, just beyond this threshold, the network is marginally stable and is described by the physics of the rigidity percolation. In principle, the network can be dominated either by bending (low concentration) or stretching modes (high concentration). When the polymer length is increased further, we enter the nonaffine bending regime, and subsequently crossover to the affine regime. At low concentrations the distance between crosslinks is large, and thus the entropic stretching modulus of the semiflexible polymers is softer than the enthalpic stretching modulus. Thus, there is a low-concentration affine thermal regime, and a high concentration affine mechanical regime \cite{Head2003b}.

In panel b) of Fig.~\ref{fig: phase diagrams}, we sketch the expected regimes for an athermal fiber network in 3D with binary crosslinks as a function of $\ell/\ell_c$ and $\ell_b/\ell_c$, where $\ell_b^2$ is set by the ratio between the bending and stretching rigidity of a fiber. When increasing $\ell/\ell_c$, the system transitions from a solution state ($G=0$), to a marginally stable network. When the fiber length is further increased, it starts dominating the nonaffine deformations, and thus we enter the length-controlled bending regime. However, no matter how soft the fibers are to bending, the systems always crosses over over to an affine regime at high $\ell/\ell_c$. As long as the network is mechanically stable ($G>0$), the system is bend-dominated at low $\ell_b/\ell_c$, and stretch dominated at $\ell_b/\ell_c$.

\subsubsection{Is the affine limit stable?}\label{sec:affinestable}
The discussion above was limited to the simple athermal fiber
limit, in which the fiber segments are characterized by a 1D Young's modulus
that is independent of length. By contrast, thermal semiflexible polymers
have an entropic stretch modulus that depends sensitively on length
(see Sec~\ref{sec:force extension}), and this may have important implications for
the macroscopic elastic response of real semiflexible polymer
networks; typically, such systems can be expected to
exhibit polydispersity in the
length of segments between crosslinks, and thus also a
polydispersity in the stretching moduli of these segments.

It has been argued that the pure affine limit in such networks is
not strictly stable~\cite{Heussinger2007a,Mao2013b}. This can
be understood in terms of the local force balance in the
network. Consider two consecutive segments along the
a single filament (1) somewhere in a network of straight filaments. These two segments are
separated by a crosslink to another filament (2) crossing at some angle.
Assuming a purely affine network deformation, the force due to the stretching
of filament 1 on each side of the crosslink will be proportional to the Young's
modulus of the respective segments. If these are different, there is a net force
on the crosslink that must be balanced by filament 2. As this crosses at an angle, some resulting bending energy is expected and the network deformation must be locally nonaffine.
Thus in networks with polydispersity, the affine limit is
stable for athermal simple elastic fibers, but not for thermal
semiflexible filaments or other systems where the 1D Young's modulus is
not constant.

How important is this lack of local force balance and the resulting local instability
in networks with polydisperse Young's moduli? Will the necessary bending energy generated in such systems under strain be dominant over stretching, or will this result in merely a quantitative correction
to an otherwise still stretch-dominated response? On the one hand, it has been argued based on scaling
and simulations of 2D Mikado network architectures that
regimes can arise where the mechanical response depends on both
stretching and bending energies~\cite{Heussinger2007a}.  On the other hand, as
argued in the previous section, if the response is purely bend-dominated
with small or vanishing stretch response, then the bend elastic energy must increase with molecular weight $\ell$. Thus, in the limit of high molecular weight, a purely bend-dominated behavior may not be possible. Thus, the question as to whether real, disordered networks are stretch- or bend-dominated remains, particularly in the limit of high molecular weight.

\subsection{Nonaffinity and nonlinear elasticity of athermal fiber networks}\label{sec:nonaffinenonlinear}

In Sec.\ref{sec:affinemodel} we discussed the nonlinear network response of the affine thermal model. As filaments in the network undergo large affine deformations, the thermal undulations in the polymer get ``stretched out", giving rise to a dramatic entropic stiffening response, reflected by a 10-1000 fold increase of the networks differential shear modulus at large deformations~\cite{MacKintosh1995,Kroy1996,Morse1998b,Storm2005,Gardel2004a,Lin2010b,Yao2010}. However, as discussed above, semiflexible polymer networks can be nonaffine, and dominated by athermal filament bending deformations. Thus, the question arises: what is the elastic response under large imposed shear deformations of an athermal fiber network dominated by nonaffine fiber bending deformation modes? Naively, one might not expect a nonlinear stiffening response for athermal networks that are composed of purely linear elastic elements. Strikingly however, it was shown that athermal fiber networks also strain stiffen, with a dramatic increases of the differential shear modulus at moderate deformations. 

Onck et al.\ employed 2D Mikado networks to study the effects of large strains in filamentous networks, with the additional feature that static, intrinsic curvatures could be build into the filaments~\cite{Onck2005}. These ``frozen-in" undulations were sampled from a thermal equilibrium distribution for semiflexible filaments without tension, although the network was otherwise treated as athermal and fully mechanical. They found that networks that were dominated by soft bending modes for small strains crossed over to a high-strain elastic regime dominated by stiffer stretching modes. The authors argued that such a strain-induced bend-to-stretch crossover is due to filament reorientations, which is reflected as a peak in the nonaffinity parameter at strain values near the transition. The additional frozen-in undulations were not found to be responsible for the transition, although these tended to increase the strain threshold for the stiffening transition (we will return to the point of frozen-in curvature below). Indeed, a similar stiffing response was observed in 2D athermal networks even without curvature defects along the filaments in~\cite{Chandran2006,Heussinger2007c,Conti2009}. Similar results have also been seen in 3D~\cite{Broedersz2012}.

Despite the numerous numeric studies demonstrating nonlinear strain stiffening originating in nonaffine network deformations, very little analytical progress was made initially to provide insight into this behavior.
The floppy mode theory, which provided a description for the linear regime, also has implications for the onset of the nonlinear behavior~\cite{Heussinger2007c,Heussinger2006b}. 
An essential point in understanding the nonlinear elastic response of athermal networks, is that fibers can not undergo large bending deformations without stretching at all. This can be understood from a simple geometric argument. We consider one of the filaments crosslinked at a length scale $\ell_c$ in a deformed network, and suppose that there is a transverse displacement $\delta u_{\perp}$ at one of the crosslinks. This transverse displacement not only results in a curvature of the filament, but also in an axial deformation $\ell_c+\delta$. This axial deformation can be related to the transverse bend deformation, $\ell_c^2+\delta u_{\perp}^2\sim(\ell_c+\delta)^2$, where $\delta u_{\perp}\sim \gamma \ell$ is assumed in the floppy mode model. For moderate deformations one finds to leading order $\delta\sim\delta u_{\perp}^2/\ell_c$. It was argued in~\cite{Lieleg2007} that the floppy mode description
for the linear regime only remains valid as long as the axial fiber stretch $\delta$ is small compared to the available thermal 
excess length $\ell_c/\ell_p$ (see Eq.~\eqref{eq: AFM critical strain and stress}). Thus, this implies a critical strain set by $\delta u_{\perp}^2/\ell_c^2\sim \delta_c/\ell_c$, yielding
\begin{equation}
\gamma_c\sim \ell_c^{3/2}/( \ell^2 \ell_p)^{1/2}.\label{eq:CritStrainFloppy}
\end{equation} 
This is an argument for a thermal semiflexible polymer  that is dominated by nonaffine bending mechanics in the linear regime, but that stiffens entropically under shear.

From simulations, we know that athermal networks of linear elastic fibers also stiffen. We can build on the argument in the previous section to provide some insight in this behavior~\cite{Broedersz2012}. As in the floppy mode model,  length-controlled nonaffine deformations are assumed $\delta u_{\perp}\sim \gamma \ell$. This deformation results in a bend with an amplitude $\sim\gamma \ell$, and a corresponding bending energy $\delta E_B \sim \kappa \ell^2 \gamma^2/\ell_c^3 $. We know from the geometric argument in the previous paragraph, that there is also a higher order axial stretch $\delta\sim\delta u_{\perp}^2/\ell_c$ of the filament, but what is the energy associated to this stretching deformation? The axial strain associated to this stretch is $\epsilon\sim(\gamma \ell/\ell_0)^2+O(\gamma^4)$, which amounts to a stretch energy  $\delta E_S \sim \mu (\gamma \ell)^4/\ell_0^3$. Thus, we expect that these higher order stretch contributions start dominating the elastic response at a strain where  $\delta E_B\approx\delta E_S$, resulting in a stiffening of the network's shear modulus. This implies a critical strain,
\begin{equation}\label{eq:nonlinear}
\gamma_c\sim \frac{\ell_b}{\ell},
\end{equation}
where $\ell_b=\sqrt{\kappa/\mu}$.
Interesting, both Eqs.\ \eqref{eq:CritStrainFloppy} and \eqref{eq:nonlinear} show a characteristic strain for the onset of nonlinear behavior that vanishes in the limit of increasing molecular weight $\ell$. Thus, both models can be said to have an absent or vanishing linear response regime in this limit. The argument that led to this last result assumed that the bending energy scales with $\ell^2$. However, the floppy mode model for the Mikado network predicted a slightly stronger scaling, which would then lead to $\gamma_c\sim (\ell_b/\ell)(\ell/\ell_c)^{(\mu-5)/2}$, with $\mu=5$ for the 3D Phantom model and $\mu \approx 6.67$ for the 2D Mikado model.

In the discussion above, we have assumed that the nonaffine deformations are governed by filament length. Near isostatic connectivity thresholds (See section \ref{sec:isostaticity}), where the network is marginally stable, we know that nonaffine deformations can be dominated by the proximity of network connectivity to the isostatic point. Indeed, for such cases, arguments like the one discussed in the previous paragraph taken together with nonaffine deformations that follow the form in Eq.~\eqref{eq:Gammamax}~\cite{Wyart2008,Broedersz2011a}, imply a critical strain for the onset of stiffening
\begin{equation}
\gamma_c\sim \left(\frac{\ell_b}{\ell_c}\right)^{\lambdacf/\phi+1}\label{eq:gammaCritical},
\end{equation}
where $\lambdacf$ is a critical exponent associated with the nonaffine fluctuations, and $\phi$ is a crossover exponent, both of which are discussed in detail in sections \ref{sec:crossover} and \ref{sec:fluctuations}.

It is interesting to note that the various results in Eqs.\ \eqref{eq:CritStrainFloppy}, \eqref{eq:nonlinear} and \eqref{eq:gammaCritical} make rather different predictions for the dependence of the critical strain on $\ell_c$. Since this is a parameter that is expected to depend on network concentration $c$, roughly as $\ell_c\sim c^{-1/2}$. Thus Eqs.\ \eqref{eq:CritStrainFloppy} and \eqref{eq:gammaCritical} both predict a decrease in the critical strain with increasing polymer concentration. Qualitatively, such a decrease is observed for many biopolymer networks, including actin \cite{Gardel2004a,Gardel2004b,Tharmann2007}, fibrin \cite{Piechocka2010} and intermediate filaments \cite{Lin2010a}. However, the experimentally observed exponents are more consistent the predictions of Eqs.\ \eqref{eq: AFM critical strain and stress} and \eqref{eq:lcAff}. By contrast, the prediction of the athermal nonaffine model far from the isostatic point in Eq.\ \eqref{eq:nonlinear} is that the critical strain should be independent of concentration. This is consistent with reports for collagen networks \cite{Piechocka2011}, which are expected to be athermal. 

\begin{figure}[t]
\includegraphics[width=\columnwidth]{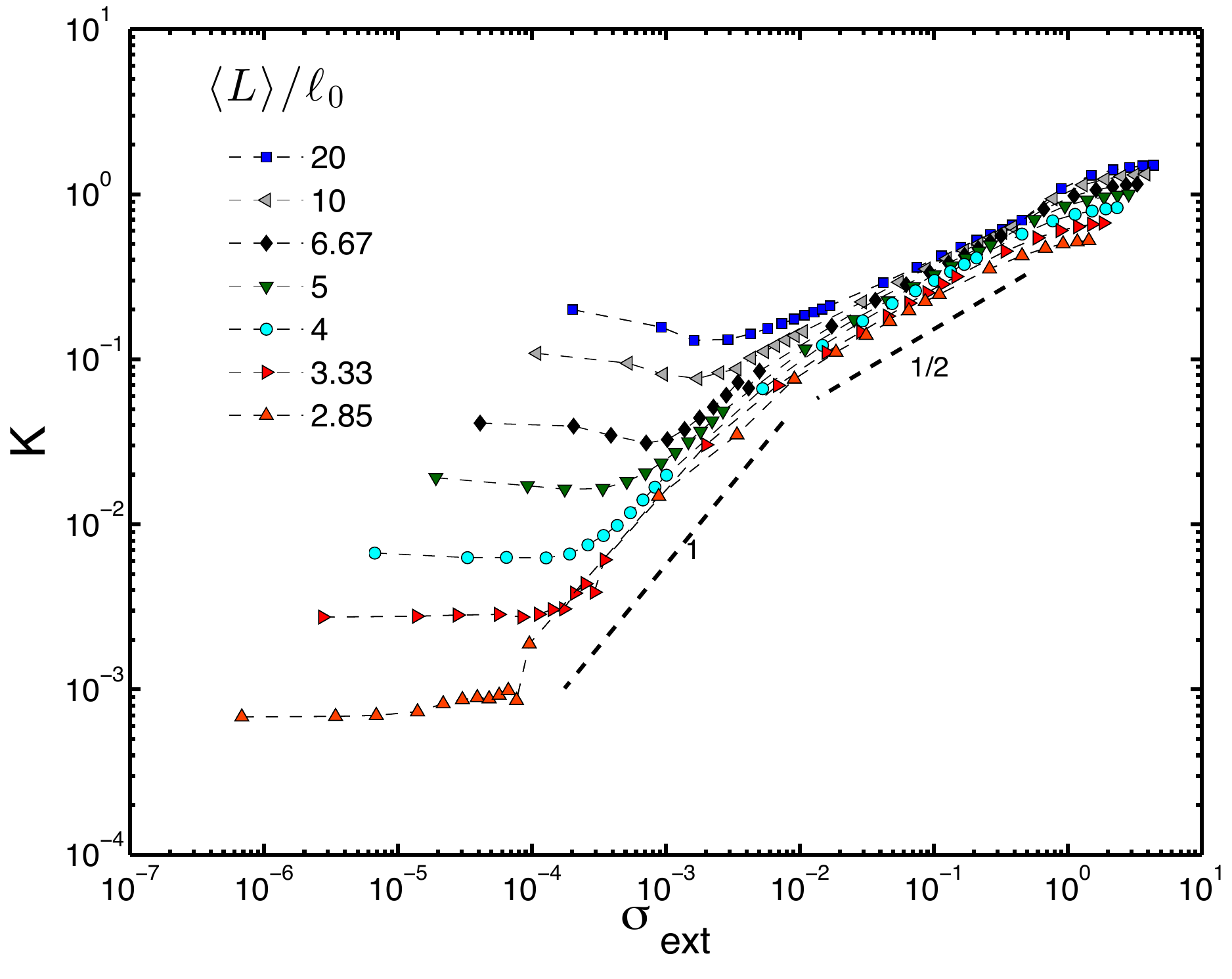}
\caption{Nonlinear elasticity of a fiber network on a diluted 2D Phantom lattice. The differential shear modulus $K=\d \sigma/\d \gamma$ as a function of the applied external stress $\sigma_{\rm ext}$ for various values of $\langle \ell \rangle$ at fixed bending rigidity $\kappa=10^{-3}$.  $K$ and $\sigma_{\rm ext}$ are measured in units of $\mu/\ell_0$. It is not completely clear whether definite powerlaw regimes exist, but the stiffening curves for $\langle \ell \rangle\lesssim 5$ initially show a stiffening behavior of approximately $K \sim \sigma$ that crosses over to a regime $K \sim \sigma^{1/2}$ at large shear, as shown by the dashed lines that indicate a slope of $1$ and $1/2$. For longer filaments, only the second, weaker stiffening response is apparent. From~\cite{Broedersz2011b}.}
\label{2Dphantomstiffening}
\end{figure}

The arguments above only address the strain at which stiffening sets in. Numerical data also clearly indicate the presence of a stiffening regime in athermal fiber networks \cite{Onck2005}, but such results do not identify a specific functional form of the stiffening response. Beyond the critical strain (or critical stress), the network stiffens gradually as the system crosses over from bending dominated to stretching dominated elasticity. At very large deformations, the differential shear modulus of linear elastic fibers is expected to asymptotically approach the affine high-strain prediction, and no longer stiffen. But, what happens during the crossover? Recent lattice-based and Mikado simulations not only show the expected asymptotic affine behavior, but also suggest a regime where $K\sim  \sigma^{1/2}$ at higher stress, as well as an initial stiffening regime with an approximate $K\sim \sigma^\alpha$ with $\alpha\approx 1$~\cite{Conti2009,Broedersz2011b}, as shown in Fig.~\ref{2Dphantomstiffening}. However, the origin of this form of stiffening behavior, and even whether it represents a genuine power-law regime remains unclear. It has been argued that such nonlinearity can be viewed as a transition from predominantly bending to stretching behavior \cite{Onck2005}. But, pure spring networks without bending have been shown to exhibit shear stresses $\sigma$ that increase quadratically with strain $\gamma$ \cite{Wyart2008}, suggesting that only the onset of the $K=d\sigma/d\gamma\sim\sigma^{1/2}$ regime corresponds to a transition to stretching-dominated behavior. (See also Sec.~\ref{sec:StabilizeMarginal}.) Thus, whether the nonlinear response of fiber networks can be generally described by a crossover from bending to stretching remains unclear \cite{LicupUnpub}. 

\subsubsection{More on the role of intrinsic curvature}
We would like to return briefly to the point of intrinsic curvature. 
An interesting perspective on the role of intrinsic curvatures was provided in~\cite{Kabla2006}. Their model captures the geometrical effects of the quenched disorder in the intrinsic curvatures of the fibers. When a single such filament is stretched, it will first unbend as the natural curvatures are ironed out, after which the presumedly stiffer back bone may stretch. They considered an inextensible, weakly curved filament with random curvatures chosen from a distribution that represents the curvatures of fibers in felt. Using this, they calculated a force extension curve, which for large forces diverges as
\begin{equation}
f\sim\frac{1}{\sqrt{1-\epsilon}},
\end{equation}
where $\epsilon$ measures the extension relative to the fully extended state.
Interestingly, this is in contrast to the predicted divergence of the force-extension behavior of a semiflexible polymer, for which $f\sim\frac{1}{(1-\epsilon)^2}$~\cite{Marko1995,Gardel2004a,MacKintosh1995,Fixman1973}.
Thus, even though both mechanisms for a nonlinear response have their origin in stretching out fluctuations, quenched fluctuations and thermal fluctuations can give rise to
a quantitatively different form for the divergence. Kabla and Mahadevan used a unit-cell approach in their quenched fiber model to describe the elastic and plastic behavior of felt networks.

\subsubsection{Negative normal stress in athermal networks}
Recently, biopolymer networks were shown to exhibit an additional form of nonlinear elastic response known as \emph{negative normal stress} or a negative/anomalous \emph{Poynting effect}. When most materials are subjected to simple shear, they tend to expand in the strain gradient direction. This is an effect first observed by Poynting a little more than one hundred years ago \cite{Poynting1909}. Poynting performed careful experiments twisting wires, which he showed resulted in an axial extension of the wires. This Poynting effect can also be expressed in terms of the positive (compressive) stresses that would develop axially if such a wire is not allowed to elongate. This effect is fundamentally nonlinear, in that it cannot change sign under twisting in the reverse direction: while shear stresses are \emph{odd} in twist or shear strain, normal stresses must be \emph{even}. Thus, no linear normal stress response is expected at small strain. 

Interestingly, rheological studies of a wide variety of biopolymer gels have been shown to exhibit the opposite effect: they develop tensile stresses or contract in the axial direction, which shows up as a negative thrust in a rheometer \cite{Janmey2007}. 
Moreover, the normal stresses were shown to become as large in magnitude as the shear stresses. It was shown in that same work that the affine thermal model can account for the unexpected sign and large magnitude of the normal stresses. The stiffening response of Mikado networks has also been shown to coincide with additional nonlinearities, such as the appearance of negative normal stresses and a softening response due to buckling \cite{Kang2009,Onck2005,Conti2009,Heussinger2007c}. Although both affine and nonaffine models can account for negative normal stresses, they predict a qualitatively different dependence of the normal stress as a function of shear stress~\cite{Conti2009}. Experiments on fibrin networks appear to be in better agreement with the non-affine predictions~\cite{Kang2009}.

Normal stresses are frequently studied in other soft matter systems, in connection with such phenomena as the Weissenberg effect, in which a viscoelastic fluid tends to climb a rod that is rotated in the liquid \cite{LarsonBook}. This can also be seen in the stirring of bread dough, as a network of gluten begins to form. Normal stresses appear in the stress tensor along the diagonal, where hydrostatic pressure also appears as a uniform contribution along the diagonal of the stress tensor. But, only spatial variations in the pressure can affect the flow and deformation of incompressible materials. Thus, the stress tensor for such materials is only defined up to an additive isotropic (pressure-like) term, with the result that rheological measurements are only sensitive to \emph{normal stress differences} among the various diagonal terms in the stress tensor. 

It is important to note, however, that this only follows for incompressible materials. And, while systems such as biopolymer network can usually be considered to be incompressible by virtue of the solvent they are imbedded in, the two-component character of such systems can lead osmotic compressibility of the network \cite{Brochard1977,Gittes1997,Levine2001,MacKintosh2008}. This means that rheological measurements in such effectively compressible materials can, in principle, measure individual normal stress components of the tensor. As was argued by Janmey et al., \cite{Janmey2007}, this can be expected especially for biopolymer systems with their relatively open meshworks, where the mesh size can be as large a several micrometers, e.g., in the case of collagen and fibrin gels. On long time scales, the solvent can be expected to move relatively freely through such a porous network. Only on shorter time scales, or for finer meshworks, where this motion is impeded by hydrodynamics, will the network be expected to \emph{inherit} the incompressibility of the solvent: here, a strong hydrodynamic coupling of network and solvent is expected. For this reason, normal stress measurements in biopolymer gels have been reported in terms of the single, axial normal stress component \cite{Janmey2007,Kang2009}.

\section{Marginal stability and critical phenomena in fiber networks}\label{sec:isostaticity}

The importance of network connectivity, and concepts such as isostaticity and criticality have long been recognized in the fields of rigidity percolation and jamming phenomena. As we will discuss here, many of these concepts have also proven to be helpful in understanding interesting aspects of fibrous networks.
 
Maxwell introduced an analysis of the mechanical stability of spring networks highlighting the importance of  connectivity~\cite{Maxwell1865}. Spring-like bonds give rise to central forces, i.e., forces which only depend on the relative distance between two connected network nodes. Maxwell's constraint counting argument predicts that such spring networks are mechanically rigid at connectivities exceeding $\zcf=2 d$. At this central force (CF) isostatic point the number of constraints arising from the central-force interactions $N z/2$ precisely balances the number of internal degrees of freedom $N d$. This prediction for the isostatic connectivity is remarkably accurate for jammed systems and is reasonably accurate in percolation networks~\cite{Thorpe1983,Thorpe1985,He1985,Schwartz1985,Feng1984a}. (See \cite{Hecke2010,Liu2010} for recent reviews on this subject.) These, and many other studies have also demonstrated that a variety of systems, including network glasses and jammed systems exhibit a rich mechanical behavior that is controlled by the proximity of network connectivity to the isostatic connectivity.

What is the role of connectivity and isostaticity in fiber networks? Clearly this is more subtle than for spring networks, since fibers resist both stretching and bending; while fiber stretching can be modeled with spring-like central-force interactions, fiber bending requires non-central, 3-point interactions. Thus, bending interactions add constraints of a different nature that can stabilize the systems at connectivities below the central-force isostatic point. Indeed, various studies on network glasses and jammed systems have illustrated how additional interactions can stabilize networks below the CF threshold~\cite{Garboczi1986,Wyart2008}. More recently, various studies looked at the role of other stabilizing quantities, such as contractile stresses, viscous interaction and temperature, and we will return to these studies  below.

Filamentous networks such as biological gels typically have average connectivities between three and four, positioning them well below the CF isostatic threshold in 3D \cite{Lindstrom2010}. Thus, their rigidity can be be strongly influenced or even controlled by other stabilizing effects, such as bending rigidity. However, although the network stability may rely on fiber bending rigidity, this does not necessarily imply that the network mechanics is governed by fiber bending, as evidenced by the length controlled bend-to-stretch crossover discussed in Sec.\ \ref{sec:phantom}. The role of network disorder and nonaffinity is also presumed to become more important in such under-connected networks. Indeed, the precise role of bending interactions in biological fiber networks has been subject of much debate~\cite{Head2003a,Head2003b,Wilhelm2003,Gardel2004a,Storm2005, Lieleg2007, Onck2005,Heussinger2006b,Heussinger2007c,Chaudhuri2007,Buxton2007,Huisman2011}

One fruitful approach to studying the role of network connectivity in 2D and 3D has been to use network architectures based on lattice structures~\cite{Das2007,Das2012,Mao2013a,Mao2013b,Broedersz2011a,Broedersz2012,Broedersz2011b,Sheinman2012b},  and we discuss these studies in the next section. 

We should pause to ask how useful such an approach might be for describing real networks. Differences in network architecture can have dramatic consequences for the network mechanics~\cite{Heussinger2007a}. The precise architecture of biological networks in different physiological contexts and \emph{in vitro} reconstituted biopolymer gels is not well understood. While the architectural variety is an interesting subject of investigation in and of itself, we now ask whether there may be simple overarching principles governing the network mechanics that do not depend sensitively on architectural details. If such principles exist, this could justify using a network architecture that is convenient from a theoretical perspective, enabling both efficient computation and tractable analytical calculations. However, one should ask whether the results of these lattice models, or other minimalistic models for that matter (see \ref{sec:Mikado}), do not rely crucially on the simplified geometry or dimensionality and still hold for more realistic network. 

\subsection{Lattice-based bond-dilution networks}\label{sec:lattice}

These networks consist of straight fibers organized on a lattice geometry. The constituent filaments resist stretch, with modulus $\mu$, and compression, with modulus $\kappa$. Thus, they generate central forces directed along the fiber segment between crosslinks; because the fibers resist bending, they also generate torques favoring parallel alignment of consecutive segments along a single fiber. The connectivity can be controlled by randomly diluting bonds between crosslinks, such that the probability for each bond to be present is $p$, as illustrated in Fig.~\ref{fig:nonaffinescheme2} for a 2D triangular lattice and  Fig.~\ref{fig: fiber network models}c for a 3D face centered cubic lattice. Thus the  connectivity in such a network, set by the average number of bonds connected to a node excluding dangling bonds, is roughly given by $z=p Z$, where $Z$ is the coordination number of the undiluted lattice.

We focus here on networks with freely-hinged bonds between fibers, in contrast with earlier studies of rigidity percolation, including studies of network glasses \cite{Thorpe1983,Schwartz1985,He1985,Sahimi1993}. The motivation for this is partly to keep the number of parameters to a minimum, but also because of the large aspect ratio of crosslink distance to molecular scale or low volume fraction of most biopolymer networks, which means that the fiber segments have a large lever-arm for bending fibers at crosslinks that fix the bond angles. Bond bending can be included, however, and it has been shown to stabilize networks to a somewhat lower connectivity threshold \cite{Das2012}. Otherwise, the qualitative features are much the same as for freely-hinged bonds. 

\begin{figure}
\includegraphics[width=\columnwidth]{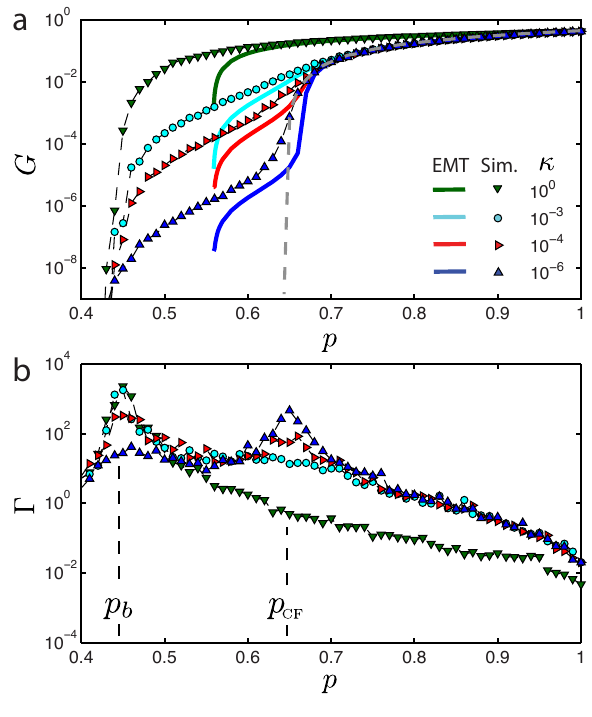}
\caption{The shear modulus  for a diluted 2D triangular lattice fiber networks. The shear modulus $G$, is shown in units of
$\mu/\ell_0$ as a function of the bond occupation
probability, $\P$, for a range of filament bending rigidities,
$\kappa$ (in units of  $\mu\ell_0^2$). The numerical results for $\kappa=0$ are shown as
dashed grey lines. The EMT calculations for a 2D triangular lattice are
shown as solid lines in (a). b) The nonaffinity measure,
$\Gamma$, is shown as a function of $\P$ for various values of
$\kappa$ for a 2D triangular lattice. Adapted from~\cite{Broedersz2011a}.}
\label{fig: lattice results}
\end{figure}

The numerical results for this lattice-based fiber model with freely freely-hinged crosslinks are shown in Fig.~\ref{fig: lattice results} for a triangular lattice in 2D. For $\kappa=0$ (dashed grey line Fig.~\ref{fig: lattice results}), the shear modulus vanishes continuously at a critical value $\pcf$ (for a 2D triangular lattice $\pcf\approx 0.651$, and for a 3D fcc lattice $\pcf\approx0.473$). In particular, it is well established that near the CF isostatic point $G\sim \mu|p-\pcf|^\fcf$ (see Table~\ref{table:exponents}). In contrast, in the limit of large $\kappa/\mu$ the shear modulus is approximately $G \sim \mu p$. Thus, the shear modules approaches the affine limit and, thus, becomes independent of $\kappa$; even if the network has a connectivity below the CF threshold, if nonaffine bending deformations are energetically more costly than stretching deformation, it is more favorable to deform through stretching.
 
For all networks with a finite bending stiffness, the shear modulus vanishes continuously at a rigidity threshold $\pb$. This behavior is governed by the bending rigidity exponent $\fb$ (see Table~\ref{table:exponents}). The value of $\pb$ independent of the bending rigidity (for a 2D triangular lattice $\pb\approx 0.445$, and for a 3D fcc lattice $\pb \approx0.268$). This bending threshold can be understood from a counting argument similar to Maxwell's analysis at the central-force threshold, but now extended to include bending constraints. 

\subsubsection{Counting argument for the ``bending" rigidity threshold}

The bending isostatic point $\pb$ of lattice-based fibrous networks can be calculated using 
Maxwell counting and mean-field arguments. Isostatic conditions require that the total number
of network constraints due to both stretching and bending are equal to the total number of degrees
of freedom. In $d$ dimension, the total number of internal degrees of freedom is equal to $dN_{c}$,
where $N_{c}$ is the number
of network crosslinks. The number of constrains due to the stretching stiffness of the bonds is $N_{b}p$, where $N_{b}$ is
the number of bonds in the undiluted network ($p=1$). In addition, the bending rigidity contributes $d-1$ constraints at any pair of neighboring coaxial bonds, and the total number of such bonds is $N_{b}p^{2}$. Thus, the rigidity percolation transition occurs when
\begin{equation}
dN_{c}=N_{b}\left(p+\left(d-1\right)p^{2}\right)
\end{equation}
or
\begin{equation}
p_{b}=\frac{\sqrt{1+\frac{4dN_{c}}{N_{b}}\left(d-1\right)}-1}{2\left(d-1\right)}.
\end{equation}
For the triangular lattice we obtain
$p_{b}=\frac{\sqrt{\frac{11}{3}}-1}{2}\simeq0.4574$, for the Kagome and square lattices
$p_{b}=\frac{\sqrt{5}-1}{2}\simeq0.618$, while for the fcc lattice $p_{b}=\frac{\sqrt{5}-1}{4}\simeq0.309$, in reasonable agreement with simulations~\cite{Das2012,Mao2013a,Mao2013b,Broedersz2011a}. A more accurate calculation of the rigidity points can be found in~\cite{Broedersz2011a}, and similar arguments for off-lattice networks have also been presented~\cite{Huisman2011}. 

\subsubsection{The critical crossover regime between stretching and bending dominated mechanical behavior}\label{sec:crossover}
Although this bending threshold, $\pb$, marks the true onset for rigidity in fiber networks, there is another important connectivity threshold for these systems:  For low  enough $\kappa$, the shear modulus crosses over at the central force isostatic point $\pcf$ from bending dominated to stretching dominated behavior, as shown in Fig.~\ref{fig: lattice results}a. This crossover in the mechanical response is accompanied by a cusp in the nonaffine fluctuations, as quantified by the nonaffinity parameter $\Gamma$ (see Fig.~\ref{fig: lattice results}b and Eq.~\eqref{eq:1pointNApar} for the definition of the nonaffinity parameter); the amplitude of this cusp diverges with vanishing $\kappa$, highlighting the critical state of the network when $p=\pcf$ and $\kappa=0$.

It is instructive to draw an analogy between these observations in fiber networks and the behavior of other well-understood  models for critical behavior at finite temperature. When a small bending rigidity is added to the network model, the system is stabilized at and below the central-force isostatic point: the shear modulus no longer vanishes and the strain fluctuations are now finite. Thus, at least qualitatively, the impact of $\kappa$ on the strain fluctuations and the shear modulus near the central-force isostatic point is  analogous to the impact of an external field or coupling parameter on the order parameter and its fluctuations near the critical temperature, as the field takes the system away from criticality.  This turns out to be more than just a qualitative analogy, but with some intriguing nuances between  athermal networks  and  thermal systems. 

The CF isostatic point plays a central role in determining the cross-over from the stretching dominated regime to the bending dominated regime~\cite{Straley1976,Garboczi1986,Wyart2008,Broedersz2011a,Buxton2007}. Only in the limit $\kappa \rightarrow 0$, the CF isostatic point is a true critical point. From the perspective of critical phenomena, the bending rigidity may be thought of as an applied field or coupling constant that  leads to a crossover from one critical system to another  (such as from the Heisenberg model to the Ising model~\cite{Fisher83}). In such thermal systems, there is a continuous evolution of the critical point as this coupling parameter is varied. Interestingly, there is no such continuous evolution with variation in $\kappa$ in athermal fiber networks, which show a discontinuous jump from $\pcf$ to $\pb$, as soon as $\kappa$ becomes nonzero. Furthermore, although the analogy between $\kappa$ and a field is insightful, there are other important formal differences. For instance, the magnetic field couples linearly to a symmetry breaking order parameter, while this not the case for $\kappa$.

\begin{figure}
\includegraphics[width=\columnwidth]{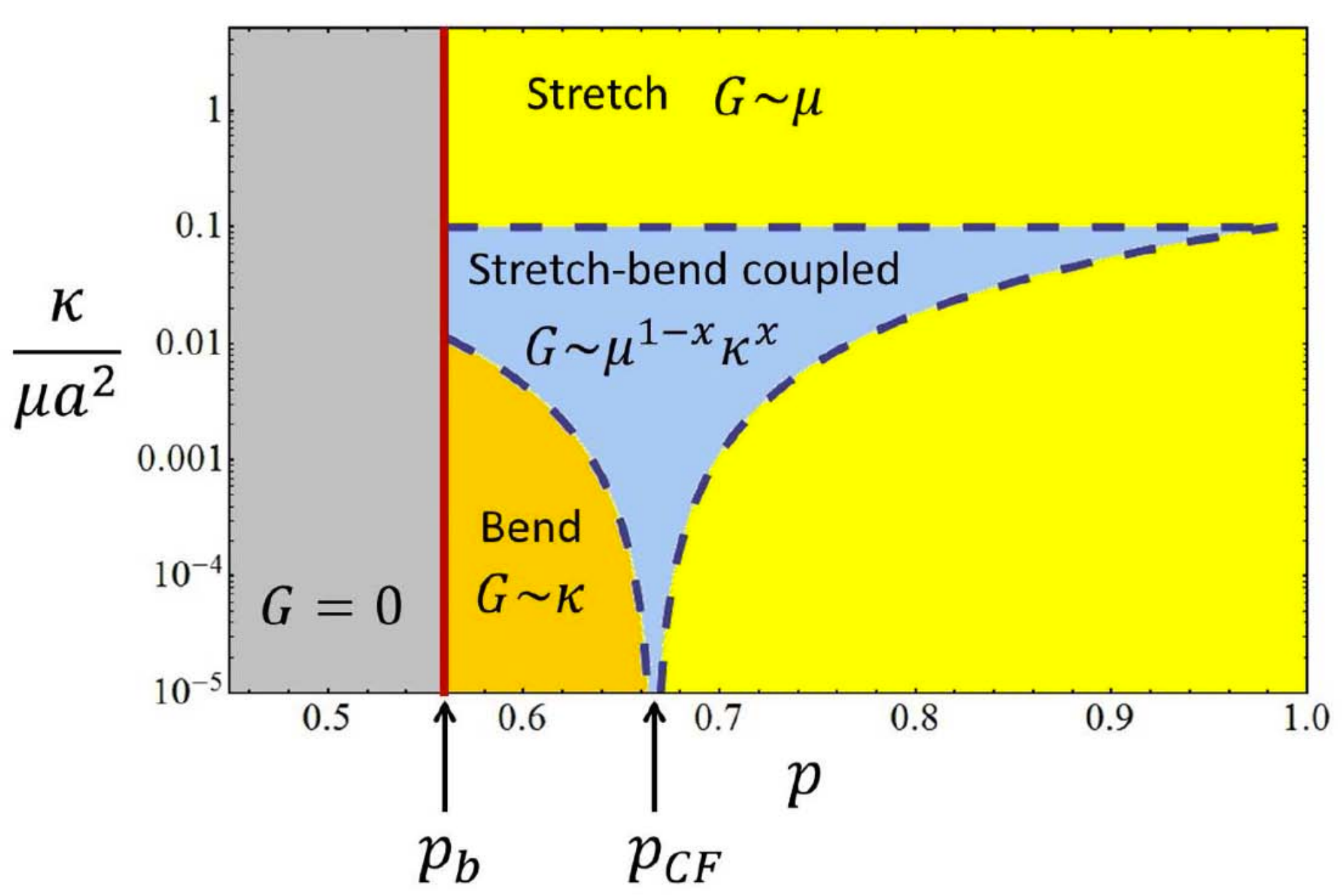}
\caption{Phase diagram for  a fibrous
network on a diluted triangular lattice. Above the rigidity percolation point $\pb$ there are three
distinct mechanical regimes: a stretching dominated regime with
$G\sim\mu$, a bending dominated regime with $G\sim\kappa$ and a regime
in which bend and stretch modes couple with $G\sim\mu^{1-x}\kappa^{x}$.
Here $x$ is related to the critical exponents $x=\fcf/\phi$. The mechanical regimes are controlled by the isostatic point
$\pcf$, which acts as a zero-temperature critical point. Adapted from~\cite{Broedersz2011a,Mao2013a}.}
\label{fig:criticalphase}
\end{figure}

These ideas about how $\kappa$ impacts the mechanical response near $\pcf$ have been formalized by constructing an effective medium theory (EMT) using the coherent potential approximation (CPA) by Mao and Lubensky~\cite{Mao2013a, Broedersz2011a} (See Sec.\ref{sec:EMT}). This model shows that the shear modulus may be written as a universal function when $\kappa/\mu \ll \Delta p$, with 
\begin{equation}
\label{eq:widomscale} G=\mu |\Delta p|^\fcf \mathcal{G}_{\pm} \Big(\frac{\kappa}{\mu} |\Delta p|^{-\phi} \Big),
\end{equation}
where $\mathcal{G}_{\pm}$ is a universal function where the two branches  apply above and below the transition. When the argument of  $\mathcal{G}_{\pm}(y)$, $y\ll 1$, $\mathcal{G}_+(y) \sim \text{const.}$ and $\mathcal{G}_-(y)\sim y$, such that $G \sim \mu|\Delta p |^\fcf$ for $\Delta p >0$ and $G\sim \kappa |\Delta p|^{\fcf-\phi}$ for $\Delta p <0$. In the opposite limit $(\kappa/\mu) |\Delta p|^{-\phi} \gtrsim 1$, i.e., in the critical regime, $G$ must become independent of $\Delta p$ since $G$ is neither zero nor infinite at $\Delta p = 0$. Thus, Eq.\ (\ref{eq:widomscale}) predicts $G \sim \kappa^{\fcf/\phi} \mu^{1-(\fcf/\phi)}$ in the vicinity of $\pcf$, yielding an anomalous mechanical regime that is governed by both the stretching and bending energies. The various mechanical regimes are summarized in a phase diagram in Fig.~\ref{fig:criticalphase}.

Interestingly, the scaling form in Eq.~\eqref{eq:widomscale} is
analogous to that for the conductivity of a random resistor
network~\cite{Straley1976} with bonds occupied with resistors of
conductance $\sigma_{>}$ and $\sigma_<$ with respective probabilities
$p$ and $(1-p)$, as well as random spring networks with soft and stiff springs~\cite{Garboczi1986, Wyart2008}.

The universal scaling function in Eq.~\eqref{eq:widomscale} is also predicted by the EMT theory when $\kappa/\mu \ll \Delta p$,  with 
\begin{eqnarray}
	\mathcal{G}_{\pm}(y) & \simeq & \frac{3}{2}\big(
		\pm 1+\sqrt{1+4\mathcal{A}y/9} \big) \nonumber\\
\end{eqnarray}
where $\mathcal{A}\simeq 2.413$, $f_{\rm EMT} = 1$ and $\phi_{\rm EMT} = 2$. Interestingly, these mean field exponents are identical to those found in central-force networks
with two types of springs \cite{Garboczi1986, Wyart2008}. However, in lattice-based fiber networks non mean-field exponents are found (See Table~\ref{table:exponents}).

\begin{table}
\caption{Critical exponents for bond-diluted lattice fiber networks in 2D (triangular lattice) and 3D (fcc lattice)~\cite{Broedersz2011a}.} 
\centering 
\begin{tabular}{c c c c c c c c} 
\hline\hline 
exponent & 2D sim &  2D EMT & 3D sim\\ [0.5ex] 
\hline 
$\fcf$ & $1.4\pm0.1$ & $1$ & $1.6\pm0.2$\\ 
$\phi$ & $3.0\pm0.2$ & $2$ & $3.6\pm0.3$\\
$\nucf$ & $1.4\pm0.2$ & & \\
$\lambdacf$ & $2.2\pm0.4$ & & \\
$\fb$ & $3.2\pm0.4$ & $1$ & $2.3\pm0.2$\\ 
$\nub$ & $1.3\pm0.2$ & & \\
$\lambdab$ & $1.8\pm0.3$ & & \\
[1ex] 
\hline 
\end{tabular}
\label{table:exponents} 
\end{table}

We will provide a more detailed discussion of the effective medium approach below. But, we will first discuss some other interesting aspects of these networks, which generalize to other marginal systems, including networks in the presence of viscous interactions, thermal fluctuations or internal stresses.

\subsubsection{What can we learn from the nonaffine  fluctuations in marginal networks?}\label{sec:fluctuations}
The analogy observed above between the mechanics of fiber networks and thermal critical phenomena begs the question as to whether other signature of criticality may also be present here. Among the most important and general aspects of critical phenomena are fluctuations and a corresponding correlation length, both of which diverge at the critical point. These features have also been shown for athermal fiber networks, as well as for particle packings near the jamming transition. Perhaps the most natural candidate for fluctuations in such systems is the nonaffinity of the deformation field, i.e., the fluctuations in the strain field. Indeed, it was found that the nonaffine fluctuations $\Gamma$ diverge as $\Gamma_{\pm} |p-\pcf|^{-\lambda_{cf}}$ for $\kappa=0$ near the CF critical point, and $\Gamma_{\pm} |p-p_{b}|^{-\lambda_b}$ for $\kappa>0$ at the rigidity percolation point (See Sec.\ref{sec:char_nonaffinity} and Eq.~\eqref{eq:1pointNApar} for more details on the definition of the nonaffinity parameter). This is similar to findings in spring networks in a jammed configuration~\cite{Wyart2008}, but with non mean-field exponents in the case of lattice fiber networks. Moreover, one can find an associated divergent length-scale $\xi=\xi_{\pm} |\Delta p|^{-\nu}$ near the respective critical points. This scaling can be determined by performing a finite size scaling analysis. The divergence of $\xi$ is limited by the system size $W$, which should and does suppress the divergence of $\Gamma$.

When $\kappa>0$, the system is no longer critical at $\pcf$, yet signatures of criticality remain near $\pcf$.  Specifically, the divergence of the nonaffinity parameter is  suppressed, but grows as $\kappa\rightarrow0$. Close to the CF isostatic point one finds a peak of $\Gamma$ that scales as 
\begin{equation}
\Gamma_{\rm max}\sim\left(\frac{\kappa}{\mu}\right)^{-\lambdacf/\phi}.
\label{eq:Gammamax}
\end{equation}
Moreover, as in ordinary critical phenomena, the diverging fluctuations are also associated with a diverging susceptibility $\chi\sim\Gamma$. This suggests that the order parameter (here, $G$) can be expressed in terms of the susceptibility and the field or coupling constant (here, $\kappa$) that takes the system away from the critical point at $\pcf$:
\begin{equation}
G\sim \kappa \Gamma_{\rm max}\sim \mu^{\lambdacf/\phi} \kappa^{1-\lambdacf/\phi}=\mu^{1-\fcf\phi} \kappa^{\fcf/\phi},
\end{equation}
which can be confirmed by simulation. The scaling behavior of the nonaffine fluctuations also has important implications for the critical strain at which these networks become nonlinear, as discussed in Sec.~\ref{sec:nonaffinenonlinear}.

\subsection{Stability of marginal networks}\label{sec:StabilizeMarginal}
We have discussed above how bending rigidity can stabilize an otherwise floppy network below the isostatic or marginal state of connectivity. This is an example of a much broader class of phenomena, in which additional interactions or \emph{fields} can change the state of a system and lead to rich critical phenomena associated with the marginal state. The basic idea goes back at least to the 1970s in the context of random resistor networks \cite{Dykne1971,Efros1976,Straley1976}. In the present context, this is also closely related to rigidity percolation studies in 1980s, e.g., in \cite{Garboczi1986}. More recently, in the context of jamming, the importance of critical fluctuations and crossover has also been shown \cite{Wyart2008}. 

Here, it is useful to draw an analogy with ferromagnetism in statistical physics. We associate the shear modulus $G$ of a network with the magnetic order parameter $m$: the \emph{ordered} phase is the stable, rigid one. In essence, much as a magnetic field $h$ can \emph{stabilize} a paramagnet, i.e., by creating a non-zero magnetization, the bending stiffness $\kappa$ above can act to stabilize an otherwise floppy network. Given the critical nature of the underlying marginal point in the absence of additional interactions, signatures of this critical point can be seen away from the critical point, as for a magnetic system: for instance, for weak applied magnetic fields, both the susceptibility and magnetic fluctuations exhibit evidence of a divergence near the critical point, although this divergence is suppressed or rendered finite by the finite magnetic field. For fiber networks, this is illustrated by the divergence of the fluctuations near $\pcf$ in Fig.\ \ref{fig: lattice results}b, which is suppressed by bending stiffness $\kappa$. Similar effects are also seen in jamming \cite{Wyart2008}, where nonaffine fluctuations are suppressed by addition of weak springs. 

One of the signatures of criticality in magnetic systems is the relationship between the magnetization $m$ and the applied field $h$ along the critical isotherm, where $m\sim h^{1/\delta}$, where $\delta=3$ in mean-field theory. This can be seen as a consequence of the equation of state relating $h$ to $m$, much like the pressure-volume relation in a liquid-gas critical system: $h\simeq m/\chi+b m^3$, for some constant $b$. The susceptibility $\chi$ diverges at the critical point, resulting in $m\sim h^{1/3}$ along the critical isotherm. 
Here, the form of the equation of state is constrained by symmetry to have only odd terms. For bending stabilized marginal networks, one can expect a similar relationship between $G$ and $\kappa$: $\kappa\sim G/\chi_{\kappa}+b G^2$, where $\chi_{\kappa}$ represents the susceptibility of $G$ on $\kappa$. Note that this is based on the following assumptions: i) the system is operating in a regime where $G=0$ in the absence of $\kappa$; ii) $\kappa$ is a stabilizing field that renders $G$ nonzero, but small; and iii) $\kappa$ is analytic and can be expanded in powers of small $G$, in analogy with other mean-field theories. Importantly, even terms are no longer forbidden here by symmetry, resulting in $G\sim \kappa^{1/2}$ at the critical point. This suggests a simple explanation for the observation of the approximate square-root dependence of $G$ on $\kappa$ in Sec.~\ref{sec:crossover}. This mean-field argument is general, and it suggests a similar square-root dependence of $G$ in the critical regime on any stabilizing field such as $\kappa$, viscous stresses, active internal stresses, and thermal fluctuations. Such behavior is consistent with other mean-field arguments and effective medium theories~\cite{Sheinman2012b,Das2012,Wyart2008,Wyart2010,Broedersz2011a, Yucht2013,Tighe2012,Tighe2011, Mao2013a,Mao2013b}.

Both the stabilization and associated critical behavior of marginal and floppy (sub-marginal) networks have been shown for a broad class of different networks, including spring and fiber networks, and for range of stabilizing fields, including external and internal stresses \cite{Sheinman2012b,Broedersz2011b}, viscous interactions by an embedding Newtonian fluid~\cite{Lerner2012a,Lerner2012b,Lerner2013,Tighe2012,Tighe2011,Andreotti2012,Wyart2010,Yucht2013}, large external strains~\cite{Sheinman2012a}, and even thermal fluctuations \cite{Dennison2013}. Although each of these cases has interesting distinguishing features, the general critical phenomena framework and the connection between network mechanics and strain fluctuations applies to all. Moreover, these cases show approximate square-root dependence of $G$ on the corresponding stabilizing field: e.g., in the presence of thermal fluctuations, anomalous entropic elasticity is seen, in which $G\sim\sqrt{T}$ at finite temperature $T$ or $G\sim\sqrt{\sigma}$ under applied shear stress $\sigma$ (see discussion at end of Sec.~\ref{sec:nonaffinenonlinear}.)

One of the possible biological implications of the stabilizing effect of stresses is the observation that internal stress by molecular motors can stabilize and control the mechanics of intracellular networks \cite{Sheinman2012b}. This can provide a simple and general mechanism for control of cell mechanics without the need to change the amount or even the connectivity of cytoskeletal networks. Additionally, the critical nature of the model systems suggests the possibility of exquisite control of mechanics through the expected strong (mechanical) susceptibility, making such a system a highly responsive material.

\subsection{Effective medium theories}\label{sec:EMT}

We now review the effective medium approach, first formulated in~\cite{Feng1985}, in its most simplest form: for spring networks. We will then end with a brief discussion on how this EMT approach can be extended for various situations.

The effective medium network is an \emph{undiluted} network, with renormalized bond stiffness $\tilde{g}$, depending on the degree of dilution $p$ of the actual network it represents. The EMT provides a self-consistent construction to determine this renormalized bond stiffness from which the mechanical response of the effective network can be calculated.

\begin{figure}
\includegraphics[width=\columnwidth]{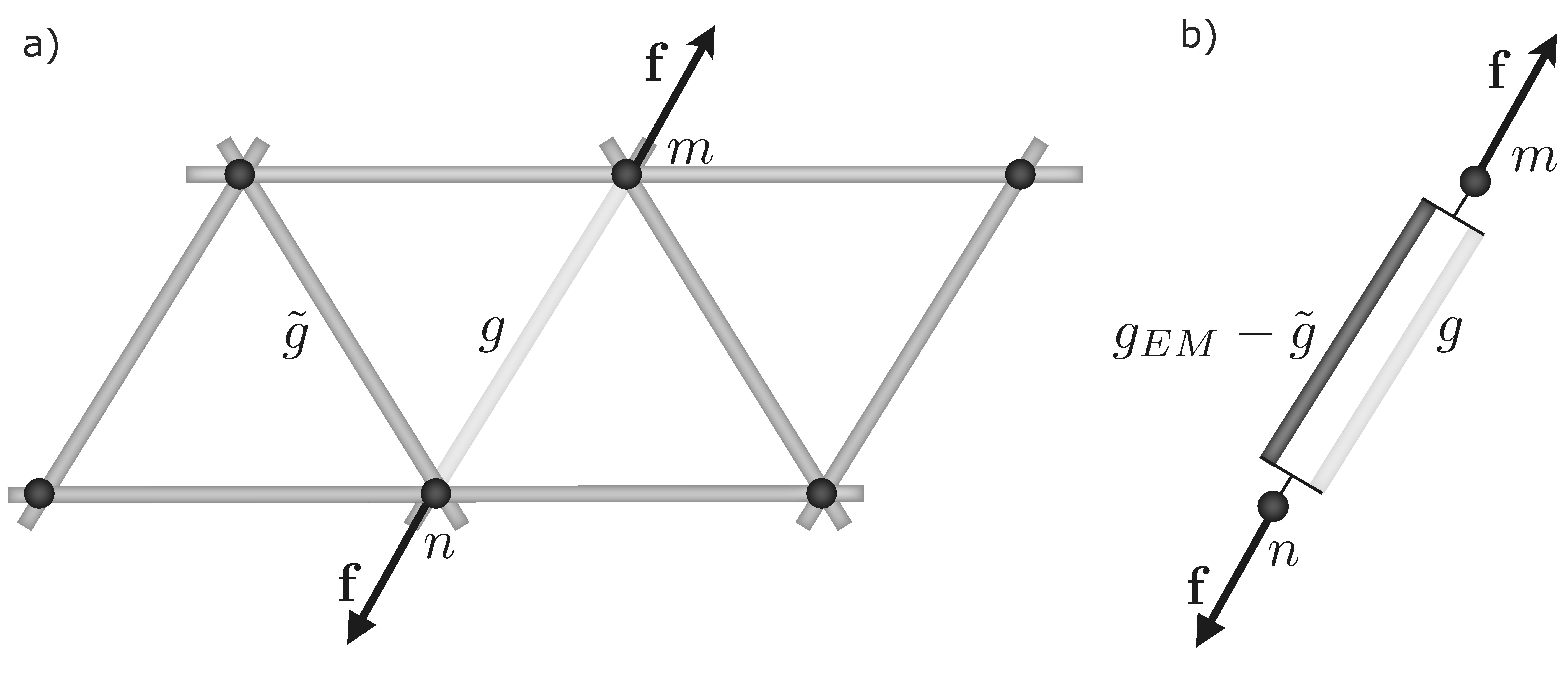}
\caption{Illustration of the effective medium framework. a) The bonds in the effective medium have a stiffness $\hat{g}$, which are calculated self-consistently. The bond between nodes $n$ and $m$ is replaced by a random bond $g$, which leads to a distortion if the effective network is under strain. However, an additional force $f$ can be applied to counter this distortion. b) Illustrates how the stiffness between nodes $n$ and $m$ can be described by two springs in parallel. This figure is a slightly altered version from~\cite{Feng1985}.}
\label{fig:EMTscheme}
\end{figure}

Suppose the effective network is subjected to a macroscopic infinitesimal  strain $\epsilon$, deforming bond $nm$ affinely by $ {\bf \hat{r}}_{nm} \epsilon$, where $ {\bf \hat{r}}_{nm}$ is the unit vector along bond $nm$. Subsequently, replacing this effective medium bond with stiffness $g$ (See Fig.~\ref{fig:EMTscheme}),  sampled from the distribution $P(g)$, gives rise to an additional, nonaffine deformation $\delta {\bf u}$. The original, (uniform) deformation can be restored by applying an additional force to the bond
\begin{equation}
{\bf f} = {\bf \hat{r}}_{nm} \epsilon(\widetilde{g} - g)
\end{equation}
Since the network is assumed to be in the linear response regime, applying this force to an unstrained network would have given the same deformation $\delta {\bf u}$, that resulted from substituting a bond in the strained effective network in the absence of the force.
If we had only removed the $nm$ bond, the effective stiffness between these two nodes due to the surrounding network is $g_{EM}-\hat{g}$, where $g_{EM} $ is the force on a
bond in the perfect effective medium network in response to a unit displacement. Then, inserting a random bond $g$ between nodes $n$ and $m$, leads to a local stiffness $g_{EM}-\hat{g}+g$ (See Fig.~\ref{fig:EMTscheme}b).
Thus, the nonaffine deformation that arose from the bond replacement in the strained effective network (without the force $f$) can be expressed as
\begin{equation}
\delta\mathbf{u}=\frac{\mathbf{\hat{r}}_{nm}\epsilon\left(\widetilde{g}-g\right)}{g_{EM}-\widetilde{g}+g},
\label{eq:DisplacementMainText}
\end{equation}
This deformation clearly depends on the stiffness of the inserted  bond chosen randomly from the distribution  $P(g)$, leading to either a contraction or dilation of the network. The stiffness of the effective medium should be chosen such that on average, we recover the macroscopically imposed deformation, and thus, these nonaffine displacements due to bond insertion should vanish on average.
Hence, the self-consistency condition requires that the local 
fluctuations in the deformation field in the deformation field must average to zero,
$\left\langle \delta\mathbf{u}\right\rangle =0$,
leading to the following implicit equation for $\widetilde{g}$, 
\begin{equation}
\int_{0}^{\infty}\frac{g-\widetilde{g}}{g_{EM}+g-\widetilde{g}}P\left(g\right)dg=0.\label{eq:EMTSCcond}
\end{equation}
This equation can be solved by first determining  $g_{EM}^{-1}$ as the displacement in response to a unit force with wave vector ${\bf k}$ between directed along nodes $n$ and $ m $, $\mathbf{f}\left(\mathbf{k}\right)=\mathbf{\hat{r}}_{nm}\left(1-e^{i\mathbf{k}\cdot\mathbf{\hat{r}}_{nm}}\right)$,  
by solving the network's equation of motion
\begin{equation}
\mathbf{u}\left(\mathbf{k}\right)=-D^{-1}\left(\mathbf{k}\right)\cdot\mathbf{f}\left(\mathbf{k}\right),
\end{equation}
where $D\left(\mathbf{k}\right)$ is the dynamical matrix of the Bravais lattice, and $\mathbf{u}\left(\mathbf{k}\right)$ is the displacement in $k$-space due to this force.
The displacement of the $nm$ bond due to a unit force, which is needed to solve~\eqref{eq:EMTSCcond}, follows from
\begin{eqnarray}
\label{eq:GEM}
g_{EM}^{-1}  =  \frac{1}{N}\mathbf{r}_{nm}\cdot\underset{\mathbf{k}}{\sum}\mathbf{u}\left(\mathbf{k}\right)\left(e^{-i\mathbf{k}\cdot\mathbf{r}_{nm}}-1\right) 
\end{eqnarray} 
where $N$ is the total number of nodes in the network.
 
For a random bond-diluted lattice with bond stiffness $\mu$ and dilutiom $p$, the self-consistency condition (Eq.~\eqref{eq:EMTSCcond}) can be written as 
\begin{equation}
\label{eq:mEff_diluted}
p\frac{\mu-\widetilde{g}}{g_{EM}+\mu-\widetilde{g}}-\left(1-p\right)\frac{\widetilde{g}}{g_{EM}-\widetilde{g}}=0,
\end{equation} 
From this it can be found that $G$ vanishes continuously at the CF isostatic point as $G\sim\mu \Delta p^\fcf$, with $\Delta p=p-\pcf$, and the mean field results $\pcf=2/3$ and $\fcf=1$. 

A more detailed discussion of this approach can be found in~\cite{Feng1985}. In this reference Feng et al. also describe an alternative, scattering approach~\cite{Lax1951} using the Coherent Potential Approximation (CPA), which leads to the same results as the ``static" approach discussed above.

\begin{figure}
\includegraphics[width=\columnwidth]{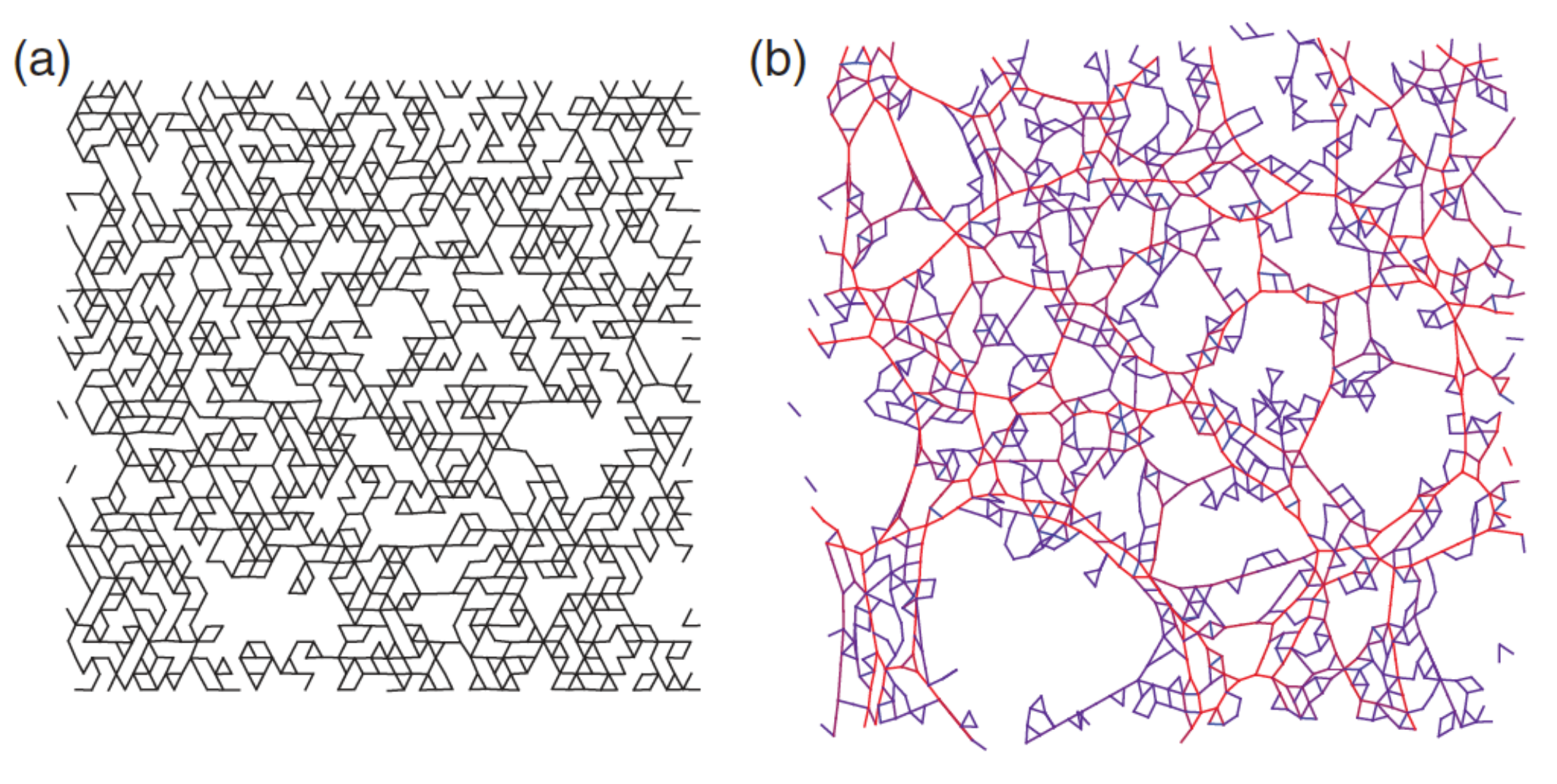}
\caption{ A small section of the undeformed (a) and
expanded (b) diluted triangular lattice. The average coordination
number in this example is $z=3$. Adapted from~\cite{Sheinman2012a}.
}
\label{fig:EMTnonlinear}
\end{figure}

The EMT framework was further developed to describe spring networks for a variety of situations. An EMT was developed to describe ``glasses" with bond-bending forces~\cite{He1985}. To describe the nonlinear elastic response of spring networks under large external isotropic strain (Fig~\ref{fig:EMTnonlinear}), a perturbative approach for infinitesimal dilution was discuss in
~\cite{Tang1988}, and, more recently, for arbitrary dilution in~\cite{Sheinman2012a}. In this approach, the EMT Hamiltonian at finite strain is expanded around a nonlinear state for small nonaffine deformations. It was found that the external strain shifts the isostatic point continuously from $\pcf$ to the (lower) conductivity threshold of the network. Networks with internal stresses~\cite{Alexander1998} garnished attention recently because of their relevance for fiber networks contracted by force-generating molecular motors~\cite{Mizuno2007,Koenderink2009} or contractile cells~\cite{Lam2010}. 
An EMT  approach developed for this scenario was described in~\cite{Sheinman2012b}. Such internal stresses can stabilize subisostatic networks mechanically, and can even poise the network in a critical state. Finally, EMT's were also developed to describe the dynamic shear modulus of a spring network embedded in a viscous medium~\cite{Wyart2010,Lerner2013,Yucht2013}.
%


EMTs for bond-diluted lattices with CF springs are, in principle, straightforward because the springs reside on an individual bond. In contrast, EMTs for lattices with 3-point bending forces are considerably more involved because such bending forces reside on two bonds, whereas the dilution procedure only removes individual bonds one at a time. Two such approaches have been developed to incorporate this effect by Das et al., \cite{Das2007,Das2012} and Mao et al., \cite{Broedersz2011a,Mao2013a,Mao2013b}. Given the complications of 3-point bending forces, there is no single, obvious way to implement the effective medium approach, and these two groups have introduced two different approximations, which will not be elaborated here. However, it is interesting to note that one of these approaches appears to be better at calculating the rigidity threshold \cite{Das2012}, while the other does a better job of capturing the magnitude of the elastic modulus far from this threshold \cite{Broedersz2011a,Mao2013a,Mao2013b}. The latter approach necessitated the inclusion of third-neighbor couplings not present in the earlier approach by Das et al.~\cite{Das2007,Das2012}. Thus, it still remains a challenge to construct an EMT for a fiber network that can accurately capture both the bending threshold and the mechanical response. Furthermore, EMT approaches to describe the nonlinear response of such networks, or their dynamic response when coupled to a viscous liquid, should be of considerable interest, but have not yet been reported.

\subsubsection{Contractile nonaffine and marginal networks}\label{sec:contractilenonaffine}

The nonlinear mechanical response of reconstituted biopolymer networks in many cases reflects the nonlinear force-extension behavior of the constituting cross-links or filaments~\cite{Gardel2004a,Gardel2004b,Storm2005,Wagner2006,Kasza2009,Broedersz2008}. For such networks, both experiments and theoretical studies show that internal stress generation by molecular motors can result in network stiffening in direct analogy to an externally applied uniform stress~\cite{Koenderink2009,Mizuno2007,MacKintosh2008,Levine2009,Liverpool2009,Head2010}. However, as discussed above, the mechanical response of semiflexible polymers is highly anisotropic and is typically much softer to bending than to stretching. In some cases, this renders the network deformation highly nonaffine with most of the energy stored in bending modes. Such nonaffinely deforming stiff polymer networks can also exhibit a nonlinear mechanical response, even when the network constituents are linear elastic fibers. Can internal stresses generated, for instance, by molecular motors or contractile cells embedded in the network, stiffen such networks?

This was studied numerically 2D networks of athermal, stiff filaments using the 2D Phantom model~\cite{Broedersz2011b} (also see Sec.~\ref{sec:phantom}). In the absence of motors, these networks can exhibit strain stiffening under an externally applied shear. This behavior has been attributed to a cross-over between two mechanical regimes; at small strains the mechanics is governed by soft bending modes and a nonaffine deformation field, while at larger strains the elastic response is governed by the stiffer stretch modes and an affine deformation field~\cite{Onck2005}. Interestingly, motors that generate internal stresses can also stiffen the network. The motors induce force dipoles leading to muscle like contractions, which  ``pull out" the floppy bending modes in the system~\cite{Broedersz2011b}. This induces a cross-over to a stiffer stretching dominated regime. Through this mechanism, motors can lead to network stiffening in nonaffine athermal fiber networks in which the constituting filaments in the network are themselves linear elements. 

To obtain a better understanding of this behavior, this was studied in more detail in 3D fiber networks based on the diluted  face centered cubic (fcc) lattice, using both simulations and an analytical approach. Networks are formed by crosslinked straight fibers with linear stretching and bending elasticity. These fibers are organized on a fcc lattice in which a certain fraction of the bonds can randomly be removed. This  allows one to explore a wide range of network connectivities, $0\le z \le 12$. Motor activity is introduced by contractile, static and strain-independent force dipoles acting between neighboring network nodes.
The shear modulus of these networks, with or without contractile stress, can be determined numerically by applying a small shear deformation.

It was found that motors can stabilize the elastic response of otherwise floppy, unstable networks. The motor stress also controls the mechanics of stable networks above a characteristic threshold in connectivity, in the vicinity of which the network exhibits critical strain fluctuations. Interestingly, the network's stiffness is controlled by a coupling of the motor induced stresses $\sigma_m$ to the strain fluctuations according to a constitutive relation,
\begin{equation}
G=G_0+\Gamma(\sigma_m)\times\sigma_m+\sigma_m,
\end{equation}
$G_0$ is the shear modulus in the absence of stress, and where prefactors in the last two terms have been left out. The linear modulus $G_0$ in the absence of stress may be zero.  $\Gamma$ is the nonaffinity parameter proportional to the susceptibility that controls the network response to stress, and which may depend on the stress $\sigma_m$. 

\begin{figure*}
\includegraphics[width=1.8 \columnwidth]{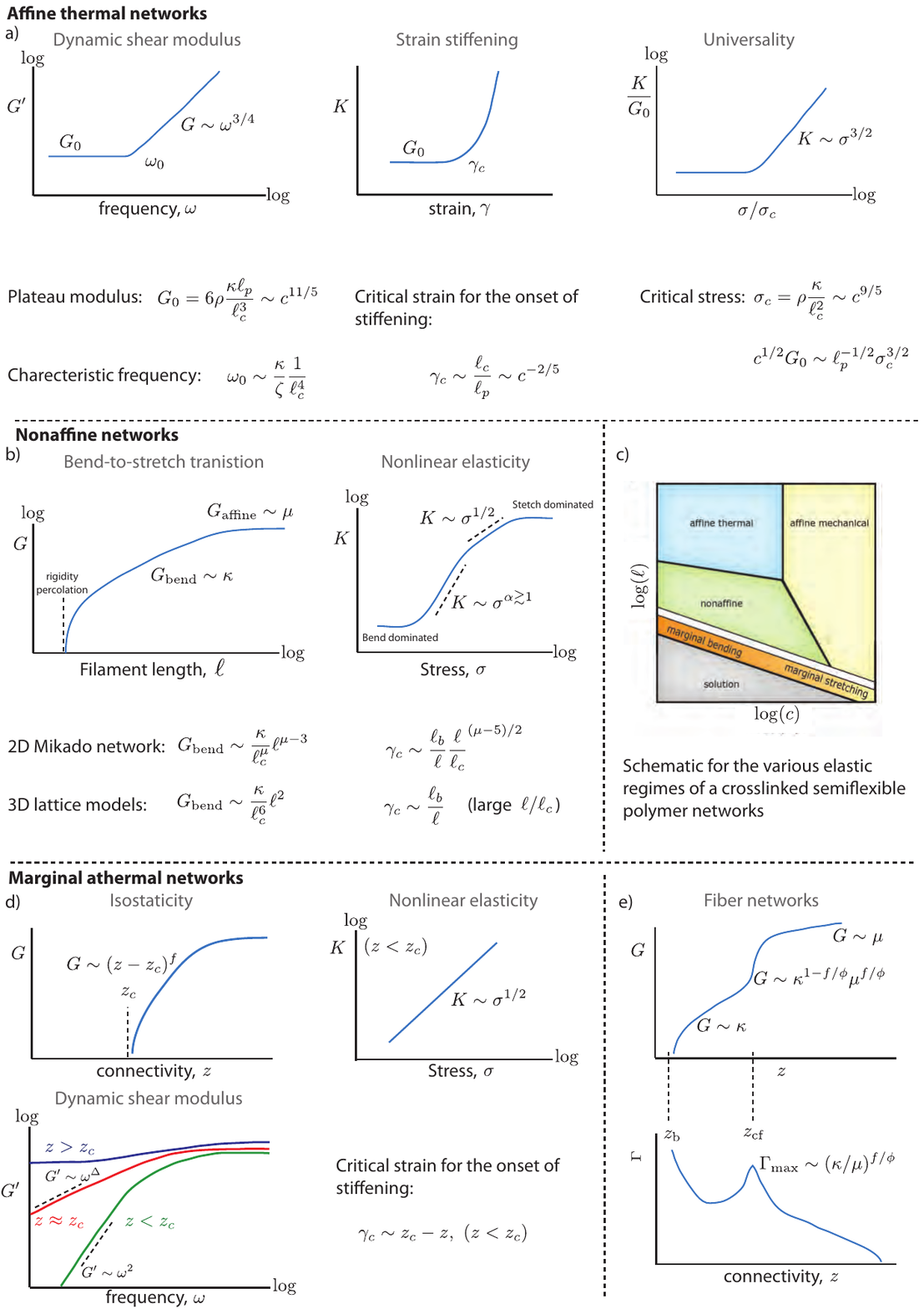}
\caption{Graphical summary for the behavior of crosslinked networks. a) Affine thermal model (Sec.~\ref{sec:affinemodel})\cite{Gardel2004a,Gardel2004b,MacKintosh1995} b,c) Nonaffine fiber networks(Sec.~\ref{sec:NAsection})~\cite{Wilhelm2003,Head2003a,Head2003b,Heussinger2007c,Heussinger2006b,Lubensky2011,Broedersz2012,Onck2005,Broedersz2011b}. d) Marginally stable spring networks (Sec.~\ref{sec:lattice})~\cite{Wyart2008,Lerner2012a,Lerner2012b,Lerner2013,Tighe2012,Tighe2011,Andreotti2012,Wyart2010,Yucht2013}. e) Marginally stable fiber networks ~\cite{Wyart2008, Broedersz2011a,Mao2013a,Mao2013b}.}
\label{fig:graphicalsummary}
\end{figure*}

The coupling between stress and the nonaffine fluctuations gives rise to anomalous regimes at the stability thresholds, at which network criticality implies divergent strain fluctuations with a power law dependences on motor stress. This is reflected as an anomalous dependence of the network's shear modulus on stress. To be in such a critical regime the network needs to be marginally stable. This can be achieved by either tuning the network connectivity such that it is close to the bending or central-force isostatic point, or by adding a near-critical density of motors to marginally stabilize an otherwise floppy network. Importantly, this critical density does not sufficiently enhance the effective connectivity of the network to bring it near to the bending or central-force isostatic point.
In such critical regimes, the shear modulus depends nonlinearly on both motor stress and single filament elasticity
\cite{Lam2010,Broedersz2011b, Chen2011}. Interestingly, this dependence on internal motor stress differs qualitatively from that of an applied external stress.

\section{Summary and outlook}
In this review, we discussed some of the main theoretical developments over roughly the last two decades on semiflexible polymers and their assemblies as bundles, solutions, and crosslinked networks. We focussed on physical and minimal approaches that have studied the basic principles of these systems, with some bias towards biopolymer systems~\cite{Bausch2006,Fletcher2010,Lieleg2010b,Kasza2007} and our personal interests. We have not discussed more realistic approaches that aim to capture some of the specific molecular details of biopolymers and architectural features of the networks they form~\cite{Kim2009}. And, we have only briefly touched upon some fascinating recent examples of applications of some of the ideas coming from various biopolymer studies to synthetic hydrogels \cite{Kouwer2013} and carbon nanotubes \cite{Fakhri2009,Fakhri2010}. Both of these examples are likely just the tip of the iceberg: it seems there is much more to be gained in translation of such ideas to new materials \cite{Bertrand2012}, and one can expect much more work in the future along these lines. 

We started off with a discussion of the properties of single semiflexible polymers. A defining characteristic of semiflexible polymers is that thermal energy only excites small bending fluctuations around their straight zero-temperature conformation. As a result, their mechanical response  is highly anisotropic: Buckling under compression, stiffening entropically under even modest extensions, while bending easily. Moreover, the entropic stretch modulus of a semiflexible polymer governed by its bending rigidity, and is inversely proportional to temperature~\cite{MacKintosh1995,Kroy1996}, in contrast to a flexible polymer which has an entropic modulus proportional to temperature~\cite{Rubinstein2003}. Thus, while flexible polymers are completely dominated by entropy, the properties of semiflexible polymers reflect a competition between the entropy and the bending energy. The important role of bending also has implications for the dynamics of semiflexible filaments, which exhibit a much stronger wavelength dependence of relaxation than for flexible polymers.

The competition between entropy and bending energy in semiflexible polymers has interesting consequences for the assemblies they form. We discussed the intriguing properties of semiflexible bundles~\cite{Claessens2006,Bathe2008,Heussinger2007b,Heussinger2010} and solutions~\cite{Isambert1996,Hinner1998,Morse1998a,Morse1998b,Gittes1997,Schnurr1997}. Various experiments indicate that semiflexible polymer solutions such as entangled actin networks~\cite{Semmrich2007} and living cells~\cite{Deng2006,Fabry2001} exhibit soft glassy behavior~\cite{Sollich1997}, e.g., a weak power-law dependence of the dynamic shear modulus on frequency. The glassy worm like chain model has offered various important insights into such behavior~\cite{Kroy2007,Kroy2008}, but in its current form, this is a phenomenological approach. Thus, the construction of a microscopic theory for glassy semiflexible polymer systems remains an important challenge. 

Networks with dynamic crossllinks have aspects of both solutions and permanently crosslinked networks~\cite{Lieleg2008,Broedersz2010b,Heussinger2011,Strehle2011}. However, these transiently crosslinked networks exhibit a dynamic rheological response distinct from solutions or permanent networks, with a surprising dependence on stress~\cite{Lieleg2010,Yao2013,Norstrom2011}. Experiment on reconstituted actin networks indicate that the onset of stress relaxation shifts to lower frequency in the presence of stress, suggesting that the crosslinks may actually become more stable under an applied force. This may have implications in biological processes such as mechanosensing~\cite{Luo2013}.

The affine model constitutes the simplest analytical approach to describe the mechanical response of a permanently crosslinked semiflexible polymer network~\cite{MacKintosh1995,Morse1998b,Morse1998c,Storm2005}, and this model captures various experiments on reconstituted biopolymer networks~\cite{Gardel2004a,Gardel2004b,Koenderink2006,Tharmann2007,Lin2010a,Lin2010b,Yao2010} and synthetic stiff polymers~\cite{Kouwer2013}. However, the low connectivity ($\lesssim4$) of many biopolymer networks implies that networks are only weakly constrained (especially in 3D), and could deform through nonaffine bending modes~\cite{Head2003a,Head2003b,Wilhelm2003}. In addition, semiflexible polymers are  softer to bending deformations than to stretching deformations, which begs the question: Why would the network not always favor nonaffine deformations? Nonaffine bending deformations can be ``leveraged" by filament length, and thus become large and  energetically less favorable than the affine stretching deformations in the high molecular weight limit~\cite{Heussinger2007c,Heussinger2006b,Head2003a,Head2003b}. Hence, even 3D networks with connectivities $\lesssim4$ can be tuned into a mechanical regime where the shear modulus is governed by affine stretching deformations~\cite{Broedersz2012,Lubensky2011}. We have summarized the main predictions of affine and nonaffine filamentous networks in Fig.~\ref{fig:graphicalsummary}a-c.

We discussed various approaches to describe the nonaffine regime, and the crossover to affine behavior in athermal filamentous networks~\cite{Head2003a,Head2003b,Wilhelm2003,Heussinger2006b,Broedersz2012,Lubensky2011}. Many, if not all, of the elastic regimes of these networks have now been identified and understood, at least at the level of scaling theory. However, a comprehensive analytical theory that describes fiber networks over the full range of behaviors, including the rigidity percolation regime, the length-controlled bending regime, and the affine regime still remains elusive. Theoretical studies have recently started exploring nonlinear~\cite{Onck2005,Conti2009,Broedersz2012,Broedersz2011b,Heussinger2007c}, dynamic~\cite{Huisman2010b}, and thermal effects in nonaffine fibrous networks \cite{Carrillo2013}, but these effects still remain poorly understood. Moreover, experiments have still provided little direct evidence for nonaffine mechanical behavior~\cite{Stein2011,Lieleg2007}, in part probably because unambiguous theoretical predictions have been lacking. To address nonaffine behavior in fibrous networks, various groups have now started combining macroscopic rheological methods with a microscopic visualization of the strain field in the network~\cite{Liu2007,Schmoller2010,Wen2012,Basu2011,Munster2013}.

Nonaffine deformations become paramount when a fiber network becomes marginally stable~\cite{Wyart2008,Broedersz2011a}. This represents and interesting and promising connection between the basic physics of elastic networks and jamming \cite{Hecke2010,Liu1998,Liu2010}. Depending on connectivity, networks can be poised near a marginally stable state analogous to that of granular packings. And, much as compression can stabilize such packings, fiber or biopolymer networks can be stabilized by various interactions or fields, including applied stress, strain, internal molecular motor activity, and even thermal fluctuations, leading to rich critical phenomena \cite{Wyart2008,Broedersz2011a,Sheinman2012a,Sheinman2012b,Tighe2012,Sun2012,Vitelli2012,Dennison2013,Sheinman2012a}. Theoretically, marginally stable fiber networks are predicted to exhibit rich critical behavior, including large, or even divergent, nonaffine strain fluctuations and anomalous elasticity (Fig.~\ref{fig:graphicalsummary}d,e), with close connections to the field of rigidity percolation~\cite{Thorpe1983,Thorpe1985,He1985,Schwartz1985,Feng1984b}, as well as jamming. However, easily tunable experimental fibrous systems to study this behavior are still lacking, and to date, there is little direct evidence on marginal or critical behavior in real biopolymer systems. Recently, however, several groups have begun to study networks in which molecular motors drive the system to effectively lower connectivity \cite{Soares2011,Kohler2011,Murrell2012}, and even to a state resembling a critical point \cite{Alvarado2013}.

\begin{acknowledgments}
This work was supported  in part by a Lewis-Sigler fellowship (CPB), and in part by FOM/NWO (FCM). We thank W.\ Bialek, C.\ Brangwynne, A.\ Bausch, E.\ Conti, M.\ Das, M.\ Dennison, M.\ Depken, B.\ Fabry, E.\ Frey, M.\ Gardel, D.\ Head, C.\ Heussinger, L.\ Huisman, P.\ Janmey, L.\ Jawerth, K.\ Kasza, G.\ Koenderink, A.\ Levine, A.\ Licup, O.\ Lieleg, Y.-C.\ Lin, T.\ Liverpool, T.\ Lubensky, B.\ Machta, X.\ Mao, S.\ M\"unster,  A.\ Sharma, M.\ Sheinman, L.\ Sander, C.\ Schmidt, C.\ Storm, M.\ Tikhonov, B.\ Tighe, D.\ Weitz,  M.\ Wigbers, M.\ Wyart, and N.\ Yao, and M.\ Yucht for many stimulating discussions.\end{acknowledgments}
\nocite{*}
\bibliography{v21.bib}


\end{document}